\documentclass[a4paper,11pt]{article}

\usepackage{amsmath,amsthm,amsbsy,amssymb,amsfonts,graphicx,mathrsfs,bm,accents,cite,xcolor,setspace}
\usepackage[unicode]{hyperref}
\usepackage[utf8]{inputenc}
\usepackage[all]{xy}

\makeatletter
\catcode`\@=11
\@addtoreset{equation}{section}

\makeatother

\renewcommand{\thefootnote}{\arabic{footnote}}

\newcommand{\Exp}[1]{\operatorname{e}^{#1}}
\newcommand{\abs}[1]{\lvert{#1} \rvert}
\newcommand{\rmd}{{\mathrm{d}}}
\newcommand{\nn}{\nonumber}
\newcommand{\Lie}{\pounds}
\newcommand{\gLie}{\hat{\pounds}}

\newcommand{\cB}{\mathcal B}
\newcommand{\cC}{\mathcal C}\newcommand{\cD}{\mathcal D}
\newcommand{\cE}{\mathcal E}\newcommand{\cF}{\mathcal F}
\newcommand{\cG}{\mathcal G}\newcommand{\cH}{\mathcal H}

\newcommand{\cK}{\mathcal K}\newcommand{\cL}{\mathcal L}
\newcommand{\cM}{\mathcal M}
\newcommand{\cP}{\mathcal P}
\newcommand{\cR}{\mathcal R}
\newcommand{\cS}{\mathcal S}

\newcommand{\cZ}{\mathcal Z}

\makeatletter
\newcommand*{\rmT}{{\mathpalette\@transpose{}}}
\newcommand*{\@transpose}[2]{\raisebox{\depth}{$\m@th#1\intercal$}}
\makeatother

\newcommand{\GL}{\text{GL}}
\newcommand{\OO}{\text{O}}
\newcommand{\SO}{\text{SO}}
\newcommand{\SU}{\text{SU}}
\newcommand{\diag}{\mathrm{diag}}
\newcommand{\AdS}{\text{AdS}}
\newcommand{\rmS}{\text{S}}
\newcommand{\TT}{\text{T}}

\newcommand{\cg}{\mathfrak{g}}
\newcommand{\ket}[1]{\lvert {#1}\rangle}
\newcommand{\bra}[1]{\langle {#1} \rvert}
\newcommand{\Pp}{P}
\newcommand{\Pm}{\bar{P}}
\newcommand{\wA}{A}
\newcommand{\bX}{{\bm X}}

\newcommand{\sfd}{\mathsf{d}}
\newcommand{\sfg}{\mathsf{g}}
\newcommand{\sfB}{\mathsf{B}}
\newcommand{\sfE}{\mathsf{E}}
\newcommand{\sfH}{\mathsf{H}}
\newcommand{\sfP}{\mathsf{P}}

\newcommand{\gga}{\mathrm{a}}
\newcommand{\ggb}{\mathrm{b}}
\newcommand{\ggc}{\mathrm{c}}
\newcommand{\ggd}{\mathrm{d}}
\newcommand{\gge}{\mathrm{e}}
\newcommand{\ggf}{\mathrm{f}}

\newcommand{\sla}[1]{\setbox0=\hbox{$#1$} 
\dimen0=\wd0 \setbox1=\hbox{/} \dimen1=\wd1 
\ifdim\dimen0>\dimen1 \rlap{\hbox to \dimen0{\hfil/\hfil}} #1 
\else\rlap{\hbox to \dimen1{\hfil$#1$\hfil}} / \fi}

\newcommand{\signEpsilon}{1}
\if1\signEpsilon
 \def\EPSpos{-}
 \def\EPSneg{}
 \def\EPSplus{-}
 \def\EPSminus{+}
\fi

\allowdisplaybreaks[3]

\begin{document}

\begin{titlepage}
\renewcommand{\thefootnote}{\fnsymbol{footnote}}

\vspace*{1cm}

\centerline{\Large\textbf{Type II DFT solutions from Poisson--Lie $T$-duality/plurality}}%

\vspace{2.0cm}

\centerline{
{\large Yuho Sakatani}%
}

\vspace{0.2cm}

\begin{center}
{\it Department of Physics, Kyoto Prefectural University of Medicine,}\\
{\it Kyoto 606-0823, Japan}\\
{\small\texttt{yuho@koto.kpu-m.ac.jp}}
\end{center}

\vspace*{2mm}

\begin{abstract}
String theory has the $T$-duality symmetry when the target space has Abelian isometries. A generalization of the $T$-duality, where the isometry group is non-Abelian, is known as the non-Abelian $T$-duality, which works well as a solution-generating technique in supergravity. In this paper, we describe the non-Abelian $T$-duality as a kind of $\OO(D,D)$ transformation when the isometry group acts without isotropy. We then provide a duality transformation rule for the Ramond--Ramond fields by using the technique of double field theory (DFT). We also study a more general class of solution-generating technique, the Poisson--Lie (PL) $T$-duality or $T$-plurality. We describe the PL $T$-plurality as an $\OO(n,n)$ transformation and clearly show the covariance of the DFT equations of motion by using the gauged DFT. We further discuss the PL $T$-plurality with spectator fields, and study an application to the $\AdS_5\times\rmS^5$ solution. The dilaton puzzle known in the context of the PL $T$-plurality is resolved with the help of DFT. 
\end{abstract}

\thispagestyle{empty}
\end{titlepage}

\setcounter{footnote}{0}

\begin{spacing}{1.17}
\tableofcontents
\end{spacing}

\section{Introduction}

The $T$-duality was discovered in \cite{Kikkawa:1984cp} as a symmetry of string theory compactified on a torus. 
The mass spectrum or the partition function of string theory on a $D$-dimensional torus was studied for example in \cite{Sakai:1985cs,Nair:1986zn,Ginsparg:1986wr,Giveon:1988tt,Shapere:1988zv} and the $T$-duality was identified as an $\OO(D,D;\mathbb{Z})$ symmetry. 
It was further studied from a different approach \cite{Buscher:1987sk,Buscher:1987qj}, and the transformation rules for the background fields (i.e.~metric, the Kalb--Ramond $B$-field, and the dilaton) under the $T$-duality were determined. 
In \cite{Cecotti:1988zz,Duff:1989tf}, the $T$-duality was understood as an $\OO(D,D)$ symmetry of the classical equations of motion of string theory. 
The classical symmetry was clarified in \cite{hep-th/9110053} by using the gauged sigma model, and this approach has proved quite useful, for example when we discuss the global structure of the $T$-dualized background \cite{hep-th/9309039}. 
The transformation rules for the Ramond--Ramond (R--R) fields and spacetime fermions were determined in \cite{hep-th/9504081,hep-th/9504148,hep-th/9601150,hep-th/9907152}. 
This well-established symmetry of string theory is called the Abelian $T$-duality since it relies on the existence of Killing vectors which commute with each other (see \cite{hep-th/9401139,hep-th/9410237} for reviews). 

An extension of the $T$-duality to the case of non-commuting Killing vectors was explored in \cite{hep-th/9210021} (see \cite{Fridling:1983ha,Fradkin:1984ai} for earlier works), and this is known as the non-Abelian $T$-duality (NATD). 
Various aspects have been studied in \cite{hep-th/9308154,hep-th/9308112,hep-th/9309039,hep-th/9402031,hep-th/9403155,hep-th/9404063,hep-th/9406082,hep-th/9409011,hep-th/9411242,hep-th/9502065,hep-th/9503045,hep-th/9507014,hep-th/9510092,hep-th/9602179,1408.1715}, but unlike the Abelian $T$-duality, there are still many things to be clarified. 
For example, the partition function in the dual model is not the same as that of the original model (see \cite{1805.03657} for a recent study), and NATD may rather be regarded as a map between two string theories. 
The global structure of the dual geometry is also not clearly understood \cite{hep-th/9309039}. 
However, NATD at least generates many new solutions of supergravity, and it can be utilized as a useful solution-generating technique. 

Under NATD, the isometries are generally broken, and naively we cannot recover the original model from the dual model. 
However, this issue was resolved by relaxing the condition for the dualizability \cite{hep-th/9502122}. 
The generalized duality is called the Poisson--Lie (PL) $T$-duality \cite{hep-th/9509095}, and it can be performed even in the absence of the usual Killing vectors. 
The PL $T$-duality is based on a pair of groups with the same dimension, $G$ and $\tilde{G}$\,, that form a larger Lie group known as the Drinfel'd double $\mathfrak{D}$\,. 
The PL $T$-duality is a symmetry that exchanges the role of the subgroups $G$ and $\tilde{G}$\,. 
Conventional NATD can be reproduced as a special case where one of the two groups is an Abelian group. 
Aspects of the PL $T$-duality and generalizations have been studied in \cite{hep-th/9512040,hep-th/9605212,hep-th/9602162,hep-th/9610198,hep-th/9611199,hep-th/9710163,hep-th/9904188,hep-th/0106211,1105.0162}, and concrete applications are given, for example, in \cite{hep-th/9509095,hep-th/9803175,hep-th/9903152,hep-th/0210095,1308.0153}. 

Low-dimensional Drinfel'd doubles were classified in \cite{hep-th/0110139,math/0202209,math/0202210}, and it was stressed that some Drinfel'd double $\mathfrak{d}$ can be decomposed into several different pairs of subalgebras $\cg$ and $\tilde{\cg}$\,, $(\mathfrak{d},\,\cg,\,\tilde{\cg}) \cong (\mathfrak{d},\,\cg',\, \tilde{\cg}') \cong \cdots$\,. 
The decomposition is called the Manin triple, and each Manin triple corresponds to a sigma model. 
The existence of several decompositions suggests that many sigma models are related through a Drinfel'd double. 
This idea was explicitly realized in \cite{hep-th/0205245} and the classical equivalence of the sigma models was called the PL $T$-plurality (see \cite{hep-th/0403164,hep-th/0408126,hep-th/0601172,hep-th/0608069,1201.5939} for more examples). 
Various aspects of the PL $T$-plurality were discussed in \cite{hep-th/0608133,1811.12235} and in particular quantum aspects of the PL $T$-duality/plurality were studied in \cite{hep-th/9509123,hep-th/9512025,hep-th/9803019,hep-th/0205245,hep-th/0304053,0902.1459,0904.4248,0910.0431,0910.1345,1212.5936}. 

Recent developments in NATD were triggered by \cite{1012.1320}, which provided the transformation rule for the R--R fields under NATD. 
Although the analysis was limited to the case where the isometry group acts freely, that restriction was relaxed in \cite{1104.5196}. 
By exploiting the techniques, NATD for an $\SU(2)$ isometry were extensively studied in \cite{1205.2274,1212.1043,1212.4840,1301.6755,1302.2105,1305.7229,1310.1264,1310.1609,1311.4842,1312.4945,1402.3294,1408.0912,1408.6545,1409.7406,1410.2650,1411.7433,1503.00553,1503.07527,1507.02659,1507.02660,1508.06568,1509.04286,1511.00269,1511.05991,1609.09061,1701.01643,1703.00417} (mainly in the context of AdS/CFT correspondence) and many novel solutions were constructed. 
Subsequently, the transformation rules that can also be applied to the fermionic $T$-duality were obtained in \cite{1806.04083}. 

More recently, NATD has received much attention in the context of integrable deformations of string theory, since a class of integrable deformation called the homogeneous Yang--Baxter deformation was shown to be a subclass of NATD \cite{1609.02550,1609.09834,1611.08020,1706.10169}. 
Other integrable deformations such as the $\lambda$-deformation and the $\eta$-deformations can also be understood as a subclass of the so-called $\cE$-model \cite{1508.05832}, which was developed in the PL $T$-duality \cite{hep-th/9512040,hep-th/9502122}. 
Moreover, as discussed in \cite{1504.06303,1504.07213,1506.05784,1508.05832}, the $\lambda$-deformation and the $\eta$-deformations are related by a PL $T$-duality and an analytic continuation. 
Thus, there is a close relationship between the PL $T$-duality and integrable deformations (see \cite{1606.03016,1706.08912,1809.01614,1903.00439} for recent studies on the $\cE$-model). 

Another approach to the $T$-duality has been developed in \cite{hep-th/9302036,hep-th/9305073,hep-th/9308133,0904.4664,0908.1792,1003.5027,1006.4823,1009.2624,1011.1324,1011.4101,1012.2744,1105.6294,1106.5452,1107.0008,1108.4937,1112.5296,1206.3478,1207.4198,1210.5078} and is called the double field theory (DFT). 
This manifests the Abelian $\OO(D,D)$ $T$-duality symmetry at the level of supergravity by formally doubling the dimensions of the spacetime. 
Several formulations of DFT have been proposed, such as the flux formulation (or the gauged DFT) \cite{1109.0290,1109.4280,1201.2924,1304.1472} and DFT on group manifolds (or DFT$_{\text{WZW}}$) \cite{1410.6374,1502.02428,1509.04176}. 
Recently, by applying the idea of DFT$_{\text{WZW}}$, a formulation of DFT which manifests the PL $T$-duality was proposed in \cite{1707.08624} and the transformation of the R--R fields under the PL $T$-duality was discussed for the first time. 
The idea was developed in \cite{1810.11446} and applications to various integrable deformations were studied (see also \cite{1803.03971} for discussion on the PL $T$-duality, $\OO(D,D)$ symmetry, and integrable deformations). 
The covariance of the supergravity equations of motion under the PL $T$-duality was also shown in \cite{1708.04079,1810.07763} using mathematical approaches. 

In this paper, we revisit the traditional NATD in a general setup where the non-vanishing $B$-field and the R--R fields are included. 
By assuming that the isometry group acts freely on the target space, we describe the NATD as a kind of $\OO(D,D)$ rotation of the supergravity fields. 
From the obtained $\OO(D,D)$ matrix, we can easily determine the transformation rule for the R--R fields by using the technique of DFT. 
Indeed, by using the information of given isometry generators, we provide simple duality transformation rules for bosonic fields. 

We then demonstrate the efficiency of the formula by studying some concrete examples. 
Since many examples have already been studied in the literature, in this paper, we will basically consider the cases where the isometry group is non-unimodular $f_{\gga\ggb}{}^\gga\neq 0$\,. 
This type of NATD is not well studied because the resulting dual geometry does not satisfy the supergravity equations of motion \cite{hep-th/9308112,hep-th/9403155,hep-th/9409011}. 
However, as pointed out in \cite{1710.06849,1801.09567}, the dual geometry in fact satisfies the generalized supergravity equations of motion (GSE) \cite{1511.05795,1605.04884}. 
When the target space satisfies the GSE, string theory has the scale invariance \cite{Hull:1985rc,1511.05795} and the $\kappa$-symmetry \cite{1605.04884}. 
The conformal symmetry may be broken, but recently, a local counterterm that cancels out the Weyl anomaly was constructed in \cite{1811.10600} (see also \cite{1703.09213}), and string theory may be consistently defined even in the generalized background. 
Even if it is not the case, NATD for a non-unimodular algebra still works as a solution-generating technique in supergravity, because an arbitrary GSE solution can be mapped to a solution of the usual supergravity \cite{1508.01150,1511.05795,1611.05856,1703.09213} by performing a (formal) $T$-duality. 
Then, combining the NATD with $f_{\gga\ggb}{}^\gga\neq 0$ and the formal $T$-duality, we can generate a new supergravity solution. 

We also study the PL $T$-plurality with the R--R fields. 
In fact, the PL $T$-plurality can be regarded as a constant $\OO(n,n)$ transformation acting on ``untwisted fields'' $\{\hat{\cH}_{AB},\,\hat{d},\,\hat{\cF}\}$\,. 
By requiring the untwisted fields to satisfy the dualizability condition or the $\cE$-model condition of \cite{1811.10600}, we show that the DFT equations of motion in the original and the transformed background are covariantly related by an $\OO(n,n)$ transformation. 
This shows that if the original background satisfies the DFT equations of motion, the transformed background is also a solution of DFT. 
We also discuss the PL $T$-plurality with spectator fields. 
Again, by requiring certain conditions for the untwisted fields, we show that the DFT equations of motion are satisfied in the dual background. 
By using the proposed duality rules, we study an example of the PL $T$-plurality with the R--R fields. 

In studies of the PL $T$-plurality the so-called dilaton puzzle has been discussed in \cite{hep-th/0205245,hep-th/0403164,hep-th/0408126,hep-th/0601172}. 
Under a PL $T$-plurality transformation, a dual-coordinate dependence (i.e.~dependence on the coordinates of the dual group $\tilde{G}$) can appear in the dilaton. 
When such coordinate dependence appears, the background does not have the usual supergravity interpretation, and we are forced to disallow such transformation. 
However, in DFT we can treat the dual coordinates and the usual coordinates on an equal footing and we do not need to worry about the dilaton puzzle. 
As discussed in \cite{1611.05856,1703.09213}, a DFT solution with a dual-coordinate-dependent dilaton can be regarded as a solution of GSE, and by performing a further formal $T$-duality, we can obtain a linear dilaton solution of the usual supergravity. 
In this way, the issue of the dilaton puzzle is totally resolved and we can consider an arbitrary PL $T$-plurality transformation. 

This paper is organized as follows. 
In section \ref{sec:DFT-review}, we briefly review DFT and GSE. 
In section \ref{sec:NATD}, we begin with a review of the traditional NATD, and translate the results into the language of DFT. 
We then provide a general transformation rule for the R--R fields. 
Examples of NATD without and with the R--R fields are studied in section \ref{sec:NATD-example-NS-NS} and \ref{sec:NATD-example-R-R}. 
In section \ref{sec:T-plurality}, we study the PL $T$-plurality in terms of DFT and determine the transformation rules from the DFT equations of motion. 
As an example of the PL $T$-plurality, in section \ref{sec:PL-AdS5}, we study the PL $T$-plurality transformation of $\AdS_5\times \rmS^5$ solution. 
Section \ref{sec:conclusion} is devoted to conclusions and discussions. 

\section{A review of DFT and GSE}
\label{sec:DFT-review}

\subsection*{Generalized-metric formulation of DFT}

There are several equivalent formulations of DFT, but the generalized-metric formulation \cite{1006.4823,1011.1324,1105.6294,1112.5296} may be the most accessible; we thus utilize it as much as possible in this paper. 
In this formulation, the fundamental fields are a symmetric tensor, called the generalized metric $\cH_{MN}(x)$\,, and a scalar density $\Exp{-2\,d(x)}$ called the DFT dilaton. 
The Lagrangian of DFT is given by
\begin{align}
\begin{split}
 &\cL_{\text{DFT}} = \Exp{-2\,d}\cS \,,
\\
 &\cS \equiv \frac{1}{8}\,\cH^{MN}\,\partial_M \cH^{PQ}\partial_N \cH_{PQ} -\frac{1}{2}\, \cH^{PQ}\,\partial_Q\cH^{MN}\,\partial_N \cH_{PM} 
 + 4\,\partial_M d\,\partial_N\cH^{MN} 
\\
 &\qquad -4\,\cH^{MN}\,\partial_M d\,\partial_N d
 - \partial_M\partial_N\cH^{MN} 
 +4\,\cH^{MN}\,\partial_M\partial_N d\,. 
\end{split}
\label{eq:DFT-action}
\end{align}
Here, the fields are supposed to depend on the generalized coordinates $(x^M)=(x^m,\,\tilde{x}_m)$ ($M=1,\dotsc,2\,D$\,, $m=1,\dotsc,D$), and we raise or lower the indices $M,N$ by using the $\OO(D,D)$-invariant metric $\eta_{MN}$ and its inverse $\eta^{MN}$\,:
\begin{align}
 \eta_{MN}=\begin{pmatrix} 0 & \delta_m^n \\ \delta^m_n & 0 \end{pmatrix},\qquad
 \eta^{MN}=\begin{pmatrix} 0 & \delta^m_n \\ \delta_m^n & 0 \end{pmatrix}.
\end{align}
The generalized metric $\cH_{MN}$ is defined to be an $\OO(D,D)$ matrix,
\begin{align}
 \cH_M{}^P\,\cH_N{}^Q\,\eta_{PQ} = \eta_{MN} \,,
\end{align}
and this property allows us to define projection operators as
\begin{align}
 P^{MN} &\equiv \frac{1}{2}\,\bigl(\eta^{MN}+\cH^{MN}\bigr)\,,\qquad 
 \bar{P}^{MN}\equiv \frac{1}{2}\,\bigl(\eta^{MN}-\cH^{MN}\bigr)\,,
\end{align}
which satisfy $P_{M}{}^N +\bar{P}_M{}^N =\delta_M^N$\,. 
For consistency, we assume that arbitrary fields or gauge parameters $A(x)$ and $B(x)$ satisfy the so-called section condition,
\begin{align}
 \eta^{MN}\,\partial_M \partial_N A =0 \,,\qquad 
 \eta^{MN}\,\partial_M A\, \partial_N B =0 \,.
\end{align}
According to this requirement, none of the fields can depend on more than $D$ coordinates. 
Under the section condition, the DFT action is invariant under the generalized Lie derivative
\begin{align}
\begin{split}
 \gLie_V \cH_{MN} &\equiv V^P\,\partial_P \cH_{MN} + \bigl(\partial_M V^P- \partial^P V_M\bigr)\,\cH_{PN} + \bigl(\partial_N V^P- \partial^P V_N\bigr)\,\cH_{MP}\,,
\\
 \gLie_V d &\equiv V^M\,\partial_M d - \frac{1}{2}\,\partial_M V^M \,. 
\end{split}
\end{align}
Namely, the generalized Lie derivative generates the gauge symmetry of DFT, known as the generalized diffeomorphisms. 
Under the section condition, we can also check that the generalized Lie derivative is closed by means of the C-bracket $[\gLie_V,\,\gLie_W] = \gLie_{[V,\,W]_{\text{C}}}$\,, where
\begin{align}
\begin{split}
 [V_1,\,V_2]_{\text{C}}^M &\equiv \frac{1}{2}\,\bigl(\gLie_{V_1} V_2^M - \gLie_{V_2} V_1^M\bigr) 
\\
 &= V_1^N\,\partial_N V_2^M - V_2^N\,\partial_N V_1^M - V_{[1}^N\, \partial^M V_{2]N}\,.
\end{split}
\end{align}
In particular, when the gauge parameters $V_\gga^M$ satisfy $\eta_{MN}\,V_\gga^M\,V_\ggb^N=2\,c_{\gga\ggb}$ ($c_{\gga\ggb}$ a constant)\,, we can show that the C-bracket coincides with the generalized Lie derivative,
\begin{align}
 [V_\gga,\,V_\ggb]_{\text{C}}^M = \gLie_{V_\gga}V_\ggb^M = -\gLie_{V_\ggb}V_\gga^M\,,
\end{align}
similar to the case of the usual Lie derivative $\Lie_{v_\gga}v_\ggb^m=[v_\gga,\,v_\ggb]^m$\,. 

In fact, the scalar $\cS$ in \eqref{eq:DFT-action} can be understood as the generalized Ricci scalar curvature
\begin{align}
 \cS \equiv \frac{1}{2}\,\bigl(P^{MK}\,P^{NL}-\bar{P}^{MK}\,\bar{P}^{NL}\bigr)\, S_{MNKL}\,,
\end{align}
where the (semi-covariant) curvature $S_{MNPQ}$ is defined by
\begin{align}
\begin{split}
 S_{MNPQ} &\equiv R_{MNPQ} + R_{PQMN} - \Gamma_{RMN} \,\Gamma^R{}_{PQ} \,, 
\\
 R_{MNPQ} &\equiv \partial_M \Gamma_{NPQ} -\partial_N \Gamma_{MPQ} 
 +\Gamma_{MPR}\,\Gamma_N{}^R{}_Q - \Gamma_{NPR}\,\Gamma_M{}^R{}_Q\,. 
\end{split}
\end{align}
If we use the curvature $\cS$, the invariance of the DFT action under generalized diffeomorphisms is manifest. 
Then, the DFT action can be understood as a natural generalization of the Einstein--Hilbert action. 

The equations of motion are also summarized in a covariant form as\footnote{They are summarized as $\cG_{MN}\equiv \cS_{MN}-\frac{1}{2}\,\cS\,\cH_{MN} =0$\,. Here, the generalized Einstein tensor $\cG_{MN}$ satisfies the Bianchi identity $\nabla^M \cG_{MN}=0$ \cite{1507.07545}, where $\nabla_M$ is the covariant derivative for the connection $\Gamma_{MNP}$\,.}
\begin{align}
 \cS = 0 \,,\qquad \cS_{MN}=0 \,,
\label{eq:DFT-eom-Ricci}
\end{align}
where the generalized Ricci tensor is defined by
\begin{align}
 \cS_{MN} \equiv \bigl(P_M{}^P\,\bar{P}_N{}^Q+\bar{P}_M{}^P\,P_N{}^Q \bigr)\, S_{RPQ}{}^R \,. 
\end{align}
For concrete computation, the following expression may be more useful:
\begin{align}
 \cS_{MN}&= -2\,\bigl(P_M{}^P\,\bar{P}_N{}^Q + \bar{P}_M{}^P\,P_N{}^Q\bigr)\, \cK_{PQ} \,,
\\
 \cK_{MN} &\equiv \frac{1}{8}\,\partial_M \cH^{PQ}\,\partial_N \cH_{PQ} 
 - \frac{1}{2}\,\partial_{(M|} \cH^{PQ}\,\partial_P \cH_{|N)Q}
 + 2\,\partial_M \partial_N d 
\nn\\
 &\quad + \bigl(\partial_P - 2\,\partial_P d\bigr)\,\Bigl( 
        \frac{1}{2}\,\cH^{PQ}\,\partial_{(M} \cH_{N)Q} 
        + \frac{1}{2}\,\cH^Q{}_{(M|}\,\partial_Q \cH^P{}_{|N)} 
        -\frac{1}{4}\,\cH^{PQ}\,\partial_Q \cH_{MN} \Bigr) \,.
\label{eq:DFT-eom}
\end{align}

When we make the connection to conventional supergravity, we suppose $\tilde{\partial}^m=0$ and parameterize the generalized metric and the DFT dilaton as
\begin{align}
 (\cH_{MN}) = \begin{pmatrix} g_{mn}-B_{mp}\,g^{pq}\,B_{qn} & B_{mp}\,g^{pn} \\
  -g^{mp}\,B_{pn} & g^{mn}
 \end{pmatrix}\,, \qquad 
\Exp{-2\,d}= \Exp{-2\,\Phi}\sqrt{\abs{g}}\,,
\end{align}
by using the standard NS--NS fields $\{g_{mn},\,B_{mn},\,\Phi\}$\,. 
Then, $\cS$ and $\cS_{MN}$ reduce to
\begin{align}
\begin{split}
 \cS &=R + 4\,D^m \partial_m \Phi - 4\,D^m \Phi\,D_m\Phi - \frac{1}{12}\,H_{mnp}\,H^{mnp} \,,
\\
 (\cS_{MN})&= \begin{pmatrix}
 2\,g_{(m|k}\,s^{[kl]} \,B_{l|n)} - s_{(mn)} 
 - B_{mk}\,s^{(kl)}\,B_{ln}\quad & B_{mk}\,s^{(kn)} - g_{mk}\,s^{[kn]} \\
 s^{[mk]}\,g_{kn} -s^{(mk)}\,B_{kn}\quad & s^{(mn)}
 \end{pmatrix} \,,
\\
 s_{mn} &\equiv R_{mn}-\frac{1}{4}\,H_{mpq}\,H_n{}^{pq} + 2 D_m \partial_n \Phi 
 - \frac{1}{2}\,D^k H_{kmn} + \partial_k\Phi\,H^k{}_{mn} \,,
\end{split}
\end{align}
and the standard supergravity Lagrangian
\begin{align}
 \cL = \sqrt{\abs{g}}\Exp{-2\Phi}\Bigl(R + 4\,D^m \partial_m \Phi - 4\,D^m \Phi\,D_m\Phi - \frac{1}{12}\,H_{mnp}\,H^{mnp}\Bigr)\,,
\end{align}
and the equations of motion are reproduced,
\begin{align}
 R + 4\,D^m \partial_m \Phi - 4\,D^m \Phi\,D_m\Phi - \frac{1}{12}\,H_{mnp}\,H^{mnp}=0\,,\quad
 s_{(mn)}=0\,,\quad s_{[mn]}=0\,.
\end{align}

We can also introduce the R--R fields in a manifestly $\OO(D,D)$ covariant manner. 
However, the treatment of the R--R fields is slightly involved, and we will not write out the covariant expression explicitly here (see Appendix \ref{app:DFT}, and also \cite{1703.09213,1803.05903} for the detail). 
In the following, aimed at readers who are not familiar with DFT, we will try to describe the R--R fields as the usual $p$-form fields as much as possible. 

\subsection*{Gauged DFT}

When we manifest the covariance under the PL $T$-plurality, it is convenient to rewrite the DFT equations of motion \eqref{eq:DFT-eom-Ricci} by using the technique of the gauged DFT \cite{1109.0290,1109.4280,1201.2924,1304.1472}. 

Suppose that the generalized metric $\cH_{MN}$ has the form
\begin{align}
 \cH_{MN}(x) = \bigl[U(x)\,\hat{\cH}\,U^{\rmT}(x)\bigr]_{MN}\,, \qquad
 U \equiv (U_M{}^A)\,,
\label{eq:twist-matrix}
\end{align}
where the $\hat{\cH}_{AB}$ is a constant matrix, which we call the untwisted metric. 
In this case, it is useful to define $\cF_{ABC}$ and $\cF_A$\,, called the gaugings or the generalized fluxes, as
\begin{align}
\begin{split}
 &\cF_{ABC} \equiv 3\,\Omega_{[ABC]}\,, \qquad 
 \cF_A \equiv \Omega^B{}_{AB} + 2\, \cD_A d \,,
\\
 &\Omega_{ABC} \equiv - \cD_A U_B{}^M\, U_{MC} = \Omega_{A[BC]} \,,\quad \cD_A \equiv U_A{}^M\,\partial_M \,,\quad 
 U_A{}^M \equiv (U^{-1})_A{}^M\,. 
\end{split}
\label{eq:generalized-fluxes}
\end{align}
They behave as scalars under generalized diffeomorphisms. 

By using the generalized fluxes, we can show that the DFT equations of motion \eqref{eq:DFT-eom-Ricci}, under the section condition, are equivalent to
\begin{align}
 \cR = 0\,,\qquad \cG^{AB} = 0 \,, 
\label{eq:EOM-flux}
\end{align}
where
\begin{align}
\begin{split}
 \cR &\equiv -2\,\Pm^{AB}\, \bigl(2\,\cD_A\cF_B - \cF_A\,\cF_B\bigr) 
 - \frac{1}{3}\,\Pm^{ABCDEF}\,\cF_{ABC}\,\cF_{DEF} \,,
\\
 \cG^{AB} &\equiv -4\,\Pm^{D[A}\, \cD^{B]}\cF_D + 2\,(\cF_D-\cD_D)\,\check{\cF}^{D[AB]} - 2\,\check{\cF}^{CD[A}\,\cF_{CD}{}^{B]} \,. 
\end{split}
\end{align}
Here, we have defined
\begin{align}
\begin{split}
 (\eta_{AB}) &\equiv \begin{pmatrix} 0 & \delta_\gga^\ggb \\ \delta^\gga_\ggb & 0 \end{pmatrix}, \qquad 
 (\eta^{AB}) \equiv \begin{pmatrix} 0 & \delta^\gga_\ggb \\ \delta_\gga^\ggb & 0 \end{pmatrix}, \qquad 
 \check{\cF}^{ABC}\equiv \Pm^{ABCDEF}\,\cF_{DEF}\,,
\\
 P_{AB}&\equiv \frac{1}{2}\,\bigl(\eta_{AB}+\hat{\cH}_{AB}\bigr)\,,\qquad 
 \bar{P}_{AB}\equiv \frac{1}{2}\,\bigl(\eta_{AB}-\hat{\cH}_{AB}\bigr)\,,
\\
 \Pm^{ABCDEF} &\equiv 
   \Pm^{AD}\,\Pm^{BE}\,\Pm^{CF}
 + \Pp^{AD}\,\Pm^{BE}\,\Pm^{CF}
 + \Pm^{AD}\,\Pp^{BE}\,\Pm^{CF}
 + \Pm^{AD}\,\Pm^{BE}\,\Pp^{CF}
\\
  &= \tfrac{1}{4}\,\bigl(\hat{\cH}^{AD}\,\hat{\cH}^{BE}\,\hat{\cH}^{CF}
   - \hat{\cH}^{AD}\,\eta^{BE}\,\eta^{CF}
   - \eta^{AD}\,\hat{\cH}^{BE}\,\eta^{CF}
   - \eta^{AD}\,\eta^{BE}\,\hat{\cH}^{CF}\bigr)
\\
 &\quad + \tfrac{1}{2}\,\eta^{AD}\,\eta^{BE}\,\eta^{CF} \,,
\end{split}
\label{eq:P-check}
\end{align}
and the indices $A,\,B$ are raised or lowered with $\eta_{AB}$ and $\eta^{AB}$\,. 
Under the section condition we can check that $\cR=\cS$\,. 
The equivalence between $\cS_{MN}=0$ and $\cG^{AB}=0$ is slightly more non-trivial, but it is concisely explained in \cite{1304.1472} (see also Appendix \ref{app:DFT}). 

In the flux formulation of DFT \cite{1304.1472}, we take the untwisted metric $\hat{\cH}_{AB}$ as a diagonal Minkowski metric, and then $E_A{}^M \equiv U_A{}^M$ is regarded as the generalized vielbein. 
The fundamental fields are $E_M{}^A$ and $d$\,, and the equations of motion \eqref{eq:EOM-flux} can be derived from
\begin{align}
 \cL = \Exp{-2\,d}\, \cR \,. 
\end{align}
On the other hand, in this paper, we rather interpret \eqref{eq:twist-matrix} as a reduction ansatz and the equations of motion \eqref{eq:EOM-flux} are just rewritings of \eqref{eq:DFT-eom}, similar to the gauged DFT \cite{1109.0290,1109.4280,1201.2924}. 
For our purpose, it is enough to consider the cases where the generalized fluxes are constant. 
In that case, the equations of motion are simple algebraic equations,
\begin{align}
 &\cR = \frac{1}{12}\,\cF_{ABC}\,\cF_{DEF} \,\bigl(3\,\hat{\cH}^{AD}\,\eta^{BE}\,\eta^{CF}- \hat{\cH}^{AD}\,\hat{\cH}^{BE}\,\hat{\cH}^{CF}\bigr) - \hat{\cH}^{AB}\, \cF_A\,\cF_B = 0 \,,
\label{eq:flux-dilaton-eom}
\\
 &\cG^{AB} = \frac{1}{2}\,\bigl(\eta^{CE}\,\eta^{DF} - \hat{\cH}^{CE}\,\hat{\cH}^{DF} \bigr)\,\hat{\cH}^{G[A}\,\cF_{CD}{}^{B]}\,\cF_{EFG} + 2\,\cF_D \,\check{\cF}^{D[AB]} = 0 \,, 
\label{eq:flux-vielbein-eom}
\end{align}
where we have again used the section condition. 

In general, the untwisted metric and the DFT dilaton may depend on the coordinates $y^\mu$ on the uncompactified external spacetime. 
In this case, we denote the extended coordinates as $(x^M)=(y^\mu,\,x^i,\,\tilde{y}_\mu,\,\tilde{x}_i)$ and consider
\begin{align}
 \cH_{MN} = \bigl[U(x^I)\,\hat{\cH}(y^\mu)\,U^{\rmT}(x^I)\bigr]_{MN}\,, \qquad 
 d = \hat{d}(y^\mu) + \sfd(x^I)\,,
\label{eq:SS-ansatz}
\end{align}
where $(x^I)\equiv (x^i,\,\tilde{x}_i)$\,. 
By following \cite{1201.2924}, we assume that $\hat{\cH}_{AB}(y)$ and $\hat{d}(y)$ satisfy
\begin{align}
 \cD_A \hat{\cH}_{BC}(y) = \partial_A \hat{\cH}_{BC}(y)\,,\qquad
 \cD_A \hat{d}(y) = \partial_A \hat{d}(y) \qquad (\partial_A\equiv \delta_A^M\,\partial_M)\,,
\label{eq:external-condition}
\end{align}
and then the generalized Ricci scalar (under the section condition) becomes \cite{1201.2924} [see \eqref{eq:cR-decomp}]
\begin{align}
\begin{split}
 \cS &= \hat{\cS} + \frac{1}{12}\,\cF_{ABC}\,\cF_{DEF} \,\bigl(3\,\hat{\cH}^{AD}\,\eta^{BE}\,\eta^{CF}- \hat{\cH}^{AD}\,\hat{\cH}^{BE}\,\hat{\cH}^{CF}\bigr) - \hat{\cH}^{AB}\, \cF_A\,\cF_B
\\
 &\quad - \frac{1}{2}\,\cF^A{}_{BC}\,\hat{\cH}^{BD}\,\hat{\cH}^{CE}\,\cD_D\hat{\cH}_{AE} + 2\,\cF_A\,\cD_B\hat{\cH}^{AB} - 4\,\cF_A\,\hat{\cH}^{AB}\,\cD_B \hat{d}\,,
\end{split}
\label{eq:generalized-Ricci-gauged}
\end{align}
where $\hat{\cS}$ denotes the generalized Ricci scalar associated with $\{\hat{\cH}_{AB},\,\hat{d}\}$\,, and the fluxes $\cF_{A}$ and $\cF_{ABC}$ are now made of $\{U_M{}^A(x),\, \sfd(x)\}$\,. 
It is important to note that the equation of motion $\cS=0$ is invariant under a constant $\OO(D,D)$ rotation
\begin{align}
 \hat{\cH}_{AB} \to (C\,\hat{\cH}_{AB}\,C^{\rmT})_{AB}\,,\qquad 
 U_A{}^M \to C_A{}^B\,U_B{}^M\,,
\label{eq:gauged-ODD}
\end{align}
which also transforms the generalized fluxes covariantly. 
This transformation looks similar to the PL $T$-plurality discussed in section \ref{sec:T-plurality}, but they are totally different transformations since \eqref{eq:gauged-ODD} does not change $\cH_{MN}$ while the PL $T$-plurality changes $\cH_{MN}$\,. 

\subsection*{GSE from DFT}

As already explained, if we choose a section $\tilde{\partial}^m=0$\,, the DFT equations of motion reproduce the usual supergravity equations of motion. 
On the other hand, we can derive the GSE by choosing another solution of the section condition \cite{1611.05856,1703.09213},
\begin{align}
 &\cH_{MN} = \cH_{MN}(x^m)\,,\qquad d = d_0(x^m) + I^m \,\tilde{x}_m \qquad (I^m \text{ a constant})\,,
\end{align}
where the DFT dilaton has a linear dependence on the dual coordinates. 
In order to satisfy the section condition, we require the vector field $I^m$ to satisfy
\begin{align}
 \gLie_\bX \cH_{MN} = \gLie_\bX d = 0 \,,\qquad (\bX^M) \equiv \begin{pmatrix} I^m \\ 0 \end{pmatrix} ,
\label{eq:fixed-X}
\end{align}
which are equivalent to
\begin{align}
 \bX^P\,\partial_P \cH_{MN} = \bX^P\,\partial_P d = \bX^P\,\partial_P d_0 = 0\,, 
\end{align}
and indeed ensure the section condition,
\begin{align}
 \eta^{PQ}\,\partial_P \cH_{MN}\,\partial_Q d = \bX^P\,\partial_P \cH_{MN} = 0\,,\qquad 
 \eta^{MN}\,\partial_M d\,\partial_N d = 2\,\bX^P\,\partial_P d_0 = 0\,. 
\end{align}
If we choose this section and parameterize $\cH_{MN}$ as usual in terms of $\{g_{mn},\,B_{mn}\}$ and $d_0$ as $\Exp{-2\,d_0}=\Exp{-2\,\Phi}\sqrt{\abs{g}}$\,, the DFT equations of motion (without R--R fields) become
\begin{align}
 &R + 4\,D^m \partial_m \Phi - 4\,\abs{\partial \Phi}^2 - \frac{1}{2}\,\abs{H_3}^2 - 4\,\bigl(I^m I_m+U^m U_m + 2\,U^m\,\partial_m \Phi - D_m U^m\bigr) =0 \,,
\nn\\
 &R_{mn}-\frac{1}{4}\,H_{mpq}\,H_n{}^{pq} + 2 D_m \partial_n \Phi + D_m U_n +D_n U_m = 0 \,,
\\
 &-\frac{1}{2}\,D^k H_{kmn} + \partial_k\Phi\,H^k{}_{mn} + U^k\,H_{kmn} + D_m I_n - D_n I_m = 0 \,,
\nn
\end{align}
where $U_m \equiv I^n\,B_{nm}$\,. 
They are precisely the GSE studied in \cite{Hull:1985rc,1511.05795,1605.04884}. 
When $I^m=0$ (where the Killing equations are trivial), they reduce to the usual supergravity equations of motion. 

Another way to derive the GSE is to make the modification
\begin{align}
 \partial_M d \to \partial_M d + \bX_M\qquad (\bX_M \text{ a generalized vector})\,,
\label{eq:mDFT-shift}
\end{align}
everywhere in the DFT equations of motion \cite{1611.05856}. 
As long as $\bX^M$ satisfies
\begin{align}
 \gLie_\bX \cH_{MN} = \gLie_\bX d = 0 \,,\qquad \eta_{MN}\,\bX^M\,\bX^N = 0\,,
\end{align}
we can choose a gauge such that $\bX^M$ takes the form \eqref{eq:fixed-X} \cite{1703.09213}. 
In terms of the generalized flux, obviously this modification corresponds to
\begin{align}
 \cF_A \to \cF_A + 2\, \bX_A \,,\qquad \bX_A \equiv U_A{}^M\,\bX_M \,. 
\end{align}
Even in the presence of the R--R fields, this replacement is enough to derive the type II GSE, although we additionally need to require the isometry condition for the R--R fields,
\begin{align}
 \Lie_I F = 0 \,.
\end{align}

\subsection*{A formal $T$-duality}

In generalized backgrounds, where the supergravity fields satisfy the GSE, the string theory may not have conformal symmetry. 
Accordingly, when we obtain a generalized background as a result of NATD, it is usually regarded as a problematic example, and such backgrounds have not been considered seriously. 
However, as discussed in \cite{1508.01150,1511.05795,1611.05856,1703.09213}, by performing a formal $T$-duality, we can always transform a generalized background to a linear-dilaton solution of the usual supergravity. 
Here, we review what the formal $T$-duality is. 

The DFT equations of motion are covariant under a constant $\OO(D,D)$ transformation,
\begin{align}
 x^M \to \Lambda^M{}_N\,x^N\,,\qquad 
 \cH_{MN} \to \bigl(\Lambda\,\cH\,\Lambda^{\rmT}\bigr)_{MN}\,, \qquad 
 \partial_M d \to \partial_M d \,. 
\label{eq:formal-Odd}
\end{align}
In particular, if we consider an $\OO(D,D)$ matrix,
\begin{align}
 \Lambda = \begin{pmatrix} \bm{1} - e_z & e_z \\ e_z & \bm{1} - e_z
\end{pmatrix},\qquad e_z \equiv \diag(0,\dotsc,0,\underbrace{1}_{x^z},0,\dotsc,0)\,,
\label{eq:factorized}
\end{align}
it corresponds to the (factorized) $T$-duality along the $x^z$-direction. 
For a given GSE solution with $d=d_0 + I^z\,\tilde{x}_z$\,, the $\OO(D,D)$ rotation \eqref{eq:formal-Odd} with \eqref{eq:factorized} exchanges the coordinates $x^z$ and $\tilde{x}_z$\,, and the dilaton becomes $d=d_0 + I^z\,x^z$\,. 
According to the Killing equation, the generalized metric is independent of $x^z$\,, and the dual coordinate $\tilde{x}_z$ does not appear in the resulting background. 
This means that the GSE background is transformed to a solution of the usual supergravity with a linear dilaton $d=d_0 + I^z\,x^z$\,. 

The reason we call this $\OO(D,D)$ transformation a ``formal'' $T$-duality is as follows. 
The usual Abelian $T$-duality in the presence of $D$ Abelian isometries is an $\OO(D,D)$ transformation,
\begin{align}
 \cH_{MN} \to \Lambda_M{}^P\,\Lambda_N{}^Q\,\cH_{PQ}\,, \qquad 
 \partial_M d \to \partial_M d \,.
\label{eq:T-dual}
\end{align}
The difference from \eqref{eq:formal-Odd} is whether the coordinates are transformed or not. 
If we transform the coordinates, \eqref{eq:formal-Odd} is always a symmetry of the DFT equations of motion even without isometries. 
In the presence of Abelian isometries, due to the coordinate independence, the transformation $x^M \to \Lambda^M{}_N\,x^N$ is trivial and the formal $T$-duality reduces to the usual $T$-duality \eqref{eq:T-dual}. 
To stress the difference, when we perform the transformation \eqref{eq:formal-Odd} with \eqref{eq:factorized} along a non-isometric direction, we call it a formal $T$-duality. 

\section{Non-Abelian $T$-duality}
\label{sec:NATD}

In this section we study the traditional NATD in general curved backgrounds. 
We begin with a review of NATD for the NS--NS sector. 
We then describe the duality as a kind of local $\OO(D,D)$ rotation and provide the general transformation rule for the R--R fields by employing the results of DFT. 
To provide a closed-form expression for the duality rule, we restrict our discussion to the case where we can take a simple gauge choice, $x^i(\sigma)=\text{const.}$

\subsection{NS--NS sector}

In the case of the Abelian $T$-duality, the dual action is obtained with the procedure of \cite{Buscher:1987qj,hep-th/9110053}. 
When a target space has a set of Killing vector fields $v_{\gga}^m$ that commute with each other $[v_{\gga},\,v_{\ggb}]=0$\,, the sigma model has a global symmetry generated by $x^m(\sigma)\to x^m(\sigma) + \epsilon^{\gga}\,v_{\gga}^m(\sigma)$\,. 
This global symmetry can be made a local symmetry by introducing gauge fields $\wA^{\gga}(\sigma)$ and replacing $\rmd x^m \to Dx^m\equiv \rmd x^m - v_\gga^m\,\wA^\gga$\,. 
We also introduce the Lagrange multipliers $\tilde{x}_{\gga}(\sigma)$, which constrain the field strengths to vanish. 
Then, by integrating out the gauge fields $\wA^\gga$ we obtain the dual action, where the Lagrange multipliers $\tilde{x}_{\gga}$ become the embedding functions in the dual geometry. 
In \cite{hep-th/9210021}, this procedure was generalized to the case of non-commuting Killing vectors. 
It was further developed later, and in the following we review NATD in a general setup as discussed in \cite{hep-th/9403155,1408.1715}. 

We consider a target space with $n$ generalized Killing vectors $V_\gga$ ($\gga=1,\dotsc,n$) satisfying
\begin{align}
 \gLie_{V_\gga} \cH_{MN} =0 \,, \qquad 
 [V_\gga,\, V_\ggb]_{\text{C}} = f_{\gga\ggb}{}^\ggc\,V_\ggc \,,\qquad
 \eta_{MN}\,V_\gga^M\,V_\ggb^N = 2\,c_{\gga\ggb}\,, \qquad 
 f_{\gga\ggb}{}^\ggd\,c_{\ggd\ggc} = 0\,. 
\label{eq:setup1}
\end{align}
Here, $c_{\gga\ggb}$ is a constant symmetric matrix. 
If we choose a section $\tilde{\partial}^m=0$ and parameterize the generalized Killing vectors as
\begin{align}
 (V_\gga^M) \equiv \begin{pmatrix} v_\gga^m \\ \tilde{v}_{\gga m} \end{pmatrix} 
 \equiv \begin{pmatrix} v_\gga^m \\ \hat{v}_{\gga m}+B_{mn}\,v_\gga^n \end{pmatrix} ,
\end{align}
these conditions reduce to\footnote{We can easily show $f_{\ggb\ggc}{}^\ggd\,c_{\ggd\gga}+f_{\ggc\gga}{}^\ggd\,c_{\ggd\ggb}=0$ and then the last condition can be expressed as $f_{[\gga\ggb}{}^\ggd\,c_{\ggc]\ggd} = 0$\,. We can further rewrite the same condition as $\frac{1}{3}\,\iota_{v_{\gga}}\iota_{v_{\ggb}}\iota_{v_{\ggc}}H_3 + \iota_{v_{[\gga}}\,f_{\ggb\ggc]}{}^{\ggd}\,\hat{v}_\ggd =0$\,, which was used in \cite{1408.1715}.}
\begin{align}
\begin{split}
 &\Lie_{v_\gga} g_{mn}=0\,,\qquad \iota_{v_\gga} H_3 + \rmd \hat{v}_\gga = 0 \,,\qquad 
 v_{(\gga}\cdot\hat{v}_{\ggb)} = c_{\gga\ggb}\,,
\\
 &\Lie_{v_\gga} v_\ggb = f_{\gga\ggb}{}^\ggc\,v_\ggc\,,\qquad 
 \Lie_{v_\gga} \hat{v}_\ggb = f_{\gga\ggb}{}^\ggc\,\hat{v}_\ggc \,,\qquad 
 f_{\gga\ggb}{}^\ggd\,c_{\ggd\ggc} = 0 \,,
\label{eq:setup2}
\end{split}
\end{align}
where the dot denotes a contraction of the index $m$\,. 
They are precisely the requirements to perform NATD \cite{hep-th/9403155,1408.1715} (see \cite{Hull:1989jk,Hull:1990ms} for the origin of the conditions). 

Under the setup, we consider the gauged action by following the standard procedure \cite{Buscher:1987qj,hep-th/9110053}. 
Ignoring the dilaton term, the gauged action takes the form \cite{Hull:1989jk,Hull:1990ms,hep-th/9403155,1408.1715}
\begin{align}
\begin{split}
 S &\equiv \frac{1}{4\pi\alpha'}\int_\Sigma \bigl(\EPSneg g_{mn}\,D x^m\wedge *\,D x^n 
 -2\,\wA^\gga\wedge \hat{v}_\gga + B_{\gga\ggb}\,\wA^\gga\wedge \wA^\ggb \bigr) 
 +\frac{1}{2\pi\alpha'} \int_\cB H_3
\\
 &\quad + \frac{1}{4\pi\alpha'} \int_\Sigma \bigl(2\,\wA^\gga\wedge\rmd \tilde{x}_\gga + f_{\gga\ggb}{}^\ggc\,\tilde{x}_\ggc\,\wA^\gga\wedge \wA^\ggb \bigr)\qquad 
 (\partial\cB=\Sigma)\,,
\end{split}
\label{eq:gauged-NATD}
\end{align}
where we have introduced gauge fields $\wA^\gga(\sigma)\equiv \wA^\gga_a(\sigma)\,\rmd\sigma^a$ $(a=0,1)$ and have defined
\begin{align}
 D_a x^m\equiv \partial_a x^m - \wA^\gga_a\,v_\gga^m \,, \quad
 F^\gga \equiv \rmd \wA^\gga + \frac{1}{2}\,f_{\ggb\ggc}{}^\gga\,\wA^\ggb\wedge \wA^\ggc \,,\quad
 B_{\gga\ggb} \equiv \hat{v}_{[\gga}\cdot v_{\ggb]} \,. 
\end{align}
Under the conditions \eqref{eq:setup1}, this action is invariant under the local symmetry,
\begin{align}
\begin{split}
 \delta_\epsilon x^m(\sigma) &= \epsilon^\gga(\sigma)\,v_\gga^m(x)\,,\qquad 
 \delta_\epsilon \wA^\gga(\sigma) = \rmd \epsilon^\gga(\sigma) + f_{\ggb\ggc}{}^\gga\,\wA^\ggb(\sigma)\,\epsilon^\ggc(\sigma) \,,
\\
 \delta_\epsilon \tilde{x}_\gga(\sigma) &= c_{\gga\ggb}\,\epsilon^{\ggb}(\sigma) - f_{\gga\ggb}{}^\ggc \, \epsilon^\ggb(\sigma) \, \tilde{x}_\ggc(\sigma) \,.
\end{split}
\label{eq:coordinate-gauge}
\end{align}

If we first use the equations of motion for the Lagrange multipliers $\tilde{x}_\gga$\,, the field strengths $F^\gga$ are constrained to vanish and the gauge fields will become a pure gauge. 
Then, at least locally, we can choose a gauge $\wA^\gga=0$ and the original theory will be recovered,
\begin{align}
 S_0 = \EPSneg \frac{1}{4\pi\alpha'}\int_\Sigma g_{mn}\,\rmd x^m\wedge *\,\rmd x^n 
 +\frac{1}{2\pi\alpha'} \int_\cB H_3 \,.
\end{align}
On the other hand, by using the equations of motion for $\wA^\gga$ first, we obtain the dual model. 
For this purpose, it is convenient to rewrite the action as
\begin{align}
 S &= \EPSneg \frac{1}{4\pi\alpha'}\int_\Sigma g_{mn}\,\rmd x^m\wedge *\,\rmd x^n 
 + \frac{1}{2\pi\alpha'} \int_\Sigma B_2 
\nn\\
 &\quad + \frac{1}{4\pi\alpha'}\int_\Sigma \bigl[
 2\,\wA^\gga \wedge \nu_{\gga} 
 \EPSminus g_{\gga\ggb}\, \wA^\gga \wedge * \wA^\ggb + (B_{\gga\ggb} + f_{\gga\ggb}{}^\ggc\,\tilde{x}_{\ggc})\,\wA^\gga\wedge \wA^\ggb \bigr] \,,
\end{align}
where
\begin{align}
 \nu_{\gga} \equiv \rmd \tilde{x}_{\gga} \EPSplus v_{\gga}^m\,g_{mn}\, *\rmd x^n - \hat{v}_{\gga} \,,\qquad 
 g_{\gga\ggb} \equiv g_{mn}\,v_{\gga}^m\,v_{\ggb}^n\,.
\end{align}
Then, the equations of motion for $\wA^\gga$ become\footnote{They can also be expressed as
\begin{align*}
 \rmd \tilde{x}_{\gga} - (c_{\gga\ggb} - f_{\gga\ggb}{}^\ggc\,\tilde{x}_{\ggc}) \, \wA^\ggb
 = v_{\gga}^m\,\bigl(\EPSneg g_{mn}\, *D x^n + B_{mn}\,D x^n\bigr) + \tilde{v}_{\gga m}\,D x^m\,,
\end{align*}
and reduce to the standard self-duality relation when $\tilde{v}_\gga=0$ and $f_{\gga\ggb}{}^\ggc=0$\,.}
\begin{align}
 \nu_{\gga} = \EPSpos g_{\gga\ggb}\, *\wA^\ggb - (B_{\gga\ggb}+f_{\gga\ggb}{}^\ggc\,\tilde{x}_{\ggc})\, \wA^\ggb \,,
\end{align}
and this can be solved for $\wA^\gga$ as
\begin{align}
 \wA^{\gga} = \EPSpos N^{(\gga\ggb)}\, * \nu_\ggb - N^{[\gga\ggb]}\, \nu_\ggb \,,
\end{align}
where we have defined
\begin{align}
 (N^{\gga\ggb}) \equiv (E_{\gga\ggb}+f_{\gga\ggb}{}^\ggc\,\tilde{x}_{\ggc})^{-1} \,. 
\end{align}
After eliminating the gauge fields, the action becomes
\begin{align}
 S &= \frac{1}{4\pi\alpha'}\int_\Sigma \bigl(\EPSneg g_{mn}\,\rmd x^m\wedge *\,\rmd x^n 
 + B_{mn}\,\rmd x^m\wedge \rmd x^n 
 \EPSminus N^{(\gga\ggb)}\, \nu_\gga\wedge * \nu_{\ggb} + N^{[\gga\ggb]}\, \nu_{\gga} \wedge \nu_\ggb \bigr) 
\nn\\
 &= -\frac{1}{4\pi\alpha'}\int_\Sigma\rmd^2\sigma \sqrt{-\gamma}\,(\gamma^{ab}\EPSplus \varepsilon^{ab})\,\bigl(E_{mn}\,\partial_a x^m\,\partial_b x^n + N^{\gga\ggb}\, \nu_{\gga a}\, \nu_{\ggb b}\bigr) \,,
\label{eq:dual-action1}
\end{align}
where $E_{mn} \equiv g_{mn} + B_{mn}$\,. 
In the above computation, we have assumed that the matrix $(E_{\gga\ggb}+f_{\gga\ggb}{}^\ggc\,\tilde{x}_{\ggc})$ is invertible,\footnote{Note that the invertibility is not ensured even in the Abelian case $f_{\gga\ggb}{}^{\ggc}=0$\,.} but other than that the computation is general. 

Now, a major difference from the Abelian case appears. 
In the Abelian case, by choosing the adapted coordinates $v_{\gga}^m=\delta_{\gga}^m$ we can always realize a gauge $x^{\gga}(\sigma)=0$\,. 
However, in the non-Abelian case, such a gauge choice is not always possible since we cannot realize $v_{\gga}^m=\delta_{\gga}^m$\,. 
In order to provide a closed-form expression for the duality transformation rule, in this paper we assume that the gauge symmetries can be fixed as $x^i(\sigma)=c^i$ ($c^i$ constant) under a suitable decomposition of spacetime coordinates $(x^m)=(y^\mu,\,x^i)$\,. 
This gauge choice removes $n$ coordinates $x^i$ and instead introduces $n$ dual coordinates $\tilde{x}_{\gga}$\,. 
Then, the situation is the same as the Abelian case. 

Under the gauge choice $x^i(\sigma)=c^i$\,, the action \eqref{eq:dual-action1} reproduces the dual action for the dual coordinates $x'^m=(y^\mu,\,\tilde{x}_{\gga})$\,,
\begin{align}
 S &= -\frac{1}{4\pi\alpha'}\int_\Sigma\rmd^2\sigma \sqrt{-\gamma}\,(\gamma^{ab}\EPSplus \varepsilon^{ab})\, E'_{mn}\,\partial_a x'^m\,\partial_b x'^n \,,
\\
 \bigl(E'_{mn}\bigr) &\equiv \begin{pmatrix}
 E_{\mu\nu} - \bigl(v_{\gga \mu} - \hat{v}_{\gga \mu} \bigr)\,N^{\gga\ggb}\, \bigl(v_{\ggb \nu} + \hat{v}_{\ggb \nu} \bigr) & \bigl(v_{\ggc \mu} - \hat{v}_{\ggc \mu} \bigr)\,N^{\ggc\ggb} \\
 - N^{\gga\ggc}\,\bigl(v_{\ggc \nu} + \hat{v}_{\ggc \nu} \bigr) & N^{\gga\ggb} 
\end{pmatrix}\Biggr\rvert_{x^i=c^i} .
\label{eq:dual-BG}
\end{align}
Then, the NATD can be understood as a transformation of the target space geometry,
\begin{align}
 E_{mn} \quad \rightarrow \quad E'_{mn}\,.
\end{align}
Regarding the transformation rule for the dilaton, we employ the result of \cite{hep-th/9210021},
\begin{align}
 \Exp{-2\Phi'} = \frac{1}{\abs{\det (N^{\gga\ggb})}} \Exp{-2\Phi} . 
\label{eq:dual-dilaton}
\end{align}

\subsection{NATD as $\OO(D,D)$ transformation}

In order to show a general transformation rule for the R--R fields, it is convenient to describe NATD as $\OO(D,D)$ rotations. 
Starting with the original background,
\begin{align}
 (E_{mn}) = \begin{pmatrix} E_{\mu\nu} & E_{\mu j} \\ E_{i\nu} & E_{ij} \end{pmatrix},
\end{align}
we construct the dual background \eqref{eq:dual-BG} through the following three steps. 

\begin{enumerate}
\item We first perform a $\GL(D)$ transformation,
\begin{align}
 E \ \to \ E^{(1)} = \Lambda_v\,E\,\Lambda_v^{\rmT} \,,\qquad 
 \Lambda_v \equiv \begin{pmatrix} \delta_\mu^\nu & 0 \\ v_{\gga}^\nu & v_{\gga}^j \end{pmatrix} .
\end{align}
As we have assumed, we can fix the gauge symmetry $\delta_\epsilon x^i = \epsilon^\gga \,v_\gga^i$ such that $x^i(\sigma)=c^i$ is realized. 
For this to be possible, $\det(v_\gga^i)\neq 0$ should be satisfied and the $\GL(D)$ matrix $\Lambda_v$ is invertible. 
We then obtain
\begin{align}
 E^{(1)} = \begin{pmatrix} E_{\mu\nu} & E_{\mu n}\,v^n_{\ggb} \\ v_\gga^{m}\,E_{m\nu} & v_\gga^m\,v_\ggb^n\,E_{mn} \end{pmatrix} 
 = \begin{pmatrix} E_{\mu\nu} & (v_{\ggb \mu}-\hat{v}_{\ggb \mu})+\tilde{v}_{\ggb \mu} \\ (v_{\gga \nu}+\hat{v}_{\gga \nu})-\tilde{v}_{\gga \nu} & E_{\gga\ggb} + v_{[\gga} \cdot \tilde{v}_{\ggb]} \end{pmatrix} ,
\end{align}
where we have used
\begin{align}
 v_{\gga}^m\,B_{m\nu} = \hat{v}_{\gga \nu} - \tilde{v}_{\gga \nu}\,,\qquad B_{\gga\ggb}=\hat{v}_{[\gga}\cdot v_{\ggb]} \,. 
\end{align}

\item We next perform a $B$-transformation,
\begin{align}
 E^{(1)} \ \to\ E^{(2)} \equiv E^{(1)} + \Lambda_f \,,\qquad 
 \Lambda_f\equiv \begin{pmatrix} 0 & -\tilde{v}_{\ggb \mu} \\ \tilde{v}_{\gga \nu} & f_{\gga\ggb}{}^{\ggc}\,\tilde{x}_\ggc - v_{[\gga} \cdot \tilde{v}_{\ggb]} \end{pmatrix} ,
\end{align}
and obtain
\begin{align}
 E^{(2)} = \begin{pmatrix} E_{\mu\nu} & (v_{\ggb \mu}-\hat{v}_{\ggb \mu}) \\ (v_{\gga \nu}+\hat{v}_{\gga \nu}) & E_{\gga\ggb} +f_{\gga\ggb}{}^{\ggc}\,\tilde{x}_\ggc \end{pmatrix} .
\end{align}

\item Finally, we perform a $T$-duality transformation,
\begin{align}
\begin{split}
 E^{(2)} \ &\to\ E^{(3)} \equiv \bigl(\tilde{\Lambda}_{\mathsf{T}} + \Lambda_{\mathsf{T}}\,E^{(2)}\bigr)\,\bigl(\Lambda_{\mathsf{T}}+ \tilde{\Lambda}_{\mathsf{T}}\, E^{(2)}\bigr)^{-1} \,,
\\
 \Lambda_{\mathsf{T}} &\equiv \begin{pmatrix} \bm{1}_{d-n} & \bm{0} \\ \bm{0} & \bm{0} \end{pmatrix},\qquad
 \tilde{\Lambda}_{\mathsf{T}}\equiv \begin{pmatrix} \bm{0} & \bm{0} \\ \bm{0} & \bm{1}_n\end{pmatrix}, 
\end{split}
\end{align}
and obtain
\begin{align}
 E^{(3)} &= \begin{pmatrix} E_{\mu\rho} & (v_{\ggc \mu}-\hat{v}_{\ggc \mu}) \\ 0 & \bm{1} \end{pmatrix}
 \begin{pmatrix} \bm{1} & \bm{0} \\ (v_{\ggc \nu}+\hat{v}_{\ggc \nu}) & E_{\ggc\ggb} +f_{\ggc\ggb}{}^{\ggd}\,\tilde{x}_\ggd \end{pmatrix}^{-1}
\nn\\
 &= \begin{pmatrix}
 E_{\mu\nu} - \bigl(v_{\gga \mu} - \hat{v}_{\gga \mu} \bigr)\,N^{\gga\ggb}\, \bigl(v_{\ggb \nu} + \hat{v}_{\ggb \nu} \bigr) & \bigl(v_{\ggc \mu} - \hat{v}_{\ggc \mu} \bigr)\,N^{\ggc\ggb} \\
 - N^{\gga\ggc}\,\bigl(v_{\ggc \nu} + \hat{v}_{\ggc \nu} \bigr) & N^{\gga\ggb} 
\end{pmatrix} .
\end{align}
By choosing the gauge $x^i=c^i$\,, this precisely reproduces the dual background \eqref{eq:dual-BG}. 
\end{enumerate}

Of course, each step is not a symmetry of supergravity, but this decomposition is useful when we determine the transformation rule of the R--R fields. 
In terms of the generalized metric $\cH_{MN}$, the above NATD is expressed as a local $\OO(D,D)$ transformation,
\begin{align}
\begin{split}
 \cH_{MN} &\to \cH'_{MN} = (h \,\cH \, h^\rmT)_{MN}\bigr\rvert_{x^i=c^i} \,,
\\
 (h_M{}^N) &\equiv 
 \begin{pmatrix}
 \Lambda_{\mathsf{T}} & \tilde{\Lambda}_{\mathsf{T}} \\
 \tilde{\Lambda}_{\mathsf{T}} & \Lambda_{\mathsf{T}}
\end{pmatrix} 
 \begin{pmatrix}
 \bm{1} & \Lambda_f \\
 0&\bm{1}
\end{pmatrix}
 \begin{pmatrix}
 \Lambda_v &0 \\
 0&(\Lambda_v)^{-\rmT} 
\end{pmatrix},
\end{split}
\label{eq:NATD-Odd}
\end{align}
and the $\OO(D,D)$ matrix $h_M{}^N$ can be straightforwardly constructed from the given set of generalized Killing vectors $V_{\gga}=(v_{\gga}^m,\,\tilde{v}_{\gga m})$\,. 

Under a general $\OO(D,D)$ rotation,
\begin{align}
\begin{split}
 \cH_{MN} &\to \cH'_{MN}=(h\,\cH\,h^{\rmT})_{MN}\,,\qquad h_M{}^N \equiv \begin{pmatrix} p_{m}{}^{n} & q_{mn} \\ r^{mn} & s^{m}{}_{n} \end{pmatrix} ,
\\
 E_{mn} &\to E'_{mn} = [(q+p\,E)\,(s+r\,E)^{-1}]_{mn} = [(s^\rmT-E\,r^\rmT)^{-1}\,(- q^\rmT + E\,p^\rmT)]_{mn}\,, 
\end{split}
\end{align}
the determinant of the metric transforms as (see for example \cite{hep-th/9201040})
\begin{align}
 \sqrt{\abs{g}} \to \sqrt{\abs{g'}} = \abs{\det(s+r\,E)}^{-1} \sqrt{\abs{g}} \,. 
\label{eq:detg-formula}
\end{align}
Therefore, under the NATD \eqref{eq:NATD-Odd} we obtain
\begin{align}
 \sqrt{\abs{g'}} &= \abs{\det(\Lambda_{\mathsf{T}}+ \tilde{\Lambda}_{\mathsf{T}}\, E^{(2)})}^{-1}\,\abs{\det(\Lambda_v)} \sqrt{\abs{g}}\ \bigr\rvert_{x^i=c^i}
\nn\\
 &= \abs{\det (N^{\gga\ggb})} \,\abs{\det(v_\gga^i)} \sqrt{\abs{g}}\ \bigr\rvert_{x^i=c^i} \,. 
\end{align}
Combining this with \eqref{eq:dual-dilaton}, we obtain the transformation rule for the DFT dilaton:
\begin{align}
 \Exp{-2\,d'} &= \abs{\det(v_\gga^i)} \Exp{-2\,d}\ \bigr\rvert_{x^i=c^i}\,. 
\label{eq:NATD-DFT-dilaton}
\end{align}
This shows that the DFT dilaton $\Exp{-2\,d}$ transforms covariantly under the $\OO(D,D)$ rotation. 

\subsection{R--R sector}
\label{sec:RR-NATD}

Since the NS--NS fields are transformed covariantly under NATD, it is natural to expect that the R--R fields are also transformed covariantly under the same $\OO(D,D)$ rotation. 
Indeed, as we see from many examples, under NATD $\cH_{MN} \to \cH'_{MN} = (h\,\cH\,h^{\rmT})_{MN}\bigr\rvert_{x^i=c^i}$\,, the generalized Ricci tensors are always transformed covariantly,
\begin{align}
 \cS'_{MN} = (h\,\cS\,h^{\rmT})_{MN}\bigr\rvert_{x^i=c^i}\,,\qquad \cS' = \cS\bigr\rvert_{x^i=c^i} \,. 
\end{align}
This shows that the R--R fields should also transform covariantly, in order to satisfy the equations of motion of type II DFT (see Appendix \ref{app:DFT}),
\begin{align}
 \cS_{MN} = \cE_{MN}\,,\qquad \cS = 0 \,,
\end{align}
where $\cE_{MN}$ is an $\OO(D,D)$-covariant energy--momentum tensor that contains the R--R fields. 

In DFT, the R--R fields are initially studied in \cite{1012.2744}, based on an earlier work \cite{1009.2624} that reproduces the NS--NS part of the DFT action along the line of the $E_{11}$ conjecture \cite{hep-th/0104081,hep-th/0307098}.
Subsequently, two equivalent approaches to describe the R--R fields are developed. 
One treats the R--R fields as an $\OO(D,D)$ spinor \cite{1107.0008}, based on the earlier work \cite{hep-th/9907132}, and the other treats them as an $\OO(D)\times\OO(D)$ bi-spinor \cite{1206.3478}, which is based on the approach of \cite{hep-th/9912236,hep-th/0103149}.

\subsubsection*{R--R fields as a polyform}

We first explain the former because it is simpler. 
Since the treatment of the $\OO(D,D)$ spinor can be rephrased in terms of the differential form, here we treat the R--R field strength as the usual polyform (see Appendices \ref{app:conventions} and \ref{app:DFT} for our convention),
\begin{align}
 F = \sum_{p:\text{even/odd}} \frac{1}{p!}\,F_{m_1\cdots m_p}\,\rmd x^{m_1}\wedge\cdots\wedge\rmd x^{m_p}\qquad (\text{type IIA/IIB})\,. 
\end{align}
Let us summarize the behavior of an $\OO(D,D)$ spinor in terms of the polyform.
\begin{enumerate}
\item 
Under a $\GL(D)$ subgroup of $\OO(D,D)$ transformation,
\begin{align}
 (h_M{}^N) = \begin{pmatrix}
 M &0 \\
 0& M^{-\rmT} 
\end{pmatrix},\qquad M\in \GL(D)\,,
\end{align}
a polyform $F$ transforms as a $\GL(D)$ tensor,
\begin{align}
\begin{split}
 &F' = F^{(M)}\equiv \sum_p \frac{1}{p!}\,F^{(M)}_{m_1\cdots m_p}\,\rmd x^{m_1}\wedge\cdots\wedge\rmd x^{m_p}\,,
\\
 &F^{(M)}_{m_1\cdots m_p} \equiv M_{m_1}{}^{n_1}\cdots M_{m_p}{}^{n_p}\,F_{n_1\cdots n_p}\,.
\end{split}
\end{align}
\item 
Under the $B$-transformation,
\begin{align}
 (h_M{}^N) = \begin{pmatrix}
 \bm{1}_d & \omega \\
 0& \bm{1}_d 
\end{pmatrix},
\end{align}
a polyform $F$ transforms as
\begin{align}
 F' = \Exp{\omega\wedge}F \equiv F+\omega\wedge F+\frac{1}{2!}\,\omega\wedge \omega\wedge F+\cdots \,.
\end{align}
\item 
Under the (factorized) $T$-duality along the $x^m$-direction, it transforms as
\begin{align}
 F' = F \cdot \mathsf{T}_{x^m} \,,\qquad F \cdot \mathsf{T}_{x^m} \equiv F \wedge \rmd \tilde{x}_m + F \vee \rmd x^m \,,
\end{align}
where $\tilde{x}_m$ is the coordinate dual to $x^m$\,, and $\vee \rmd x^m$ denotes the interior product acting from the right. 
\item 
An arbitrary $\OO(D,D)$ transformation can be decomposed into the above three types of transformations, but for later convenience, we also show that under the $\beta$-transformation,
\begin{align}
 (h_M{}^N) = \begin{pmatrix}
 \bm{1}_d & 0 \\
 \chi & \bm{1}_d 
\end{pmatrix},
\end{align}
the transformation rule is given by
\begin{align}
 F' = \Exp{\chi\vee}F \equiv F+\chi\vee F+\frac{1}{2!}\,\chi\vee\chi\vee F+\cdots \,,\qquad 
 \chi\vee F \equiv \frac{1}{2}\,\chi^{mn}\,\iota_m\iota_n\,.
\end{align}
\end{enumerate}

By using the rules, the general formula for the R--R fields under the NATD \eqref{eq:NATD-Odd} becomes
\begin{align}
 F' = \bigl[\,\Exp{\bm{\Lambda_f}\wedge}\, F^{(\Lambda_v)}\, \bigr] \cdot \mathsf{T}_{y^1}\cdots\mathsf{T}_{y^n} \bigr\rvert_{x^i=c^i}\,,\qquad 
 \bm{\Lambda_f} \equiv \frac{1}{2}\,(\Lambda_f)_{mn}\,\rmd x^m\wedge\rmd x^n\,,
\end{align}
where the order of $\mathsf{T}_{y^1}\cdots\mathsf{T}_{y^n}$ is not important since the overall sign flip is a trivial symmetry. 

Note that the field strength $F=\rmd A$ is known as the field strength in the A-basis \cite{hep-th/0103233} (which is sometimes called the Page form). 
Another definition, $G\equiv \rmd C + H_3\wedge C$\,, is known as the C-basis (see Appendix \ref{app:conventions}). 
In the dual background, $G$ can be obtained as
\begin{align}
 G' = \Exp{-B'_2\wedge} F' \,.
\end{align}

We also note that the approach of \cite{1310.1264} based on the Fourier--Mukai transformation (see also \cite{1511.00269} for an application) will be closely related to the procedure explained here. 

\subsubsection*{R--R fields as a bi-spinor}

Next, let us also explain the treatment of the R--R fields as a bi-spinor $\bm{\cG}^\alpha{}_\beta$\,. 
Starting with a polyform $G$\,, by using a vielbein $e_a^m$ associated with $g_{mn}$\,, we define the flat components as $G_{a_1\cdots a_p}=e_{a_1}^{m_1}\cdots e_{a_p}^{m_p}\,G_{m_1\cdots m_p}$ and then define the bi-spinor $\bm{\cG}$ as
\begin{align}
 \bm{\cG} = \sum_{p} \frac{\Exp{\Phi}}{p!}\,G_{a_1\cdots a_p}\,\gamma^{a_1\cdots a_p}\,,
\end{align}
where $\gamma^{a_1\cdots a_p}\equiv \gamma^{[a_1} \cdots \gamma^{a_p]}$ and $(\gamma^a)^\alpha{}_\beta$ is the usual gamma matrix satisfying $\{\gamma^a,\,\gamma^b\}=2\,\eta^{ab}$\,. 
According to \cite{hep-th/9912236,hep-th/0103149,1206.3478} (see also \cite{1803.05903}), under a general $\OO(D,D)$ rotation
\begin{align}
 \cH_{MN} \to (h\,\cH\,h^\rmT)_{MN}\,,\qquad h = \begin{pmatrix} p & q \\ r & s \end{pmatrix},
\end{align}
the bi-spinor transforms as
\begin{align}
 \bm{\cG} \to \bm{\cG}\,\Omega^{-1}\,,
\end{align}
where $\Omega$ is a spinor representation of the Lorentz transformation $\Lambda^a{}_b$\,,
\begin{align}
 \Omega^{-1}\,\bar{\gamma}^a\,\Omega = \Lambda^a{}_b\,\bar{\gamma}^b\,,\qquad 
 \Lambda^a{}_b \equiv \bigl[e^\rmT\,(s+r\,E)^{-1}(s-r\,E^\rmT)\,e^{-\rmT}\bigr]^a{}_b\,,
\end{align}
and $\bar{\gamma}^a\equiv \gamma^{11}\,\gamma^a$\,. 
In particular, under a $T$-duality along a (spatial) $x^z$-direction, we have
\begin{align}
 \Omega = \Omega^{-1} = \frac{e_z^a\,\gamma_a}{\sqrt{g_{zz}}} \,. 
\end{align}
When the vielbein $e_m^a$ has a diagonal form, $\Omega$ is just the gamma matrix $\Omega = \gamma_z$\,. 
The $\Omega$ corresponding to the $\beta$-transformation
\begin{align}
 h = \begin{pmatrix} 1 & 0 \\ \chi & 1 \end{pmatrix},
\end{align}
was obtained in \cite{1803.05903} as
\begin{align}
 \Omega = [\det(\cE'\,\cE)_e{}^f]^{-\frac{1}{2}}\, \AE\bigl(\tfrac{1}{2}\,\beta'^{ab}\,\gamma_{ab}\bigr)\,\AE\bigl(-\tfrac{1}{2}\,\beta^{ab}\,\gamma_{ab}\bigr)\,,
\end{align}
where $\AE$ is similar to an exponential function defined in \cite{hep-th/9912236}
\begin{align}
 \AE \bigl(\tfrac{1}{2}\,\beta^{ab}\,\gamma_{ab}\bigr) \equiv \sum^5_{p=0}\frac{1}{2^{p}\,p!}\, \beta^{a_1a_2}\cdots\beta^{a_{2p-1}a_{2p}}\,\gamma_{a_1\cdots a_{2p}}\,,
\end{align}
the position of the indices $a,\,b$ are changed with $\eta_{ab}$\,, and we have also defined
\begin{align}
\begin{split}
 \cE^{ab} &\equiv e^{am}\,e^{bn}\,E^\rmT_{mn}\,,\quad 
 \beta^{ab} \equiv - \cE^{[ab]}\,,\quad \tilde{e}_m{}^a\equiv e_m{}^b\,(\cE^\rmT)_b{}^a\,,
\\
 \cE'^{ab} &\equiv \tilde{e}^a_m\,\tilde{e}^b_n\,(E^{mn}+\chi^{mn}) \,,\quad 
 \beta'^{ab} \equiv - \cE'^{[ab]}\,. 
\end{split}
\end{align}

Now, let us consider the NATD \eqref{eq:NATD-Odd}. 
Since it is not easy to find a general expression for $\Omega$\,, let us truncate the $B$-field and restrict ourselves to a simple background,
\begin{align}
 (E_{mn}) = \begin{pmatrix} g_{\mu\nu} & 0 \\ 0 & e_i^\gga\,e_j^\ggb\,\eta_{\gga\ggb} \end{pmatrix}.
\end{align}
We also suppose the generalized Killing vectors have simple forms $V_\gga = v_\gga^i\,\partial_i$ $\bigl(v_\gga^i\,e_i^\ggb=\delta_\gga^\ggb\bigr)$\,. 
Then, the vielbein $e_m^a$ has the block-diagonal form
\begin{align}
 (e_m^a) = \begin{pmatrix} \hat{e}_\mu^{\hat{\gga}} & 0 \\ 0 & e_i^\gga \end{pmatrix}, 
\end{align}
and using this, we define the R--R bi-spinor as
\begin{align}
 \bm{\cG} = \sum_p \frac{\Exp{\Phi}}{p!}\,G_{a_1\cdots a_p}\,\gamma^{a_1\cdots a_p}\,, \qquad
 G_{a_1\cdots a_p} \equiv e_{a_1}^{m_1}\cdots e_{a_p}^{m_p}\,G_{m_1\cdots m_p}\,. 
\end{align}

Under the first $\GL(D)$ transformation, $\bm{\cG}$ is invariant while the internal part of the vielbein becomes an identity matrix $e_i^\gga=\delta_i^\gga$\,. 
We next perform the $B$-transformation and $T$-dualities, but it is more useful to perform the $T$-dualities first, because the vielbein is now just an identity matrix. 
Namely, we rewrite the $B$-transformation and $T$-dualities as $T$-dualities and the $\beta$-transformation with parameter $\chi^{\gga\ggb}\equiv f_{\gga\ggb}{}^\ggc\,\tilde{x}_\ggc$\,,
\begin{align}
 \begin{pmatrix}
 \Lambda_{\mathsf{T}} & \tilde{\Lambda}_{\mathsf{T}} \\
 \tilde{\Lambda}_{\mathsf{T}} & \Lambda_{\mathsf{T}}
\end{pmatrix} 
 \begin{pmatrix}
 \bm{1} & \Lambda_f \\
 0&\bm{1}
\end{pmatrix}
 = \begin{pmatrix}
 \bm{1} & \bm{0} \\
 \Lambda_f &\bm{1}
\end{pmatrix}
 \begin{pmatrix}
 \Lambda_{\mathsf{T}} & \tilde{\Lambda}_{\mathsf{T}} \\
 \tilde{\Lambda}_{\mathsf{T}} & \Lambda_{\mathsf{T}}
\end{pmatrix} ,\qquad
 \Lambda_f= \begin{pmatrix} 0 & 0 \\ 0 & \chi^{\gga\ggb} \end{pmatrix} .
\end{align}
Under the $T$-dualities and the $\beta$-transformation, the bi-spinor is transformed as
\begin{align}
 \bm{\cG} \to \bm{\cG}\,\Omega^{-1}\,,\qquad
 \Omega^{-1} = [\det(\delta_{\ggc}^\ggd+\chi_\ggc{}^{\ggd})]^{-\frac{1}{2}}\,\AE\bigl(\tfrac{1}{2}\, \chi^{\gga\ggb}\,\gamma_{\gga\ggb}\bigr) \, \prod_{\gga=1}^n \gamma_{\gga}\,.
\end{align}
This appears to be consistent with the formula given in Eq.~(3.8) of \cite{1104.5196} up to convention. 

If we need to consider the spacetime fermions such as the gravitino and the dilatino, they are also transformed by this $\Omega$\,, and this approach will be important. 
However, in order to determine the transformation rule for the R--R fields, the first approach will be more useful.

\section{Examples without R--R fields}
\label{sec:NATD-example-NS-NS}

In this section we study examples of NATD without the R--R fields. 
In the absence of the R--R fields, our setup is basically the same as the standard one. 
In order to find new solutions, we consider NATD for non-unimodular algebras $f_{\ggb\gga}{}^{\ggb}\neq 0$\,. 

As found in \cite{hep-th/9308112}, in non-unimodular cases, the dual geometry does not solve the supergravity equations of motion. 
However, as recently found in \cite{1710.06849}, the dual geometry is a solution of GSE. 
Additional examples were discussed in \cite{1801.09567}, and there, by using the result of \cite{hep-th/9409011}, it was shown that the Killing vector $I$ in GSE is given by a simple formula,
\begin{align}
 I= f_{\ggb\gga}{}^\ggb\,\tilde{\partial}^{\gga} \,. 
\label{eq:I-in-NATD}
\end{align}
As we reviewed in section \ref{sec:DFT-review}, an arbitrary solution of GSE can be regarded as a solution of DFT with linear dual-coordinate dependence. 
Then, through a formal $T$-duality in DFT, the GSE solution can be mapped to a solution of the conventional supergravity. 
In this section, we generate new solutions of supergravity by combining the NATD for a non-unimodular algebra and the formal $T$-duality. 

In fact, by allowing for non-unimodular algebras, we can perform a rich variety of NATD. 
In order to demonstrate that, we consider several non-Abelian $T$-dualities of a single solution, the $\AdS_3\times\rmS^3\times\TT^4$ background with the $H$-flux. 

\subsection{\texorpdfstring{$\AdS_3 \times \rmS^3\times \TT^4$}{AdS\textthreeinferior{x}S\textthreesuperior{x}T\textfoursuperior}: Example 1}

In the first example, we introduce the coordinates as
\begin{align}
\begin{split}
 \rmd s^2 &= \frac{2\,\rmd x^+\,\rmd x^- + \rmd z^2}{z^2} + \rmd s^2_{\rmS^3} + \rmd s^2_{\TT^4}\,,\qquad 
 B_2 = \frac{\rmd x^+\wedge \rmd x^-}{z^2} + \omega_2 \,,
\\
 \rmd s^2_{\rmS^3}&\equiv \frac{1}{4}\, \bigl[ \rmd\theta^2 + \sin^2 \theta \, \rmd \phi^2 + (\rmd \psi + \cos\theta \, \rmd \phi)^2 \bigr] \,, \qquad
 \omega_2 \equiv - \frac{1}{4}\,\cos\theta\,\rmd\phi\wedge\rmd\psi \,. 
\end{split}
\label{eq:AdS3-S3-LC}
\end{align}
We then consider the generalized isometries generated by two generalized Killing vectors
\begin{align}
\begin{split}
 V_1 &\equiv (v_1,\,\tilde{v}_1)\equiv \bigl(-(x^+)^2\,\partial_+ + \tfrac{z^2}{2}\,\partial_- - x^+\,z\,\partial_z\,,\ \rmd x^+ -\tfrac{x^+}{z}\,\rmd z\bigr)\,,
\\
 V_2 &\equiv (v_2,\,\tilde{v}_2) \equiv \bigl(-x^+\,\partial_+ -\tfrac{z}{2}\,\partial_z\,,\ -\tfrac{1}{2\,z}\,\rmd z\bigr)\,,
\end{split}
\end{align}
which satisfy the algebra $[V_1,\,V_2]_{\text{C}} = V_1$\,. 
The structure constant has the non-vanishing trace $f_{\ggb 2}{}^\ggb = f_{12}{}^1 = 1$\,, and the dual background will be a solution of GSE. 

The $B$-field is not isometric along the $v_1$ direction, $\Lie_{v_1}B_2 \neq 0$\,, and the dual component $\tilde{v}_1$ is necessary to satisfy the generalized Killing equations $\Lie_{v_1}B_2+\rmd\tilde{v}_1=0$\,. 
Moreover, in order to realize $[V_1,\,V_2]_{\text{C}} = V_1$\,, the dual component of $V_2$ is also necessary. 
In this case, we find
\begin{align}
 (c_{\gga\ggb})=\begin{pmatrix} 0 & 0 \\ 0 & \frac{1}{2} \end{pmatrix}\neq 0\,,
\end{align}
but the requirement $f_{\gga\ggb}{}^{\ggd}\,c_{\ggd\ggc}=0$ in \eqref{eq:setup1} is not violated and we can perform the NATD. 
The gauge symmetry associated with the generalized Killing vector $V_2$
\begin{align}
 \delta_{\epsilon^2} x^+(\sigma) = \epsilon^2 \,v_2^+(x) = - \epsilon^2(\sigma)\,x^+(\sigma) \,,
\end{align}
can be fixed by realizing $x^+(\sigma)=1$\,. 
Similarly, the gauge symmetry associated with $V_1$
\begin{align}
 \delta_{\epsilon^1} z(\sigma) = \epsilon^1 \,v_1^z(x)\bigr\vert_{x^+=1} = -\epsilon^1(\sigma)\,z(\sigma) \,,
\end{align}
can be also fixed as $z(\sigma)=1$\,. 

The AdS parts of the matrices in \eqref{eq:NATD-Odd} (before the gauge fixing) become
\begin{align}
 (\Lambda_v) = \begin{pmatrix}
 -(x^+)^2 & \frac{z^2}{2} & - x^+\,z \\
 0 & 1 & 0 \\
 -x^+ & 0 & -\tfrac{z}{2}
 \end{pmatrix},\qquad 
 (\Lambda_f) = \begin{pmatrix}
 0 & 0 & 0 \\
 0 & 0 & \tilde{x}_+ -\frac{x^+}{2} \\
 -\tilde{x}_+ + \frac{x^+}{2} & 0 & 0
 \end{pmatrix} ,
\end{align}
and under the gauge $x^+=1$ and $z=1$\,, the dual background becomes
\begin{align}
\begin{split}
 \rmd s'^2 &= \frac{\rmd \tilde{x}_+^2 + 2\,(1-4\,\tilde{x}_+)\,\rmd \tilde{x}_+\,\rmd x^-}{4\,\tilde{x}_+^2} + \frac{2\,\rmd x^-\,\rmd \tilde{z}}{\tilde{x}_+} +\rmd s^2_{\rmS^3\times\TT^4}\,,\qquad 
 \Exp{-2\,\Phi'}= \tilde{x}_+^2\,,
\\
 B'_2 &= \frac{(1-4\,\tilde{x}_+)\,\rmd \tilde{x}_+\wedge\rmd x^-}{4\,\tilde{x}_+^2} - \frac{(\rmd \tilde{x}_+ +\rmd x^-)\wedge \rmd \tilde{z}}{\tilde{x}_+} + \omega_2 \,. 
\end{split}
\end{align}
As expected, this background does not solve the conventional supergravity equations of motion, but instead satisfies the GSE with the Killing vector
\begin{align}
 I' = f_{\gga \ggb}{}^\gga\,\tilde{\partial}^{\ggb} = \tilde{\partial}^z \,. 
\end{align}

Interestingly, this geometry is locally the same as the original $\AdS_3\times \rmS^3$ spacetime. 
Indeed, by changing coordinates as
\begin{align}
 x'^+ \equiv \tilde{z} - \tilde{x}_+ + \frac{1}{4}\,\ln \tilde{x}_+ \,,\qquad 
 x'^- \equiv x^-\,,\qquad 
 z' \equiv \sqrt{\tilde{x}_+}\,,
\end{align}
we obtain the expressions
\begin{align}
\begin{split}
 \rmd s^2 &= \frac{2\,\rmd x'^+\,\rmd x'^- + \rmd z'^2}{z'^2} + \rmd s^2_{\rmS^3\times\TT^4}\,,\qquad
 \Exp{-2\,\Phi} = z'^4 \,,
\\
 B_2 &= \frac{\rmd x'^+\wedge \rmd x^-}{z'^2} + \frac{2\,\rmd x'^+\wedge\rmd z'}{z'} + \omega_2 \,,\qquad
 I = \partial_+' \,. 
\end{split}
\label{eq:chiral-non-unimodular}
\end{align}
In fact, we can find a two-parameter family of solutions,
\begin{align}
\begin{split}
 \rmd s^2 &= \frac{2\,\rmd x^+\,\rmd x^- + \rmd z^2}{z^2} + \rmd s^2_{\rmS^3\times\TT^4}\,,\qquad
 \Exp{-2\,\Phi} = z^{4\,c_0\,c_1} \,,
\\
 B_2 &= \frac{\rmd x^+\wedge \rmd x^-}{z^2} + \frac{2\,c_1\,\rmd x^+\wedge\rmd z}{z} + \omega_2 \,,\qquad
 I = c_0\,\partial_+ \,,
\end{split}
\label{eq:2-param}
\end{align}
and NATD maps the original solution $(c_0,\,c_1)=(0,0)$ to the dual solution $(c_0,\,c_1)=(1,1)$. 

The metric in \eqref{eq:chiral-non-unimodular} is the same as the original one \eqref{eq:AdS3-S3-LC}, and the $B$-field is also just shifted by a closed form $B_2 \to B_2 + 2\,\rmd x^+\wedge \rmd \ln z$\,. 
The essential difference from the original background is in the dilaton and $I^m$\,. 
We note that, unlike the case of ``trivial solutions'' \cite{1803.07391}, we cannot remove the Killing vector $I^m$ in the dual geometry \eqref{eq:chiral-non-unimodular}.\footnote{According to \cite{1812.07287}, a solution of GSE is a trivial solution (namely, it also satisfies the supergravity equations of motion with $I=0$) only when $\tilde{K}^m\equiv I^n\,B_{np}\,g^{pm}$ satisfies $\Lie_{\tilde{K}}g_{mn}=0$\,, $\Lie_{\tilde{K}}\Phi +(I+\tilde{K})^2 =0$\,, and $\rmd I_1 +\iota_{\tilde{K}}H_3 = 0$ ($I_1\equiv I^m\,g_{mn}\,\rmd x^n$), but they are not satisfied here.} 

It is natural to consider performing a $B$-field gauge transformation in order to undo the shift in the $B$-field. 
However, in the standard GSE, where the only modification is given by the Killing vector $I^m$\,, the gauge symmetry for the $B$-field is already fixed and we cannot perform a $B$-field gauge transformation. 
Indeed, if we truncate the closed form in the $B$-field by hand, we find another solution:
\begin{align}
\begin{split}
 \rmd s^2 &= \frac{2\,\rmd x^+\,\rmd x^- + \rmd z^2}{z^2} + \rmd s^2_{\rmS^3\times\TT^4}\,,\qquad
 \Exp{-2\,\Phi} = z^{4\,c_1} \,,
\\
 B_2 &= \frac{\rmd x^+\wedge \rmd x^-}{z^2} + \omega_2 \,,\qquad
 I = c_0\,\partial_+ \,,
\end{split}
\end{align}
where $c_0$ is a free parameter and $c_1$ can take two values, $c_1=0$ or $c_1=1$\,. 
This is an example of the trivial solution and $c_0$ can be chosen as $c_0=0$\,. 
Then, we get two $\AdS_3\times\rmS^3\times\TT^4$ solutions of the supergravity with two different dilatons, $c_1=0$ and $c_1=1$\,. 

For an arbitrary GSE solution, by taking a coordinate system with $I=I^z\,\partial_z$ we can regard it as a DFT solution with the DFT dilaton $d=d_0 + I^z\,\tilde{x}_z$ ($\Exp{-2\,d_0}\equiv\Exp{-2\,\Phi}\sqrt{\abs{g}}$). 
Then, if we perform a formal $T$-duality that exchanges $\tilde{x}_z$ with the physical coordinate $x^z$\,, we can get a solution of the conventional supergravity where the DFT dilaton is $d=d_0 + I^z\,x^z$\,. 
In the present example \eqref{eq:2-param}, we perform a formal $T$-duality along the $x^+$-direction, and then the DFT dilaton becomes a function of the physical coordinates,
\begin{align}
 \Exp{-2\,d} = \Exp{-2\,c_0\,x^+} z^{4\,c_0\,c_1} \sqrt{\frac{\sin^2\theta}{64\,z^6}} \,. 
\end{align}
Then, the dual-coordinate dependence disappears from the background fields. 
However, in this case the AdS part of the dualized generalized metric becomes
\begin{align}
 (\cH_{MN}) 
 = \begin{pmatrix} g_{mn}-B_{mp}\,g^{pq}\,B_{qn} & B_{mp}\,g^{pn} \\
  -g^{mp}\,B_{pn} & g^{mn}
 \end{pmatrix}
 = {\tiny\left(\begin{array}{ccc|ccc}
 0 & 0 & 0 & 1 & z^2 & 0 \\[1mm]
 0 & 0 & 0 & 0 & -1 & 0 \\[1mm]
 0 & 0 & \frac{1}{z^2} & -\frac{2\,c_1}{z} & -2\,c_1\, z & 0 \\[1mm] \hline
 1 & 0 & -\frac{2\,c_1}{z} & 4\,c_1^2 & 0 & 2\,c_1\, z \\[1mm]
 z^2 & -1 & -2\,c_1\,z & 0 & 0 & 0 \\[1mm]
 0 & 0 & 0 & 2\,c_1\,z & 0 & z^2 
\end{array}\right)} ,
\end{align}
and we cannot extract the supergravity fields $\{g_{mn},\,B_{mn},\,\Phi\}$ from $\cH_{MN}$ due to $\det (g^{mn})=0$\,. 
This type of (genuinely) DFT solution is called the non-Riemannian background \cite{1307.8377}, and is studied in detail in \cite{1508.01121,1707.03713,1808.10605,1902.01867}. 
Using a parameterization given in \cite{1707.03713}, we find that
\begin{align}
\begin{split}
 (\cH_{MN}) 
 &= \begin{pmatrix} \delta_m^p & B_{mp} \\ 0 & \delta^m_p \end{pmatrix}
 \begin{pmatrix} K_{pq} & X^1_p\,Y_1^q - \bar{X}^{\bar{1}}_p\,\bar{Y}_{\bar{1}}^q \\ Y_1^p\,X^1_q - \bar{Y}_{\bar{1}}^p\,\bar{X}^{\bar{1}}_q & H^{pq} \end{pmatrix}
 \begin{pmatrix} \delta^q_n & 0 \\ -B_{qn} & \delta_q^n \end{pmatrix},
\\
 H &= \begin{pmatrix} 4\,c_1^2 & 0& 2\,c_1\,z\\ 0 & 0 & 0\\ 2\,c_1\,z& 0& z^2\end{pmatrix}, \quad K = \begin{pmatrix} \frac{1}{4\,c_1^2} & 0& 0 \\ 0 & 0 & 0\\ 0 & 0& 0 \end{pmatrix}, \quad 
 B = \begin{pmatrix} 0 & 0& -\frac{1}{2\,c_1\,z} \\ 0 & 0 & 0\\ \frac{1}{2\,c_1\,z} & 0& 0 \end{pmatrix},
\\
 X^1 &= \begin{pmatrix} -\frac{z}{2} \\ 0 \\ c_1 \end{pmatrix},\quad
 \bar{X}^{\bar{1}} = \begin{pmatrix} -\frac{z}{2} \\ \frac{1}{z} \\ c_1 \end{pmatrix},\quad
 Y_1 = \begin{pmatrix} 0 \\ -z \\ \frac{1}{c_1}\end{pmatrix},\quad
 \bar{Y}_{\bar{1}} = \begin{pmatrix} 0 \\ z \\ 0 \end{pmatrix}.
\end{split}
\end{align}
In the parameterization of \cite{1707.03713}, there are in general $n$ pairs of vectors $(X^i,\,Y^i)$ and $\tilde{n}$ pairs of vectors $(\bar{X}_{\bar{i}},\,\bar{Y}_{\bar{i}})$\,, and such a non-Riemannian background is called a ($n,\tilde{n}$) solution. 
In this classification, this background is a (1,1) solution. 

In this way, in the first example of NATD, the formal $T$-duality does not produce the usual supergravity solution, and we instead obtain a (1,1) non-Riemannian background. 

\subsection{\texorpdfstring{$\AdS_3 \times \rmS^3\times \TT^4$}{AdS\textthreeinferior{x}S\textthreesuperior{x}T\textfoursuperior}: Example 2}

In the second example, we take the coordinates
\begin{align}
 \rmd s^2 = \frac{-\rmd t^2 + \rmd x^2 + \rmd z^2}{z^2} + \rmd s^2_{\rmS^3\times\TT^4} \,,\qquad 
 B_2 = \frac{\rmd t\wedge \rmd x}{z^2} + \omega_2 \,,
\end{align}
and consider the translation and the dilatation generators as the generalized Killing vectors,
\begin{align}
 V_1 \equiv (v_1,\,\tilde{v}_1)\equiv \bigl(\partial_x\,,\ 0\bigr)\,,
\qquad
 V_2 \equiv (v_2,\,\tilde{v}_2) \equiv \bigl(t\,\partial_t + x\,\partial_x + z\,\partial_z \,,\ 0\bigr)\,,
\end{align}
which satisfy $[V_1,\,V_2]_{\text{C}} = V_1$ and $c_{\gga\ggb}=0$\,. 
Here, we fix the gauge as $x(\sigma)=0$ and $z(\sigma)=1$\,. 

The $\AdS_3$ parts of the transformation matrices are
\begin{align}
 (\Lambda_v) = \begin{pmatrix}
 1 & 0 & 0 \\
 0 & 1 & 0 \\
 t & x & z
 \end{pmatrix},\qquad 
 (\Lambda_f) = \begin{pmatrix}
 0 & 0 & 0 \\
 0 & 0 & \tilde{x} \\
 -\tilde{x} & 0 & 0
 \end{pmatrix} ,
\end{align}
and the NATD gives
\begin{align}
\begin{split}
 \rmd s'^2 &= \frac{-\tilde{x}^2\,\rmd t^2 + 2\,(1-t\,\tilde{x})\,\rmd t\,\rmd x + (1-t^2)\,\rmd \tilde{x}^2 + \rmd \tilde{z}^2}{1-2\,t\,\tilde{x}+\tilde{x}^2} + \rmd s^2_{\rmS^3\times \TT^4} \,,
\\
 B'_2 &= \frac{[\,(t-\tilde{x})\,\rmd \tilde{x} -\tilde{x}\,\rmd t\,]\wedge\rmd \tilde{z}}{1-2\,t\,\tilde{x}+\tilde{x}^2} + \omega_2 \,, \qquad
 \Exp{-2\Phi'} = 1-2\,t\,\tilde{x}+\tilde{x}^2 \,.
\end{split}
\end{align}
This satisfies the GSE by introducing the Killing vector as $I' = f_{\gga \ggb}{}^\gga\,\tilde{\partial}^{\ggb} = \tilde{\partial}^z$\,. 

Again, in order to remove the Killing vector $I$\,, let us perform a formal $T$-duality along the $\tilde{z}$-direction. 
This yields a simple linear-dilaton solution of the supergravity,
\begin{align}
\begin{split}
 \rmd s^2 &= 2\,\rmd t\,\rmd \tilde{x} + \rmd \tilde{x}^2 -2\,\tilde{x}\,\rmd t\,\rmd z + 2\,(t-\tilde{x})\,\rmd \tilde{x}\,\rmd z + \bigl(1-2\,t\,\tilde{x}+\tilde{x}^2\bigr)\,\rmd z^2 + \rmd s^2_{\rmS^3\times \TT^4}\,,
\\
 B_2 &= \omega_2 \,,\qquad \Phi = z\,,
\end{split}
\end{align}
where the AdS part of the $B$-field has disappeared. 

\subsection{\texorpdfstring{$\AdS_3 \times \rmS^3\times \TT^4$}{AdS\textthreeinferior{x}S\textthreesuperior{x}T\textfoursuperior}: Example 3}

We next use the Rindler-type coordinates,
\begin{align}
 \rmd s^2 = \frac{-x^2\,\rmd t^2 + \rmd x^2 +\rmd z^2}{z^2} + \rmd s^2_{\rmS^3\times \TT^4}\,, \qquad
 B_2 = \frac{x\,\rmd t\wedge \rmd x}{z^2} + \omega_2 \,,
\end{align}
and consider the generalized Killing vectors
\begin{align}
 V_1 \equiv (v_1,\,\tilde{v}_1)\equiv \bigl(\partial_t\,,\ 0\bigr)\,, \qquad
 V_2 \equiv (v_2,\,\tilde{v}_2) \equiv \bigl(\Exp{-t}\bigl(x^{-1}\,\partial_t + \partial_x\bigr)\,,\ 0\bigr)\,,
\end{align}
which satisfy $[V_1,\,V_2]_{\text{C}} = -V_2$ and $c_{\gga\ggb}=0$\,. 
Here, we take a gauge $t(\sigma)=0$ and $x(\sigma)=1$\,. 

The AdS parts of the transformation matrices are
\begin{align}
 (\Lambda_v) = \begin{pmatrix}
 1 & 0 & 0 \\
 \Exp{-t}x^{-1} & \Exp{-t} & 0 \\
 0 & 0 & 1
 \end{pmatrix},\qquad 
 (\Lambda_f) = \begin{pmatrix}
 0 & 0 & 0 \\
 0 & 0 & -\tilde{x} \\
 \tilde{x} & 0 & 0
 \end{pmatrix} ,
\end{align}
and the dual background, which satisfies the GSE, becomes
\begin{align}
\begin{split}
 \rmd s'^2 &= \frac{\rmd \tilde{x}^2 -2\,\rmd \tilde{t}\,\rmd\tilde{x}}{\tilde{x}\,(2-\tilde{x}\,z^2)} + \frac{\rmd z^2}{z^2} + \rmd s^2_{\rmS^3\times \TT^4}\,, \qquad
 \Exp{-2\,\Phi'} = \frac{\tilde{x}\,(\tilde{x}\,z^2-2)}{z^2} \,,
\\
 B'_2 &= \frac{1-\tilde{x}\,z^2}{\tilde{x}\,(2-\tilde{x}\,z^2)}\,\rmd \tilde{t}\wedge\rmd\tilde{x} + \omega_2 \,,\qquad 
 I' = \tilde{\partial}^t\,.
\end{split}
\end{align}

In order to obtain a solution of the supergravity, we again perform a formal $T$-duality along the $\tilde{t}$-direction. 
Again we find a non-Riemannian background,
\begin{align}
 (\cH_{MN}) ={\footnotesize\left(\begin{array}{ccc|ccc}
 \tilde{x}\,(\tilde{x}\,z^2-2) & 1-\tilde{x}\,z^2 & 0 & \tilde{x}\,z^2-1 & \tilde{x}\,(\tilde{x}\,z^2-2) & 0 \\
 1-\tilde{x}\,z^2 & z^2 & 0 & -z^2 & 1-\tilde{x}\,z^2 & 0 \\
 0 & 0 & \frac{1}{z^2} & 0 & 0 & 0 \\ \hline
 \tilde{x}\,z^2-1 & -z^2 & 0 & 0 & 0 & 0 \\
 \tilde{x} (\tilde{x}\,z^2-2) & 1-\tilde{x}\,z^2 & 0 & 0 & 0 & 0 \\
 0 & 0 & 0 & 0 & 0 & z^2 
\end{array}\right)},
\end{align}
where the $\rmS^3\times \TT^4$ part of the generalized metric is not displayed. 
This is also a (1,1) solution,
\begin{align}
\begin{split}
 (\cH_{MN}) 
 &= \begin{pmatrix} \delta_m^p & B_{mp} \\ 0 & \delta^m_p \end{pmatrix}
 \begin{pmatrix} K_{pq} & X^1_p\,Y_1^q - \bar{X}^{\bar{1}}_p\,\bar{Y}_{\bar{1}}^q \\ Y_1^p\,X^1_q - \bar{Y}_{\bar{1}}^p\,\bar{X}^{\bar{1}}_q & H^{pq} \end{pmatrix}
 \begin{pmatrix} \delta^q_n & 0 \\ -B_{qn} & \delta_q^n \end{pmatrix},
\\
 H &= \begin{pmatrix} 0 & 0& 0 \\ 0 & 0 & 0\\ 0 & 0& z^2 \end{pmatrix}, \quad K = \begin{pmatrix} 0 & 0& 0 \\ 0 & 0 & 0\\ 0 & 0& \frac{1}{z^2} \end{pmatrix}, \quad 
 B = \begin{pmatrix} 0 & -\frac{1}{2}& 0 \\ \frac{1}{2} & 0 & 0\\ 0 & 0& 0 \end{pmatrix},
\\
 X^1 &= \begin{pmatrix} \frac{\tilde{x}\,z^2}{2} \\ -\frac{z^2}{2} \\ 0 \end{pmatrix},\quad
 \bar{X}^{\bar{1}} = \begin{pmatrix} \frac{2}{z^2}-\tilde{x} \\ 1 \\ 0 \end{pmatrix},\quad
 Y_1 = \begin{pmatrix} 1 \\ \tilde{x}-\frac{2}{z^2} \\ 0 \end{pmatrix},\quad
 \bar{Y}_{\bar{1}} = \begin{pmatrix} \frac{z^2}{2} \\ \frac{\tilde{x}\,z^2}{2} \\ 0 \end{pmatrix}.
\end{split}
\end{align}

To briefly summarize, NATD works well as a solution-generating technique of DFT even if the isometry algebra is non-unimodular. 
If we additionally perform a formal $T$-duality, we usually obtain the usual supergravity solution. 
Sometimes, the parameterization of the generalized metric becomes singular and we obtain a non-Riemannian background, which does not have the usual supergravity interpretation. 
However, they are interesting backgrounds by themselves, as discussed in \cite{1508.01121,1707.03713,1808.10605,1902.01867}. 
Therefore, it is important to study NATD for non-unimodular algebras more seriously. 

\section{Examples with R--R fields}
\label{sec:NATD-example-R-R}

In this section we consider NATD with non-vanishing R--R fields. 
After reproducing a known example, we again consider examples for non-unimodular algebras. 

For convenience, let us display the summary of the duality rules. 
Under the setup
\begin{align}
 \gLie_{V_\gga} \cH_{MN} =0 \,, \qquad 
 [V_\gga,\, V_\ggb]_{\text{C}} = f_{\gga\ggb}{}^\ggc\,V_\ggc \,,\qquad
 \eta_{MN}\,V_\gga^M\,V_\ggb^N = 2\,c_{\gga\ggb}\,, \qquad 
 f_{\gga\ggb}{}^\ggd\,c_{\ggd\ggc} = 0\,,
\end{align}
where $(V_\gga^M)=(v_\gga^m,\,\tilde{v}_{\gga m})$\,, the dual background is given by
\begin{align}
\begin{split}
 \cH'_{MN} &= (h \,\cH \, h^\rmT)_{MN}\bigr\rvert_{x^i=c^i} \,, \qquad 
 \Exp{-2\,d'} = \abs{\det(v_\gga^i)} \Exp{-2\,d}\ \bigr\rvert_{x^i=c^i}\,,
\\
 F' &= \bigl[\,\Exp{\bm{\Lambda_f}\wedge}\, F^{(\Lambda_v)}\, \bigr] \cdot \mathsf{T}_{y^1}\cdots\mathsf{T}_{y^n} \bigr\rvert_{x^i=c^i}\,, \qquad 
 I= f_{\ggb\gga}{}^\ggb\,\tilde{\partial}^{\gga} \,,
\label{eq:NATD-summary1}
\end{split}
\end{align}
where
\begin{align}
 &(h_M{}^N) \equiv 
 \begin{pmatrix}
 \Lambda_{\mathsf{T}} & \tilde{\Lambda}_{\mathsf{T}} \\
 \tilde{\Lambda}_{\mathsf{T}} & \Lambda_{\mathsf{T}}
\end{pmatrix} 
 \begin{pmatrix}
 \bm{1} & \Lambda_f \\
 0&\bm{1}
\end{pmatrix}
 \begin{pmatrix}
 \Lambda_v &0 \\
 0&(\Lambda_v)^{-\rmT} 
\end{pmatrix},\qquad
 \bm{\Lambda_f} \equiv \frac{1}{2}\,(\Lambda_f)_{mn}\,\rmd x^m\wedge\rmd x^n\,,
\label{eq:NATD-summary2}
\\
 &\Lambda_v \equiv \begin{pmatrix} \delta_\mu^\nu & 0 \\ v_{\gga}^\nu & v_{\gga}^j \end{pmatrix} ,\quad
 \Lambda_f\equiv \begin{pmatrix} 0 & -\tilde{v}_{\ggb \mu} \\ \tilde{v}_{\gga \nu} & f_{\gga\ggb}{}^{\ggc}\,\tilde{x}_\ggc - v_{[\gga} \cdot \tilde{v}_{\ggb]} \end{pmatrix} ,\quad
 \Lambda_{\mathsf{T}} \equiv \begin{pmatrix} \bm{1}_{d-n} & \bm{0} \\ \bm{0} & \bm{0} \end{pmatrix},\quad
 \tilde{\Lambda}_{\mathsf{T}}\equiv \begin{pmatrix} \bm{0} & \bm{0} \\ \bm{0} & \bm{1}_n\end{pmatrix},
\nn
\end{align}
and the coordinates are transformed as $(x^m)=(y^\mu,\,x^i) \to (x'^m) =(y^\mu,\,\tilde{x}_\gga)$\,. 

\subsection{\texorpdfstring{$\AdS_3 \times \rmS^3\times \TT^4$}{AdS\textthreeinferior{x}S\textthreesuperior{x}T\textfoursuperior}}

As the first example of NATD with the R--R fields, let us review the example of \cite{1012.1320} and demonstrate that our formula gives the same result. 
The original background is
\begin{align}
\begin{split}
 \rmd s^2 &= \frac{-\rmd t^2+\rmd x^2 +\rmd z^2}{\ell^2\,z^2} + \frac{1}{4\,\ell^2}\, \bigl[ \rmd\theta^2 + \sin^2 \theta \, \rmd \phi^2 + (\rmd \psi + \cos\theta \, \rmd \phi)^2 \bigr] +\rmd s^2_{\TT^4}\,,
\\
 G_3 &= \frac{2\,\rmd t\wedge\rmd x\wedge \rmd z}{\ell^2\,z^3} - \frac{\sin\theta}{4\,\ell^2}\,\rmd\theta\wedge\rmd\phi\wedge\rmd\psi\,,
\end{split}
\end{align}
where the $\AdS_3$ and $\rmS^3$ part have the curvature $R=\mp 6\,\ell^2$\,, respectively. 

We perform NATD associated with three generalized Killing vectors on the $\rmS^3$\,,
\begin{align}
\begin{split}
 V_1 &= \bigl(\cos\psi\,\partial_\theta+\tfrac{\sin\psi}{\sin\theta}\,\partial_\phi-\tfrac{\sin\psi}{\tan\theta}\,\partial_\psi\,,\,\,0\bigr)\,,
\\
 V_2 &= \bigl(-\sin\psi\,\partial_\theta+\tfrac{\cos\psi}{\sin\theta}\,\partial_\phi-\tfrac{\cos\psi}{\tan\theta}\,\partial_\psi\,,\,\,0\bigr)\,,\qquad
 V_3 = (\partial_\psi\,,\,0)\,,
\end{split}
\end{align}
which satisfy
\begin{align}
 [V_1,\,V_2]_{\text{C}} = V_3\,,\qquad
 [V_2,\,V_3]_{\text{C}} = V_1\,,\qquad
 [V_3,\,V_1]_{\text{C}} = V_2\,. 
\end{align}
As is clear from the explicit form of the Killing vectors, we can choose a gauge
\begin{align}
 \theta(\sigma)=\frac{\pi}{2}\,,\qquad
 \phi(\sigma)=0\,,\qquad
 \psi(\sigma)=0\,.
\label{eq:RR1-gauge}
\end{align}

The $(\theta,\,\phi,\,\psi)$ parts of the transformation matrices are
\begin{align}
 (\Lambda_v) = \begin{pmatrix}
 \cos\psi & \tfrac{\sin\psi}{\sin\theta} & -\tfrac{\sin\psi}{\tan\theta}\\
 -\sin\psi & \tfrac{\cos\psi}{\sin\theta} & -\tfrac{\cos\psi}{\tan\theta}\\
 0 & 0 &1
 \end{pmatrix},\qquad 
 (\Lambda_f) = \begin{pmatrix}
 0 & \tilde{\psi} & -\tilde{\phi} \\
 -\tilde{\psi} & 0 & \tilde{\theta} \\
 \tilde{\phi} & -\tilde{\theta} & 0
 \end{pmatrix} ,
\end{align}
and the NS--NS fields in the dual background are
\begin{align}
\begin{split}
 \rmd s'^2 &= \frac{-\rmd t^2+\rmd x^2 +\rmd z^2}{\ell^2\,z^2} 
 + \frac{4\,\ell^2\,\bigl(\delta_{ij} + 16\,\ell^4\,u_i\,u_j\bigr)\,\rmd u^i\,\rmd u^j}{1+16\,\ell^4\,u_k\,u^k}
 +\rmd s^2_{T^4}\,, 
\\
 B'_2 &= -\frac{8\,\ell^4\, \epsilon_{ijk}\,u^i\,\rmd u^j\wedge \rmd u^k}{1+16\,\ell^4\,u_k\,u^k}\,,\qquad 
 \Exp{-2\Phi'} = \frac{1+16\,\ell^4\,u_k\,u^k}{64\,\ell^6}\,,
\end{split}
\end{align}
where we have denoted $(u^i)\equiv (\tilde{\theta},\,\tilde{\phi},\,\tilde{\psi})$\,, $u_i\equiv u^i$\,, and $\epsilon_{123}=1$\,. 

Now, let us consider the R--R fields. 
Under the gauge \eqref{eq:RR1-gauge}, the Page form becomes
\begin{align}
 F = \Bigl(\frac{2\,\rmd t\wedge\rmd x\wedge \rmd z}{\ell^2\,z^3} - \frac{\rmd\theta\wedge\rmd\phi\wedge\rmd\psi}{4\,\ell^2}\Bigr)\wedge\bigl[1 - \text{vol}(T^4) \bigr] \,.
\label{eq:Page-under-gauge}
\end{align}
The first $\GL(D)$ transformation is trivial, $\Lambda_v = \bm{1}$\,, under the gauge \eqref{eq:RR1-gauge}. 
We next perform the $B$-transformation $F \to \Exp{\bm{\Lambda_f}\wedge} F$ where
\begin{align}
 \bm{\Lambda_f} = u^1\,\rmd\phi\wedge\rmd\psi + u^2\,\rmd\psi\wedge\rmd\theta + u^3\,\rmd\theta\wedge\rmd\phi \,.
\end{align}
Finally, by performing $T$-dualities along the $(\theta,\, \phi,\,\psi)$-directions, we obtain
\begin{align}
 F' &= \bigl(F + \bm{\Lambda_f}\wedge F\bigr)\,(\wedge \rmd u^1 +\vee \rmd \theta)\,(\wedge \rmd u^2 +\vee \rmd \phi)\,(\wedge \rmd u^3 +\vee \rmd \psi) 
\nn\\
 &= \Bigl[\frac{1}{4\,\ell^2} - \frac{2\,\rmd t\wedge \rmd x\wedge\rmd z\wedge (u_i\,\rmd u^i - \rmd u^1\wedge\rmd u^2\wedge\rmd u^3)}{\ell^2\,z^3}\Bigr]\wedge\bigl[1- \text{vol}(T^4)\bigr] \,.
\label{eq:Page-example1}
\end{align}
From this Page form we get the R--R field strengths in the C-basis as
\begin{align}
\begin{split}
 G'_0 &= \frac{1}{4\,\ell^2}\,,\qquad 
 G'_2 = \frac{2\,\ell^2\,\epsilon_{ijk}\,u^i\,\rmd u^j\wedge\rmd u^k}{1+16\,\ell^4\,u_l\,u^l}\,,
\\
 G'_4 &= - \frac{2\,\rmd t\wedge \rmd x\wedge\rmd z\wedge u_i\,\rmd u^i}{\ell^2\,z^3}-\frac{\text{vol}(T^4)}{4\,\ell^2} \,.
\end{split}
\end{align}
These are precisely the solution of the massive type IIA supergravity obtained in \cite{1012.1320}. 

Since the R--R potential also behaves as an $\OO(D,D)$ spinor in DFT, let us also explain how to determine the R--R potential in the dual background. 
Due to the gauge fixing of \eqref{eq:RR1-gauge}, the Page form takes the form \eqref{eq:Page-under-gauge}. 
Then the R--R potential in the A-basis is
\begin{align}
 A = -\Bigl(\frac{\rmd t\wedge\rmd x}{\ell^2\,z^2} + \frac{\theta\,\rmd\phi\wedge\rmd\psi}{4\,\ell^2}\Bigr)\wedge\bigl[1 - \text{vol}(T^4) \bigr] \,,
\end{align}
where $\theta$ should not be set to $\theta=\pi/2$ in order to realize $F=\rmd A$\,. 
Similar to the field strength, $\GL(D)$ transformation is trivial, and the $B$-transformation $A \to \Exp{\bm{\Lambda_f}\wedge} A$ and $T$-dualities along the $(\theta,\, \phi,\,\psi)$-directions give
\begin{align}
 A' = \Bigl[\frac{\tilde{u}_1\,\rmd u^1}{4\,\ell^2} + \frac{\rmd t\wedge\rmd x\wedge (u_i\,\rmd u^i-\rmd u^1\wedge\rmd u^2\wedge\rmd u^3)}{\ell^2\,z^2}\Bigr]\wedge\bigl[1 - \text{vol}(T^4) \bigr]\,,
\label{eq:RR-ex1-Ap}
\end{align}
where we have denoted $\tilde{u}_1\equiv \theta$ as it is dual to $u^1=\tilde{\theta}$\,. 
Since $A$ depends on the dual coordinate explicitly, the relation between $F$ and $A$ is generalized as [see Eq.~\eqref{eq:bmd-def}]
\begin{align}
 F = \bm{d} A \,,\qquad \bm{d} \equiv \rmd x^m\wedge \partial_m + \iota_m\,\tilde{\partial}^m\,,
\end{align}
and the $A'$ in \eqref{eq:RR-ex1-Ap} correctly reproduces the $F'$ obtained in \eqref{eq:Page-example1}. 
This result is consistent with \cite{1108.4937} where the massive type IIA supergravity was reproduced from DFT by introducing a linear dual-coordinate dependence into the R--R 1-form potential. 
The potential in the C-basis can also be obtained by computing $C'=\Exp{-B'_2\wedge}A'$\,. 

\subsection{\texorpdfstring{$\AdS_5 \times \rmS^5$}{AdS\textfiveinferior{x}S\textfivesuperior}}

As the second example, let us consider a NATD of the $\AdS_5\times\rmS^5$ background associated with a non-unimodular algebra. 
The original $\AdS_5\times\rmS^5$ background is
\begin{align}
\begin{split}
 \rmd s^2 &= \frac{\eta_{\mu\nu}\, \rmd x^\mu\,\rmd x^\nu + \rmd z^2}{z^2} + \rmd s^2_{\rmS^5}\qquad (\eta_{\mu\nu})\equiv \diag(-1,\,1,\,1,\,1)\,, 
\\
 G &= 4\,\bigl(- \rmd x^0 \wedge \rmd x^1 \wedge \rmd x^2 \wedge \rmd x^3 \wedge \rmd z + \omega_5 \bigr) \,,
\end{split}
\end{align}
where
\begin{align}
\begin{split}
 \rmd s^2_{\rmS^5} &\equiv \rmd r^2 + \sin^2 r\,\rmd \xi^2 + \sin^2r\cos^2\xi\,\rmd\phi_1^2 + \sin^2r \sin^2\xi\,\rmd\phi_2^2 + \cos^2r\,\rmd \phi_3^2\,,
\\
 \omega_5 &\equiv \sin^3r \cos r\sin\xi\cos\xi\,\rmd r\wedge\rmd\xi\wedge\rmd\phi_1\wedge\rmd\phi_2\wedge\rmd\phi_3\,.
\end{split}
\end{align}

We consider a NATD associated with two Killing vectors,
\begin{align}
 V_1^M = (z\,\partial_z+x^\mu\,\partial_\mu,\,0)\,,\qquad 
 V_2^M = (\partial_1,\,0)\,,
\end{align}
which satisfy $[V_1,\,V_2]_{\text{C}} = -V_2$\,. 
The gauge symmetry can be fixed as $z(\sigma)=1$ and $x^1(\sigma)=0$\,, and the AdS parts of the transformation matrices are
\begin{align}
 (\Lambda_v) = {\footnotesize\begin{pmatrix}
 1 & 0 & 0 & 0 & 0 \\
 0 & 1 & 0 & 0 & 0 \\
 0 & 0 & 1 & 0 & 0 \\
 0 & 0 & 0 & 1 & 0 \\
 x^0 & x^1 & x^2 & x^3 & z \\
 \end{pmatrix}},\qquad 
 (\Lambda_f) = {\footnotesize\begin{pmatrix}
 0 & 0 & 0 & 0 & 0 \\
 0 & 0 & 0 & 0 & \tilde{x}_1 \\
 0 & 0 & 0 & 0 & 0 \\
 0 & 0 & 0 & 0 & 0 \\
 0 & -\tilde{x}_1 & 0 & 0 & 0 \\
 \end{pmatrix}} .
\end{align}
For simplicity, we denote $(u^\mu)\equiv (x^0,\,\tilde{x}_1,\,x^2,\,x^3)$\,; then the dual background becomes
\begin{align}
\begin{split}
 \rmd s'^2 &= \frac{\rmd \tilde{z}^2 + a_{\mu\nu}\, \rmd u^\mu\,\rmd u^\nu}{1+\eta_{\rho\sigma}\,u^\rho\,u^\sigma} + \eta_{\mu\nu}\,\rmd u^\mu\,\rmd u^\nu +\rmd s^2_{\rmS^5}\,,\qquad
 \Exp{-2\Phi'}=1+\eta_{\mu\nu}\,u^\mu\,u^\nu\,,
\\
 B'_2&= \frac{(-u^0\,\rmd u^0-u^1\,\rmd u^1+u^2\,\rmd u^2+u^3\,\rmd u^3)\wedge\rmd \tilde{z}}{1+\eta_{\rho\sigma}\,u^\rho\,u^\sigma}\,,
\end{split} 
\end{align}
where
\begin{align}
 (a_{\mu\nu})&= \begin{pmatrix}
 -u^0\,u^0 & -u^0\,u^1 & u^0\,u^2 & u^0\,u^3\\
 -u^1\,u^0 & -u^1\,u^1 & u^1\,u^2 & u^1\,u^3\\
 u^2\,u^0 & u^2\,u^1 & -u^2\,u^2 & -u^2\,u^3\\
 u^3\,u^0 & u^3\,u^1 & -u^3\,u^2 & -u^3\,u^3
\end{pmatrix}.
\end{align}

Regarding the R--R fields, the first $\GL(D)$ transformation does not change the Page form and the next $B$-transformation gives
\begin{align}
 F= 4\,\bigl(- \rmd u^0 \wedge \rmd u^1 \wedge \rmd u^2 \wedge \rmd u^3 \wedge \rmd z + \omega_5 \bigr) + 4\,u^1\,\omega_5\wedge\rmd u^1\wedge\rmd z\,. 
\end{align}
The Abelian $T$-dualities along the $z$ and $x^1$ directions give
\begin{align}
 F' = - 4\,\rmd u^0 \wedge \rmd u^2 \wedge \rmd u^3 + 4\,\omega_5\wedge \rmd \tilde{z}\wedge \rmd u^1 + 4\,u^1\,\omega_5 \,. 
\end{align}
From this Page form, we find that
\begin{align}
 G'_3 = -4\,\rmd u^0 \wedge \rmd u^2 \wedge \rmd u^3\,,
\qquad
 G'_5 = -\frac{4\,u^1\,\rmd u^0\wedge \rmd u^1 \wedge \rmd u^2 \wedge\rmd u^3\wedge \rmd \tilde{z}}{1+\eta_{\mu\nu}\,u^\mu\,u^\nu} + u^1\,\rmd\omega_4 \,.
\end{align}
Then, by introducing $I = f_{\ggb\gga}{}^\ggb\,\tilde{\partial}^\gga = \tilde{\partial}^z$\,, they satisfy the type IIB GSE. 

In order to obtain a solution of the usual supergravity, we perform a formal $T$-duality along the $\tilde{z}$-direction. 
By using the $T$-duality rule \eqref{eq:T-duality-rule}, we obtain a simple type IIA solution:
\begin{align}
\begin{split}
 \rmd s^2 &= (1+\eta_{\mu\nu}\,u^\mu\,u^\nu)\,\rmd z^2 + 2\,\bigl(-u^0\,\rmd u^0-u^1\,\rmd u^1+u^2\,\rmd u^2+u^3\,\rmd u^3\bigr)\,\rmd z
\\
 &\quad +\eta_{\mu\nu}\,\rmd u^\mu\,\rmd u^\nu + \rmd s^2_{\rmS^5}\,,\qquad \Phi=z\,,\qquad
 G_4 = 4\Exp{-z}\rmd z\wedge \rmd u^0 \wedge \rmd u^2 \wedge \rmd u^3 \,.
\end{split}\end{align}

\subsection{\texorpdfstring{$\AdS_3 \times \rmS^3\times \TT^4$}{AdS\textthreeinferior{x}S\textthreesuperior{x}T\textfoursuperior} with NS--NS and R--R fluxes}

In order to demonstrate the efficiency of our formula, let us consider a more involved example. 
We start with the $\AdS_3\times\rmS^3\times \TT^4$ solution with the NS--NS and the R--R fluxes,
\begin{align}
\begin{split}
 \rmd s^2 &= \frac{-\rmd t^2+\rmd x^2 +\rmd z^2}{z^2} + \frac{1}{4}\, \bigl[ \rmd\theta^2 + \sin^2 \theta \, \rmd \phi^2 + (\rmd \psi + \cos\theta \, \rmd \phi)^2 \bigr] +\rmd s^2_{\TT^4}\,,
\\
 B_2 &= p\,\Bigl(\frac{\rmd t\wedge\rmd x}{z^2} - \frac{\cos\theta\, \rmd\phi\wedge\rmd\psi}{4}\Bigr)\,, \quad
 G_3 = q\,\Bigl(\frac{2\,\rmd t\wedge\rmd x\wedge \rmd z}{z^3} - \frac{\sin\theta\,\rmd\theta\wedge\rmd\phi\wedge\rmd\psi}{4}\Bigr)\,,
\end{split}
\end{align}
where $p$ and $q$ are constants satisfying $p^2+q^2=1$\,. 
The Page form is
\begin{align}
 F = G_3 + F_5 - (G_3 + F_5) \wedge \text{vol}_{\TT^4} \,,
\qquad
 F_5 \equiv \rmd\Bigl(\frac{p\,q\cos\theta}{4\,z^2}\Bigr)\wedge \rmd t\wedge \rmd x\wedge \rmd \phi\wedge \rmd\psi\,. 
\label{eq:AdS3-RR-NS-Page}
\end{align}

Then, we consider two generalized Killing vectors,
\begin{align}
\begin{split}
 V_1 &\equiv (v_1,\,\tilde{v}_1) \equiv \bigl(t\,\partial_t + x\,\partial_x + z\,\partial_z \,,\ 0\bigr)\,,
\\
 V_2 &\equiv (v_2,\,\tilde{v}_2)\equiv \bigl(-2\,t\,x\,\partial_t +(-t^2 - x^2 + z^2)\,\partial_x -2\,x\,z\,\partial_z \,,\ 2\,p \,\rmd t -\tfrac{2\,p\,t}{z}\,\rmd z\bigr)\,,
\end{split}
\end{align}
which satisfy $[V_1,\,V_2]_{\text{C}} = V_2$ and $c_{\gga\ggb}=0$\,. 
The $B$-field is isometric along the dilatation generator $\Lie_{v_1} B_2 =0$\,, but it is not isometric along the special-conformal generator $\Lie_{v_2} B_2 \neq 0$ and the dual component $\tilde{v}_2$ is important. 
Here, we choose the gauge as $t(\sigma)=1$ and $x(\sigma)=1$\,. 

The AdS parts of the transformation matrices are
\begin{align}
 (\Lambda_v) = \begin{pmatrix}
 t & x & z \\
 -2\,t\,x & -t^2 - x^2 + z^2 & -2\,x\,z \\
 0 & 0 & 1
 \end{pmatrix},\qquad 
 (\Lambda_f) = \begin{pmatrix}
 0 & \tilde{x} & 0 \\
 -\tilde{x} & 0 & -\frac{2\,p}{z} \\
 \frac{2\,p}{z} & 0 & 0
 \end{pmatrix} ,
\end{align}
and the NS--NS fields and the Killing vector take the form
\begin{align}
 &\rmd s'^2 = \frac{z^2\,\rmd \tilde{t}^2 + 2\,\rmd \tilde{t}\, \rmd \tilde{x} + \rmd \tilde{x}^2 + \frac{(\tilde{x}- p)^2-1}{z^2}\, \rmd z^2 -\frac{2}{z}\,\bigl[2\,\tilde{x}\,\rmd \tilde{t} +(\tilde{x}-p)\,\rmd \tilde{x}\bigr]\, \rmd z}{z^2 + (\tilde{x} + p)^2 -1}
 +\rmd s_{\rmS^3}^2 +\rmd s_{\TT^4}^2\,,
\nn\\
 &B'_2 = \frac{-z\,(\tilde{x} + p)\,\rmd\tilde{t}\wedge\rmd\tilde{x} -\bigl[z^2 + 2\,p\,(\tilde{x} + p) -2\bigr]\,\rmd\tilde{t}\wedge\rmd z +\rmd \tilde{x}\wedge\rmd z}{z\,\bigl[z^2 + (\tilde{x} + p)^2 -1\bigr]}
 - \frac{p\,\cos\theta\, \rmd\phi\wedge\rmd\psi}{4}\,,
\nn\\
 &\Exp{-2\,\Phi'} = z^2 + (\tilde{x} + p)^2 -1\,,\qquad I' = - \tilde{\partial}^t\,. 
\end{align}

For the R--R fields, the first $\GL(D)$ transformation makes the replacement
\begin{align}
 \rmd t\wedge\rmd x\wedge\rmd z\ \rightarrow z^2\,\rmd t\wedge\rmd x\wedge\rmd z 
\end{align}
in the Page form \eqref{eq:AdS3-RR-NS-Page}, and by further acting $\Exp{\bm{\Lambda_f}\wedge}$ and $\mathsf{T}_{t}\cdot\mathsf{T}_{x}$\,, we obtain the Page form in the dual background,
\begin{align}
\begin{split}
 F'_1&= -\frac{2\,q\,\rmd z}{z}\,, \qquad
 F'_3 = \frac{2\,q}{z}\,\Bigl[p\,\rmd z\wedge \frac{\cos\theta\,\rmd \phi\wedge \rmd \psi}{4} + z\,(\tilde{x}+p)\,\omega_{\rmS^3} \Bigr]\,,
\\
 F'_5&= \frac{2\,q}{z}\,\Bigl[\rmd z \wedge \text{vol}_{\TT^4} - \bigl(z\,\rmd \tilde{t}\wedge \rmd \tilde{x}+2\,p\,\rmd \tilde{t}\wedge \rmd z\bigr)\wedge \omega_{\rmS^3}\Bigr]\,,
\\
 F'_7&= -\frac{2\,q}{z}\,\Bigl[p\,\rmd z\wedge \frac{\cos\theta\,\rmd \phi\wedge \rmd \psi}{4} + z\,(\tilde{x}+p)\,\omega_{\rmS^3} \Bigr]\wedge \text{vol}_{\TT^4}\,,
\\
 F'_9&= \frac{2\,q}{z}\,\bigl(z\,\rmd \tilde{t}\wedge \rmd \tilde{x}+2\,p\,\rmd \tilde{t}\wedge \rmd z\bigr)\wedge \omega_{\rmS^3}\wedge \text{vol}_{\TT^4}\,,
\end{split}
\end{align}
where $\omega_{\rmS^3}\equiv \frac{1}{8}\,\sin\theta\,\rmd\theta\wedge\rmd\phi\wedge\rmd\psi$\,. 
Finally, the field strength $G'=\Exp{-B'_2\wedge}F'$ becomes
\begin{align}
\begin{split}
 G'_1 &= -\frac{2\,q\,\rmd z}{z}\,,\qquad
 G'_3 = 2\,q\,(\tilde{x}+p) \, \biggl[ -\frac{z^{-1}\,\rmd \tilde{t}\wedge \rmd \tilde{x}\wedge \rmd z}{(\tilde{x}+p)^2+z^2-1} + \omega_{\rmS^3} \biggr]\,,
\\
 G'_5 &= 2\,q\,\frac{\bigl[\tilde{x}\,(z^2-2)-p\,z^2\bigr]\,\rmd \tilde{t}\wedge \rmd z - (\tilde{x}+p)\,\rmd \tilde{x}\wedge \rmd z - z\,(z^2-1)\,\rmd \tilde{t}\wedge \rmd \tilde{x}}{z\,\bigl[(\tilde{x}+p)^2+z^2-1\bigr]}\wedge \omega_{\rmS^3} 
\\
 &\quad +\frac{2\,q\,\rmd z\wedge \text{vol}_{\TT^4}}{z} \,. 
\end{split}
\end{align}
These satisfy type IIB GSE under the original constraint $p^2+q^2=1$\,. 

By performing a formal $T$-duality along the $\tilde{t}$-direction, we obtain
\begin{align}
\begin{split}
 \rmd s^2 &= \frac{(z^2+4\,p^2-4)\,\rmd z^2}{z^4}
 + \frac{2\,\bigl[(z^2 + 2\,p\,\tilde{x}+2\,p^2-2)\,\rmd t + 2\,p\,\rmd x\bigr]\,\rmd z}{z^3} 
\\
 &\quad +\frac{(z^2+\tilde{x}^2+2\,p\,\tilde{x}+p^2-1)\,\rmd t^2 + 2\,(\tilde{x}+p)\,\rmd t\,\rmd \tilde{x} + \rmd \tilde{x}^2}{z^2}
 +\rmd s_{\rmS^3}^2 +\rmd s_{\TT^4}^2\,,
\\
 B_2 &= -\frac{\rmd t\wedge\rmd\tilde{x}}{z^2} + \frac{2\,\bigl(x\,\rmd t+\rmd\tilde{x})\wedge\rmd z}{z^3}
 - \frac{p\,\cos\theta\, \rmd\phi\wedge\rmd\psi}{4}\,, \qquad
 \Exp{-2\,\Phi} = z^2\Exp{2\,t} \,,
\\
 G_2 &= \frac{2\,q\Exp{t}\rmd t\wedge \rmd z}{z}\,, \qquad
 G_4 =-\frac{2\,q \Exp{t} \bigl[z\,(\tilde{x}+p)\,\rmd t+ z\,\rmd \tilde{x} +2\,p\,\rmd z \bigr]\wedge \omega_{\rmS^3}}{z} \,,
\end{split}
\end{align}
which is a solution of type IIA supergravity. 

\subsection{Extremal black D3-brane background}

In order to show that the AdS factor is not important, let us consider an extremal black D3-brane background. 
To manifest the Bianchi type V symmetry we employ a non-standard coordinate system,
\begin{align}
\begin{split}
 \rmd s^2 &= H^{\frac{1}{2}}(r) \bigl\{-\rmd t^2 + t^2\, \bigl[\rmd x_1^2 + \Exp{2\,x^1} (\rmd x_2^2 + \rmd x_3^2) \bigr]\bigr\} + \frac{\rmd r^2}{H^2(r)}
\\
 &\quad + r^2\,\bigl(\rmd \theta^2 + \sin^2 \theta\,\rmd \xi^2 + \sin^2 \theta \cos^2 \xi\,\rmd \phi_1^2+ \sin^2 \theta \sin^2\xi\,\rmd \phi_2^2 + \cos^2 \theta\,\rmd \phi_3^2 \bigr)\,,
\\
 G_5 &= -\frac{4\,r_+^4\,t^3 \Exp{2\,x^1} \rmd t \wedge \rmd x^1 \wedge \rmd x^2 \wedge \rmd x^3 \wedge \rmd r}{r^5} 
\\
 &\quad + 4\,r_+^4\sin^3 \theta \cos \theta \sin \xi \cos \xi\,\rmd \theta \wedge \rmd \xi \wedge \rmd \phi_1 \wedge \rmd \phi_2 \wedge \rmd \phi_3\,,
\end{split}
\end{align}
where $H(r) \equiv 1-(r_+/r)^4$ and the four-dimensional metric inside the brackets $\{\cdots\}$ is flat. 
We consider the following three Killing vectors,
\begin{align}
 V_1 \equiv \bigl(\partial_1 + x^2\,\partial_2 + x^3\,\partial_3 \,,\ 0\bigr)\,, \quad
 V_2 \equiv \bigl(\partial_2 \,,\ 0\bigr)\,,\quad
 V_3 \equiv \bigl(\partial_3 \,,\ 0\bigr)\,,
\end{align}
that satisfy the algebra
\begin{align}
 [V_1,\,V_2]_{\text{C}} = -V_2 \,,\qquad 
 [V_1,\,V_3]_{\text{C}} = -V_3 \,,\qquad 
 [V_2,\,V_3]_{\text{C}} = 0 \,.
\end{align}
The $(x^1,\,x^2,\,x^3)$ parts of the matrices are
\begin{align}
 (\Lambda_v) = \begin{pmatrix}
 1 & x^2 & x^3 \\
 0 & 1 & 0 \\
 0 & 0 & 1
 \end{pmatrix},\qquad 
 (\Lambda_f) = \begin{pmatrix}
 0 & -\tilde{x}_2 & -\tilde{x}_3 \\
 \tilde{x}_2 & 0 & 0 \\
 \tilde{x}_3 & 0 & 0
 \end{pmatrix} ,
\end{align}
and the gauge symmetry is fixed as $x^i(\sigma)=0$ ($i=1,2,3$)\,. 
The dual background becomes
\begin{align}
\begin{split}
 \rmd s'^2 &= -H^{\frac{1}{2}}\,\rmd t^2 + \frac{t^4\,H\,(\rmd \tilde{x}_1^2+\rmd \tilde{x}_2^2+\rmd \tilde{x}_3^2) + \tilde{x}_3^2\,\rmd \tilde{x}_2^2 - 2\,\tilde{x}_2\,\tilde{x}_3\,\rmd \tilde{x}_2\,\rmd \tilde{x}_3 + \tilde{x}_2^2\,\rmd \tilde{x}_3^2}{t^2\,H^{\frac{1}{2}}\,\bigl(H\,t^4+\tilde{x}_2^2+\tilde{x}_3^2\bigr)} + \frac{\rmd r^2}{H^2}
\\
 &\quad + r^2\,\bigl(\rmd \theta^2 + \sin^2 \theta\,\rmd \xi^2 + \sin^2 \theta \cos^2 \xi\,\rmd \phi_1^2+ \sin^2 \theta \sin^2\xi\,\rmd \phi_2^2 + \cos^2 \theta\,\rmd \phi_3^2 \bigr)\,,
\\
 B'_2 &= \frac{\rmd \tilde{x}_1\wedge (\tilde{x}_2\,\rmd \tilde{x}_2+\tilde{x}_3\,\rmd \tilde{x}_3)}{H\,t^4 +\tilde{x}_2^2+\tilde{x}_3^2}\,, \quad
 \Exp{-2\,\Phi'} = t^2\,H^{\frac{1}{2}}\,\bigl(H\,t^4+\tilde{x}_2^2+\tilde{x}_3^2\bigr)\,,\quad 
 I'= 2\,\tilde{\partial}^1\,,
\\
 G'_2 &= - \frac{4\,r_+^4\,t^3\,\rmd t \wedge \rmd r}{r^5}\,,\qquad
 G'_4 = - \frac{4\,r_+^4\,t^3\,\rmd t \wedge \rmd \tilde{x}_1 \wedge (\tilde{x}_2\,\rmd \tilde{x}_2 +\tilde{x}_3\,\rmd \tilde{x}_3)\wedge \rmd r}{r^5\,(H\,t^4 + \tilde{x}_2^2+ \tilde{x}_3^2)}\,,
\end{split}
\end{align}
and this is a solution of type IIA GSE. 

Again, by performing a formal $T$-duality along the $\tilde{x}_1$-direction we obtain a solution of type IIB supergravity,
\begin{align}
\begin{split}
 \rmd s^2 &= H^{\frac{1}{2}}\,\bigl(-\rmd t^2+t^2\,\rmd x_1^2\bigr) + \frac{(\rmd \tilde{x}_2 - \tilde{x}_2\,\rmd x^1)^2 + (\rmd \tilde{x}_3 - \tilde{x}_3\,\rmd x^1)^2}{H^{\frac{1}{2}}\,t^2} + \frac{\rmd r^2}{H^2}
\\
 &\quad + r^2\,\bigl(\rmd \theta^2 + \sin^2 \theta\,\rmd \xi^2 + \sin^2 \theta \cos^2 \xi\,\rmd \phi_1^2+ \sin^2 \theta \sin^2\xi\,\rmd \phi_2^2 + \cos^2 \theta\,\rmd \phi_3^2 \bigr)\,,
\\
 \Exp{-2\,\Phi} &= t^4 \Exp{-4\,x^1}H(r)\,,\qquad 
 G_3 = \Exp{-2\,x^1} \frac{4\,r_+^4\,t^3\,\rmd t\wedge\rmd x^1\wedge\rmd r}{r^5}\,. 
\end{split}
\end{align}

We note that, as discussed in \cite{1811.10600}, some supergravity solutions, obtained by a combination of NATD and a formal $T$-duality can also be obtained from another route, a combination of diffeomorphisms and the Abelian $T$-dualities. 
Similarly, the solutions obtained in this paper may also be realized from such procedure. 

\section{Poisson--Lie $T$-duality/plurality}
\label{sec:T-plurality}

Here we study a more general class of $T$-duality known as the Poisson--Lie $T$-duality \cite{hep-th/9502122,hep-th/9509095} or $T$-plurality \cite{hep-th/0205245}. 
We can perform the PL $T$-duality/plurality when the target space has a set of vectors $v_\gga$ satisfying the dualizability conditions \cite{hep-th/9502122}
\begin{align}
 [v_\gga,\,v_\ggb]= f_{\gga\ggb}{}^\ggc\,v_\ggc \,,\qquad
 \Lie_{v_\gga}E_{mn}= - \tilde{f}^{\ggb\ggc}{}_\gga\,E_{mp}\,v_\ggb^p \, v_\ggc^q\,E_{qn} \,. 
\end{align}
The traditional NATD (with $\tilde{v}_{\gga}=0$) can be regarded as a special case, $\tilde{f}^{\ggb\ggc}{}_\gga=0$\,. 
We begin with a brief review of the idea and techniques, and show the covariance of the DFT equations of motion under the PL $T$-plurality. 
Namely, we show that if we start with a DFT solution, the PL $T$-dualized background is also a DFT solution. 
In some examples the Killing vector $I^m$ appears, and the dualized DFT solutions are regarded as GSE solutions. 
However, through a formal $T$-duality the GSE solutions can always be transformed into linear-dilaton solutions of the conventional supergravity. 

\subsection{Review of PL $T$-duality}

We review the PL $T$-duality as a symmetry of the classical equations of motion of the string sigma model. 
To make the discussion transparent, we first ignore spectator fields $y^\mu(\sigma)$, which are invariant under the PL $T$-duality. 
As studied in \cite{hep-th/9502122,hep-th/9509095}, it is straightforward to introduce spectators, and their treatment is discussed in section \ref{sec:spectator}. 

Let us consider a sigma model with a target space $M$, on which a group $G$ acts transitively and freely (i.e.~$M$ itself can be regarded as a group manifold),
\begin{align}
 S = \EPSneg \frac{1}{4\pi\alpha'}\int_\Sigma E_{mn}(x)\,\bigl( \rmd x^m\wedge *\,\rmd x^n \EPSminus \rmd x^m\wedge \rmd x^n\bigr)\,.
\end{align}
Under an infinitesimal right action of a group $G$\,, the coordinates $x^m$ are shifted as
\begin{align}
 g(x) \ \to\ g(x)\,(1+\epsilon^\gga\,T_\gga) \equiv g(x+\delta x)\,,\qquad
 \delta x^m = \epsilon^\gga(\sigma)\,v_\gga^m(x)\,, 
\end{align}
where $T_\gga$ ($\gga=1,\dotsc,n$) are the generators of the algebra $\cg$ satisfying
\begin{align}
 [T_\gga,\,T_\ggb]=f_{\gga\ggb}{}^\ggc\,T_\ggc\,,
\end{align}
and $v_\gga^m$ are the left-invariant vector fields satisfying
\begin{align}
 [v_\gga,\,v_\ggb] = f_{\gga\ggb}{}^\ggc\,v_\ggc\,, \qquad
 v_\gga^m\,\ell^\ggb_m = \delta_\gga^\ggb \,,\qquad \ell \equiv \ell^\gga\,T_\gga \equiv g^{-1}\,\rmd g \,. 
\end{align}
In general, the variation of the action becomes
\begin{align}
 \delta_\epsilon S = \frac{1}{2\pi\alpha'}\int_\Sigma \Bigl\{-\epsilon^\gga \,\Bigl[\rmd J_\gga \EPSplus \frac{1}{2}\,\Lie_{v_\gga}E_{mn}\,\bigl( \rmd x^m\wedge *\rmd x^n \EPSminus \rmd x^m\wedge \rmd x^n\bigr) \Bigr] 
 + \rmd \bigl(\epsilon^\gga\, J_\gga\bigr)\Bigr\}\,,
\end{align}
where
\begin{align}
 J_\gga \equiv v_\gga^m\,\bigl(\EPSneg g_{mn}\, *\rmd x^n + B_{mn}\, \rmd x^n\bigr) \,.
\end{align}
If the $v_\gga^m$ satisfy the Killing equation $\Lie_{v_\gga}E_{mn} =0$\,, equations of motion for $x^m$ can be written as
\begin{align}
 \rmd J_\gga = 0 \,.
\end{align}
In particular, if $v_\gga^m$ further satisfy $[v_\gga,\,v_\ggb]=0$ we can find a coordinate system where $v_\gga^m=\delta_\gga^m$ is realized. 
Then, the Abelian $T$-duality can be realized as the exchange of $x^m(\sigma)$ with the dual coordinates $\tilde{x}_\gga(\sigma)$\,, which are defined as
\begin{align}
 \rmd \tilde{x}_\gga \equiv J_\gga \,. 
\label{eq:abelian-dual}
\end{align}
The Bianchi identity $\rmd^2 \tilde{x}_\gga=0$ corresponds to the equations of motion in the original theory. 

The PL $T$-duality is a generalization of this duality when the vector fields $v_\gga$ satisfy
\begin{align}
 \Lie_{v_\gga}E_{mn}= - \tilde{f}^{\ggb\ggc}{}_\gga\,E_{mp}\,v_\ggb^p \, v_\ggc^q\,E_{qn} \,.
\label{eq:Drinfel'd-double-condition}
\end{align}
In this case, the variation becomes
\begin{align}
 \delta S = \frac{1}{2\pi\alpha'}\int_\Sigma \biggl[-\epsilon^\gga\,\Bigl(\rmd J_\gga - \frac{1}{2}\,\tilde{f}_\gga{}^{\ggb\ggc}\, J_\ggb\wedge J_\ggc\Bigr) + \rmd (\epsilon^\gga\,J_\gga) \biggr] \,,
\end{align}
and the equations of motion for $x^m$ become the Maurer--Cartan equation,
\begin{align}
 \rmd J_\gga - \frac{1}{2}\,\tilde{f}_\gga{}^{\ggb\ggc}\, J_\ggb\wedge J_\ggc = 0\,.
\end{align}
This suggests introducing the dual coordinates $\tilde{x}_m(\sigma)$ through a non-Abelian generalization of \eqref{eq:abelian-dual}, namely,
\begin{align}
 \tilde{r}_\gga \equiv J_\gga \qquad 
 \bigl(\ \tilde{r}\equiv \tilde{r}_\gga \, \tilde{T}^\gga \equiv \rmd \tilde{g}\,\tilde{g}^{-1} \,,\quad 
 \tilde{g} \equiv \tilde{g}(\tilde{x})\in \tilde{G}\ \bigr)\,,
\label{eq:PL-dual}
\end{align}
where $\tilde{T}^\gga$ are the generators of the dual algebra $\tilde{\cg}$ (associated with a dual group $\tilde{G}$) satisfying
\begin{align}
 [\tilde{T}^\gga,\,\tilde{T}^\ggb] = \tilde{f}^{\gga\ggb}{}_{\ggc}\,\tilde{T}^\ggc \,.
\end{align}
Then, under the equations of motion, the physical coordinates $x^m(\sigma)$ describe the motion of the string on the group $G$ while the dual coordinates $\tilde{x}_m(\sigma)$ describe the motion of the string on the dual group $\tilde{G}$\,. 

It is important to note that the condition \eqref{eq:Drinfel'd-double-condition} and the identity
\begin{align}
 [\Lie_{v_\gga},\,\Lie_{v_\ggb}]E_{mn} = \Lie_{[v_\gga,\,v_\ggb]} E_{mn} \,,
\end{align}
show the relation
\begin{align}
 f_{\gga\gge}{}^\ggc\, \tilde{f}^{\gge\ggd}{}_\ggb + f_{\gga\gge}{}^\ggd\,\tilde{f}^{\ggc\gge}{}_\ggb -f_{\ggb\gge}{}^\ggc\, \tilde{f}^{\gge\ggd}{}_\gga - f_{\ggb\gge}{}^\ggd\,\tilde{f}^{\ggc\gge}{}_\gga = f_{\gga\ggb}{}^\gge\,\tilde{f}^{\ggc\ggd}{}_\gge \,.
\label{eq:1-cocycle}
\end{align}
By considering the vector space $\tilde{\cg}$ as the dual space of $\cg$\,, $\langle T_\gga,\,\tilde{T}^\ggb \rangle=\delta_\gga^\ggb$\,, the relation gives the structure of the Lie bialgebra. 
By further introducing an $ad$-invariant bilinear form as
\begin{align}
 \langle T_A,\,T_B \rangle = \eta_{AB}\,,\qquad 
 (\eta_{AB}) = \begin{pmatrix} 0 & \delta_\gga^\ggb \\ \delta^\gga_\ggb & 0 \end{pmatrix}, \qquad 
 (T_A) \equiv (T_\gga,\,\tilde{T}^\gga)\,,
\end{align}
the commutation relations on a direct sum $\mathfrak{d}\equiv \cg\oplus\tilde{\cg}$ are determined as
\begin{align}
 [T_\gga,\,T_\ggb] = f_{\gga\ggb}{}^\ggc\,T_\ggc\,,\qquad
 [T_\gga,\,\tilde{T}^\ggb] = \tilde{f}^{\ggb\ggc}{}_\gga\,T_\ggc - f_{\gga\ggc}{}^\ggb\,\tilde{T}^\ggc\,,\qquad
 [\tilde{T}^\gga,\,\tilde{T}^\ggb] = \tilde{f}^{\gga\ggb}{}_\ggc\,\tilde{T}^\ggc\,,
\end{align}
and the pair of algebras can be regarded as that of the Drinfel'd double $\mathfrak{D}$\,. 
Given the structure of the Drinfel'd double, the differential equation \eqref{eq:Drinfel'd-double-condition} can be integrated \cite{hep-th/9502122,hep-th/9509095} as
\begin{align}
 \sfE_{\gga\ggb} \equiv v_\gga^m\,v_\ggb^n\,E_{mn} = \bigl[a^{-1} \, \hat{E} \,(a^\rmT +b^\rmT \,\hat{E})^{-1} \bigr]_{\gga\ggb} \,,
\label{eq:twist-original}
\end{align}
where the matrices $a$ and $b$ are defined by
\begin{align}
 g^{-1}\,T_A\,g = (\text{Ad}_{g^{-1}})_A{}^B\,T_B \,,\qquad \text{Ad}_{g^{-1}} = \begin{pmatrix} a_\gga{}^\ggb & 0 \\
 b^{\gga\ggb} & (a^{-\rmT})^{\gga}{}_\ggb 
\end{pmatrix} ,
\label{eq:Ad-def}
\end{align}
and $\hat{E}_{\gga\ggb}$ is an arbitrary constant matrix (that corresponds to $\sfE_{\gga\ggb}(x)$ at $g=1$). 
We can check that the $E_{mn}$ given by \eqref{eq:twist-original} indeed satisfies \eqref{eq:Drinfel'd-double-condition}.\footnote{For example, when $E_{mn}$ is invertible, we can easily check an equivalent expression $\Lie_{v_\gga}E^{mn}= \tilde{f}^{\ggb\ggc}{}_\gga\,v_\ggb^m \, v_\ggc^m$ by using the rewriting \eqref{eq:Emn-right} and $v_\ggc^m\,\partial_m\Pi^{\gga\ggb}= - (a^{-\rmT})^\gga{}_\ggd\,(a^{-\rmT})^\ggb{}_\gge\,\tilde{f}^{\ggd\gge}{}_\ggc$\,, which can be derived from \eqref{eq:Ad-def} (see \cite{hep-th/9710163}).} 

Now, we rewrite the relation \eqref{eq:PL-dual}, namely
\begin{align}
 \tilde{r}_\gga = J_\gga = \EPSneg \sfg_{\gga\ggb}\, * \ell^\ggb + \sfB_{\gga\ggb}\, \ell^\ggb \qquad 
 \bigl(\sfg_{\gga\ggb}\equiv \sfE_{(\gga\ggb)}\,,\quad \sfB_{\gga\ggb}\equiv\sfE_{[\gga\ggb]}\bigr)\,,
\end{align}
into two equivalent expressions (by following the standard trick \cite{Duff:1989tf} in the Abelian case),
\begin{align}
\begin{split}
 \ell^\gga &= \EPSpos (\sfg^{-1}\,\sfB)^{\gga}{}_{\ggb}\, * \ell^\ggb \EPSminus \sfg^{\gga\ggb}\,*\tilde{r}_\ggb \,,
\\
 \tilde{r}_\gga &= \EPSneg (\sfg-\sfB\,\sfg^{-1}\,\sfB)_{\gga\ggb}\, * \ell^\ggb \EPSminus (\sfB\,\sfg)_{\gga}{}^{\ggb}\,*\tilde{r}_\ggb \,.
\end{split}
\end{align}
They can be neatly expressed as a self-duality relation,
\begin{align}
\begin{split}
 &\sfP^A = \EPSneg \sfH^A{}_B(x)\,* \sfP^B \,,\qquad (\sfP^A) \equiv \begin{pmatrix} \ell^\gga \\ \tilde{r}_\gga \end{pmatrix}\,,
\\
 &(\sfH_{AB}) \equiv \begin{pmatrix} (\sfg -\sfB\,\sfg^{-1}\,\sfB)_{\gga\ggb} & \sfB_{\gga\ggc}\,\sfg^{\ggc\ggb} \\ -\sfg^{\gga\ggc}\,\sfB_{\ggc\ggb} & \sfg^{\gga\ggb} \end{pmatrix},
\end{split}
\label{eq:self-dual-sfH}
\end{align}
where the indices $A,B,\cdots$ are raised or lowered with $\eta_{AB}$ and its inverse $\eta^{AB}$\,. 
In terms of the metric $\sfH_{AB}$\,, the relation \eqref{eq:twist-original} can be expressed as
\begin{align}
 \sfH_{AB}(x) = (\text{Ad}_g)_A{}^C\,(\text{Ad}_g)_B{}^D\,\hat{\cH}_{CD}\,,\qquad
 (\hat{\cH}_{AB}) \equiv \begin{pmatrix} \hat{g} -\hat{B}\,\hat{g}^{-1}\,\hat{B} & \hat{B}\,\hat{g}^{-1} \\ - \hat{g}^{-1}\,\hat{B} & \hat{g}^{-1} \end{pmatrix},
\end{align}
where $\hat{g}_{\gga\ggb}\equiv \hat{E}_{(\gga\ggb)}$\,, $\hat{B}_{\gga\ggb}\equiv \hat{E}_{[\gga\ggb]}$\,, and \eqref{eq:self-dual-sfH} gives the important relation
\begin{align}
 \hat{\cP}^A = \EPSneg \hat{\cH}^A{}_B \,* \hat{\cP}^B \,,\qquad 
 \hat{\cP}(\sigma) \equiv \hat{\cP}^A\,T_A \equiv \rmd l\,l^{-1} \,, \qquad 
 l \equiv g\,\tilde{g} \,,
\label{eq:EOM-original}
\end{align}
where we have used\footnote{If we expand the right-invariant form as $\hat{\cP} = \cP^A{}_M\,\rmd x^M\,T_A$\,, we find that $\cP^A{}_M$ is not an $\OO(n,n)$ matrix:
\begin{align*}
 (\cP^A{}_M) = \begin{pmatrix}
 r^\gga_m & \Pi^{\gga\ggb}\,a_\ggb{}^\ggc\,\tilde{r}_\ggc^m \\
 0 & a_\gga{}^\ggb\,\tilde{r}_\ggb^m
\end{pmatrix}.
\end{align*}}
\begin{align}
 \hat{\cP}(\sigma) \equiv \rmd l\,l^{-1}
 = g\,\bigl(\ell^\gga\,T_\gga + \tilde{r}_\gga\,\tilde{T}^\gga \bigr)\,g^{-1} 
 = \sfP^B\,(\text{Ad}_g)_B{}^A\,T_A \,. 
\end{align}
Expressed in this form, the equations of motion are given in terms of the Drinfel'd double $\mathfrak{D}$\,; the decomposition $l= g\,\tilde{g}$ is no longer important. 

Similar to the Abelian $T$-duality we can recover the same equations of motion from the dual model, by exchanging the role of $\cg$ and $\tilde{\cg}$\,. 
Starting with the dual background $\tilde{E}_{mn}$\,, which has a set of vector fields $\tilde{v}^\gga$ satisfying
\begin{align}
 [\tilde{v}^\gga,\,\tilde{v}^\ggb]=\tilde{f}^{\gga\ggb}{}_\ggc\,\tilde{v}^\ggc\,,\qquad 
 \Lie_{\tilde{v}^\gga}E_{mn}= -f_{\ggb\ggc}{}^\gga\, \tilde{E}_{mp}\,\tilde{v}^{\ggb p} \,\tilde{v}^{\ggc q}\,\tilde{E}_{qn} \,,
\end{align}
the equations of motion can be expressed as
\begin{align}
 \hat{\cP}_A = \EPSneg \tilde{\hat{\cH}}_A{}^B \,* \hat{\cP}_B \,,\qquad 
 \hat{\cP}_A\,T^A \equiv \rmd \tilde{l}\,\tilde{l}^{-1} \,, \qquad 
 \tilde{l} \equiv \tilde{h} \,h \quad 
 (h\in \cg\,,\quad \tilde{h}\in \tilde{\cg})
\label{eq:EOM-dual}
\end{align}
by using a constant matrix $\tilde{\hat{\cH}}_A{}^B$\,. 
For the duality equivalence, we demand that \eqref{eq:EOM-original} and \eqref{eq:EOM-dual} are equivalent. 
This leads to the identifications
\begin{align}
 \hat{\cH}_{AB} = \tilde{\hat{\cH}}_{AB} \,,\qquad 
 g\, \tilde{g} = l = \tilde{l} \equiv \tilde{h} \,h\,.
\end{align}
After this identification, string theory defined on the original background $E_{mn}$ and the dual background $E'_{mn}$ give the same equations of motion, and are classically equivalent. 

In summary, in PL $T$-dualizable backgrounds the generalized metric $\cH_{MN}(x)$ is always related to a constant matrix $\hat{\cH}_{AB}$ as
\begin{align}
 \cH_{MN} = (U\,\hat{\cH}\,U^\rmT)_{MN} \,,
\label{eq:PL-LM}
\end{align}
where the matrix $U$ is defined as
\begin{align}
 U_M{}^A\equiv L_M{}^B\,(\text{Ad}_{g})_B{}^A\,,\qquad (L_M{}^A) \equiv \begin{pmatrix} \ell_m^\gga & 0 \\ 0 & v^m_\gga \end{pmatrix} .
\end{align}
By comparing this with \eqref{eq:twist-matrix}, we call the matrix $U$ the twist matrix and call the constant matrix $\hat{\cH}_{AB}$ the untwisted metric. 
The dual geometry also has the same structure, where the twist matrix is $\tilde{U}_{MA}\equiv \tilde{L}_{MB}\,(\text{Ad}_{\tilde{g}})^B{}_A$\,. 
The relation between the original and the dual background becomes
\begin{align}
 \tilde{\cH}_{MN} = (h\,\cH\,h^{\rmT})_{MN}\,,\qquad 
 h_M{}^N\equiv \tilde{U}_{MA}\,\eta^{AB}\,U_B{}^N\,. 
\end{align}
For later convenience, we rewrite the twist matrix as
\begin{align}
 U = L \, \text{Ad}_{g} = R \, \bm{\Pi} \,,
\end{align}
where we have defined
\begin{align}
\begin{split}
 &(R_M{}^A) \equiv \begin{pmatrix} r_m^\gga & 0 \\ 0 & e^m_\gga \end{pmatrix}\qquad 
 (\bm{\Pi}_A{}^B)\equiv \begin{pmatrix} \delta_\gga^\ggb & 0 \\ -\Pi^{\gga\ggb} & \delta^\gga_\ggb \end{pmatrix}, 
\\
 &r\equiv r^\gga\,T_\gga\equiv \rmd g\,g^{-1}\,,\qquad r_m^\gga\,e^m_\ggb=\delta^\gga_\ggb\,,\qquad 
 \Pi^{\gga\ggb} \equiv (b\,a^{-1})^{\gga\ggb}=-(a^{-\rmT}\,b^{\rmT})^{\gga\ggb} \,,
\end{split}
\end{align}
and used $r^\gga=(a^{-\rmT})^\gga{}_\ggb\,\ell^\ggb$\,. 
Then, in terms of $E_{mn}(x)$\,, \eqref{eq:PL-LM} can be expressed as
\begin{align}
 E_{mn}(x) = \bigl[ (\hat{E}^{-1} - \Pi)^{-1} \bigr]_{\gga\ggb}\,r^\gga_m\,r^\ggb_n\,,
\label{eq:Emn-right}
\end{align}
and, similarly, the dual background is
\begin{align}
 \tilde{E}_{mn}(\tilde{x}) = \bigl[ (\hat{E} - \tilde{\Pi})^{-1} \bigr]^{\gga\ggb}\,\tilde{r}_{\gga m}\,\tilde{r}_{\ggb n}\,. 
\end{align}

In the special case where $\tilde{f}^{\gga\ggb}{}_\ggc =0$\,, by parameterizing $\tilde{g}=\Exp{\tilde{x}_\gga\,\tilde{T}^\gga}$ we obtain $\tilde{r}=\rmd \tilde{x}_\gga\,\tilde{T}^\gga$\,, $\Pi^{\gga\ggb}=0$\,, and $\tilde{\Pi}_{\gga\ggb} = - f_{\gga\ggb}{}^\ggc\,\tilde{x}_\ggc$\,. 
This is precisely the case of NATD. 
In the dualized background, in general the isometries are broken and in the traditional NATD, we cannot recover the original model. 
However, the dual background has the form
\begin{align}
 \tilde{E}^{mn} = (\hat{E} - \tilde{\Pi})_{\gga\ggb}\,\tilde{e}^{\gga m}\,\tilde{e}^{\ggb n} = (\hat{E}_{\gga\ggb} + f_{\gga\ggb}{}^\ggc\,\tilde{x}_\ggc ) \,\tilde{v}^{\gga m}\,\tilde{v}^{\ggb n} \,,
\end{align}
where $\tilde{e}^\gga = \tilde{v}^\gga = \tilde{\partial}^\gga$\,, and we find that the dual background is $T$-dualizable,
\begin{align}
 \Lie_{\tilde{v}^\gga}E^{mn} = \tilde{\partial}^\gga E^{mn} = f_{\ggb\ggc}{}^\gga \,\tilde{v}^{\ggb m}\,\tilde{v}^{\ggc n}\,. 
\end{align}
Thus, through the PL $T$-duality we can recover the original background $E_{mn}=\hat{E}_{\gga\ggb}\,r_m^\gga\,r_n^\ggb$\,. 

As a side remark, we note that in the case of the Abelian $\OO(D,D)$ $T$-duality, the covariant equations of motion of string $\rmd x^M = \EPSneg \hat{\cH}^M{}_N \,* \rmd x^N$ \cite{Duff:1989tf} can be derived from the double sigma model (DSM) \cite{Tseytlin:1990nb,Tseytlin:1990va,hep-th/0406102,hep-th/0605149,1111.1828,1307.8377}. 
The correspondent of the DSM for the PL $T$-duality has been studied in \cite{hep-th/9512025,hep-th/9602162,hep-th/9605212,hep-th/9609112,0902.4032,1001.2479}, and this approach will be useful to manifest the PL $T$-duality. 

\subsection{PL $T$-plurality}

The Lie algebra $\mathfrak{d}$ of the Drinfel'd double $\mathfrak{D}$ can be constructed as a direct sum of two algebras $\cg$ and $\tilde{\cg}$\,, which are maximally isotropic with respect to the bilinear form $\langle \cdot,\,\cdot\rangle$\,, and the pair $(\mathfrak{d},\,\cg,\,\tilde{\cg})$ is called the Manin triple. 
In general, a Drinfel'd double has several decompositions into Manin triples, and this leads to the notion of the PL $T$-plurality \cite{hep-th/0205245}. 
More concretely, let us consider a redefinition of the generators $T_A$ of $\mathfrak{d}$\,,
\begin{align}
 T'_A \equiv C_A{}^B\,T_B\,,
\end{align}
such that the new generators also satisfy the algebra of the Drinfel'd double,
\begin{align}
 [T'_\gga,\,T'_\ggb] = f'_{\gga\ggb}{}^\ggc\,T'_\ggc\,,\qquad
 [T'_\gga,\,\tilde{T}'^\ggb] = \tilde{f}'^{\ggb\ggc}{}_\gga\,T'_\ggc - f'_{\gga\ggc}{}^\ggb\,\tilde{T}'^\ggc\,,\qquad
 [\tilde{T}'^\gga,\,\tilde{T}'^\ggb] = \tilde{f}'^{\gga\ggb}{}_\ggc\,\tilde{T}'^\ggc\,,
\end{align}
and the bilinear form is preserved,
\begin{align}
 \langle T'_A,\, T'_B\rangle = \eta_{AB} \,.
\end{align}
The latter condition shows that the matrix $C_A{}^B$ should be a certain $\OO(n,n)$ matrix. 
Since the rescaling of the generators is trivial, we choose $C_A{}^B$ as a ``volume-preserving'' $\OO(n,n)$ transformation that does not change the DFT dilaton. 

The transformation of the background fields under the $\OO(n,n)$ transformation can be found in the same manner as the PL $T$-duality. 
Starting with a background $E'_{mn}$ satisfying
\begin{align}
 [v'_\gga,\,v'_\ggb]= f'_{\gga\ggb}{}^\ggc\, v'_\ggc\,,\qquad 
 \Lie_{v'_\gga}E'_{mn}= - \tilde{f}'^{\ggb\ggc}{}_\gga\, E'_{mp}\,v'^p_{\ggb} \,v'^q_{\ggc} \,E'_{qn} \,,
\end{align}
we again obtain the same equations of motion,
\begin{align}
 \cP'^A = \EPSneg \hat{\cH}'^A{}_B \,* \cP'^B \,,\qquad 
 \cP'^A\,T'_A \equiv \rmd l'\,l'^{-1} \,, \qquad 
 l' \equiv g'\,\tilde{g}' \,. 
\end{align}
From the identification $l=l'$ we obtain
\begin{align}
 \hat{\cP}^A\, T_A = \rmd l\,l^{-1} = \rmd l'\,l'^{-1} = \hat{\cP}'^A\,T'_A = \hat{\cP}'^A\,C_A{}^B\,T_B \,, 
\end{align}
and the relation between the untwisted metrics becomes
\begin{align}
 \hat{\cH}'_{AB} = (C\,\hat{\cH}\,C^{\rmT})_{AB} \,. 
\label{eq:plural}
\end{align}
The generalized metric in the transformed frame has the form
\begin{align}
 \hat{\cH}'_{MN} = (U'\,\hat{\cH}'\,U^{\rmT})_{MN} \,,
\end{align}
and the relation between the original and the dual generalized metric is
\begin{align}
 \cH'_{MN} = (h\,\cH\,h^{\rmT})_{MN}\,,\qquad 
 (h_M{}^N)\equiv U' \,C \, U^{-1} \,. 
\label{eq:PL-T-plurality}
\end{align}
In terms of $E_{mn}(x)$\,, the original background is
\begin{align}
 E_{mn}(x) = \bigl[ (\hat{E}^{-1} - \Pi)^{-1} \bigr]_{\gga\ggb}\,r^\gga_m\,r^\ggb_n\,,
\label{eq:original-E}
\end{align}
while the dual background is
\begin{align}
 E'_{mn}(x') = \bigl[ (\hat{E}'^{-1} - \Pi')^{-1} \bigr]_{\gga\ggb}\, r'^{\gga}_m \, r'^\ggb_n \,,\qquad 
 E'_{mn} = [(\bm{q}+\bm{p}\,\hat{E})\,(\bm{s}+\bm{r}\,\hat{E})^{-1}]_{mn} \,,
\end{align}
where we parameterized the $\OO(n,n)$ matrix $C$ as
\begin{align}
 C = \begin{pmatrix} \bm{p}_{m}{}^{n} & \bm{q}_{mn} \\ \bm{r}^{mn} & \bm{s}^{m}{}_{n} \end{pmatrix} .
\end{align}
Note that the PL $T$-duality is a special case of the $T$-plurality where
\begin{align}
 C = \begin{pmatrix} 0 & 1 \\ 1 & 0 \end{pmatrix},
\end{align}
and the original background corresponds to the trivial choice $C=1$\,.

\subsubsection{Duality rule for the dilaton}

The transformation rule for the dilaton was studied in \cite{hep-th/9512025} in the context of the PL $T$-duality. 
This was improved in \cite{hep-th/0205245} in the study of the PL $T$-plurality. 
In our convention, the result is
\begin{align}
 \Exp{-2\,\Phi'} = \Exp{-2\,\bar{\Phi}} \frac{\abs{\det (\bm{q}+\bm{p}\,\hat{E})}}{\abs{\det(E'_{\gga\ggb})}\,\abs{\det a'^{-1}}}\qquad
 \bigl(E'_{\gga\ggb}\equiv e'^m_\gga\,e'^n_\ggb\,E'_{mn}\bigr)\,,
\end{align}
where $\bar{\Phi}(x)$ is an arbitrary function. 
By using the formula \eqref{eq:detg-formula}, we obtain
\begin{align}
 \sqrt{\abs{g'}} 
 &= \abs{\det(r'^\gga_m)}\,\abs{\det(\bm{1}-\Pi'\,\hat{E}')}^{-1}\, \abs{\det(\bm{s}+\bm{r}\,\hat{E})}^{-1} \sqrt{\abs{\hat{g}}} 
\nn\\
 &= \abs{\det(r'^\gga_m)}\,\abs{\det E'_{\gga\ggb}} \,\abs{\det (\bm{q}+\bm{p}\,\hat{E})}^{-1}\,\sqrt{\abs{\hat{g}}} \,, 
\end{align}
and the DFT dilaton in the dual background becomes
\begin{align}
 \Exp{-2\,d'} = \Exp{-2\,\bar{d}}\, \abs{\det(r'^\gga_m)}\, \abs{\det a'} = \Exp{-2\,\bar{d}}\, \abs{\det(\ell'^\gga_m)}\,,\qquad 
 \Exp{-2\,\bar{d}} \equiv \Exp{-2\,\bar{\Phi}} \sqrt{\abs{\hat{g}}} \,. 
\label{eq:DFT-dilaton-PL}
\end{align}
Namely, the duality rule for the DFT dilaton is
\begin{align}
 \abs{\det(v'^m_\gga)} \Exp{-2\,(d'-\bar{d})} = 1 = \abs{\det(v^m_\gga)} \Exp{-2\,(d-\bar{d})} \,. 
\label{eq:DFT-dilaton-PL1}
\end{align}
If $\bar{d}$ (or equivalently $\bar{\Phi}$) is constant, this duality rule coincides with the recent proposal \cite{1810.11446}, where the PL $T$-duality was studied by utilizing ``the DFT on a Drinfel'd double'' proposed in \cite{1707.08624}. 
There, it was shown that the dilaton transformation rule is also consistent with \cite{1708.04079}. 
Moreover, when the dual algebra is Abelian $\tilde{f}^{\gga\ggb}{}_\ggc=0$\,, we have $\abs{\det(v'^m_\gga)}=1$ and the result \eqref{eq:NATD-DFT-dilaton} known in NATD is also reproduced as a particular case. 

In fact, as demonstrated in \cite{hep-th/0205245}, the PL $T$-plurality works even if $\bar{d}$ has a coordinate dependence. 
A subtle point is that when $\Exp{-2\,\bar{d}}$ depends on the original coordinates $x^m$\,, it is not clear how to understand the $x^m$-dependence in the dual model. 
A prescription proposed in \cite{hep-th/0205245} is as follows. 
We first identify the relation between coordinates $(x^M)=(x^m,\,\tilde{x}_m)$ and $(x'^M)=(x'^m,\,\tilde{x}'_m)$ through the identification
\begin{align}
 g'(x')\,\tilde{g}'(\tilde{x}') = l = g(x)\,\tilde{g}(\tilde{x})\,.
\label{eq:Double-coordinate-change}
\end{align}
We next substitute the relation $x^M=x^M(x')$ into $\Exp{-2\,\bar{d}(x)}$ as $\Exp{-2\,\bar{d}(x)}=\Exp{-2\,\bar{d}(x(x'))}\equiv \Exp{-2\,\bar{d}(x')}$\,. 
Then, the relation \eqref{eq:DFT-dilaton-PL1} can be understood on both sides
\begin{align}
 \abs{\det(v'^m_\gga)} \Exp{-2\,[d'-\bar{d}(x')]} = 1 = \abs{\det(v^m_\gga)} \Exp{-2\,[d-\bar{d}(x)]} \,.
\label{eq:DFT-dilaton-PL2}
\end{align}

In general, $\bar{d}(x')$ may depend on the dual coordinates $\tilde{x}'_m$\,, and the background does not have the usual supergravity description. 
However, in our examples the DFT dilaton has at most a linear dependence on the dual coordinates, and it can be absorbed into the Killing vector $I^m$ in the GSE. 

\subsubsection{Covariance of equations of motion}

In the approach of \cite{1707.08624,1810.11446}, the PL $T$-duality was realized as a manifest symmetry of DFT. 
We discuss here the covariance under a more general PL $T$-plurality by using the gauged DFT. 
The approach may be slightly different from \cite{1707.08624,1810.11446} but the essence will be the same. 

In PL $T$-dualizable backgrounds, the generalized metric always has the simple form
\begin{align}
 \cH_{MN} = [U(x)\,\hat{\cH}\,U^{\rmT}(x)]_{MN}\,.
\end{align}
Since the twist matrix $U$ is explicitly determined, we can compute the generalized fluxes $\cF_{ABC}$ and $\cF_A$ defined in \eqref{eq:generalized-fluxes}. 
In fact, as shown in \cite{1810.11446}, in PL $T$-dualizable backgrounds the three-index flux is precisely the structure constant of the Drinfel'd double,
\begin{align}
 \cF_{\gga\ggb\ggc} = 0\,,\qquad 
 \cF_{\gga\ggb}{}^\ggc = f_{\gga\ggb}{}^\ggc\,,\qquad 
 \cF^{\gga\ggb}{}_{\ggc} = \tilde{f}^{\gga\ggb}{}_\ggc \,,\qquad 
 \cF^{\gga\ggb\ggc} = 0\,. 
\end{align}
We can check this by using the explicit form of the twist matrix and its inverse,
\begin{align}
 (U_M{}^A) = \begin{pmatrix} r_m^\gga & 0 \\ -e^m_\ggb\,\Pi^{\ggb\gga} & e^m_\gga \end{pmatrix} , \qquad
 (U_A{}^M) = \begin{pmatrix} e_\gga^m & 0 \\ \Pi^{\gga\ggb}\,e^m_\ggb & r^\gga_m \end{pmatrix} , 
\end{align}
and the relations $\Lie_{e_\gga}e_\ggb = -f_{\gga\ggb}{}^\ggc\,e_\ggc$\,, $\Lie_{e_\gga}r^\ggb = f_{\gga\ggc}{}^\ggb\,r^\ggc$\,, $\partial_m\Pi^{\gga\ggb}= - (a^{-\rmT})^\gga{}_\ggd\,(a^{-\rmT})^\ggb{}_\gge\,\tilde{f}^{\ggd\gge}{}_\ggf\,a^{\ggf}{}_\ggc\,r^\ggc_m$\,, and $\tilde{f}^{\gga\ggb}{}_\ggc = (a^{-\rmT})^\gga{}_\ggd\,(a^{-\rmT})^\ggb{}_\gge\,a_\ggc{}^\ggf\,\tilde{f}^{\ggd\gge}{}_\ggf - 2\,f_{\ggc\gge}{}^{[\gga}\, \Pi^{\ggb]\gge}$ (see \cite{hep-th/9710163} for useful identities). 

We can also compute the single-index flux as
\begin{align}
 \cF_A = \begin{pmatrix} 2\,e_\gga^m\,\partial_m d + e_\ggc^n\,\partial_n r^\ggc_m\,e_\gga^m \\ -(a^{-\rmT})^\gga{}_\ggb\,\tilde{f}^{\ggc\ggb}{}_\ggc + \Pi^{\gga\ggb}\,(2\,e_\ggb^m\,\partial_m d + e_\ggc^n\,\partial_n r_m^\ggc \, e_\ggb^m\bigr) + 2\,r^a_m\,\tilde{\partial}^m d
\end{pmatrix}. 
\end{align}
By using the expression for the DFT dilaton \eqref{eq:DFT-dilaton-PL}, $\Exp{-2\,d} = \Exp{-2\,\bar{d}}\, \abs{\det(r^\gga_m)}\,\abs{\det a}$\,, we find
\begin{align}
 \cF_A = U_A{}^M \, \cF_M \,,\qquad 
 \cF_M \equiv 2\,\partial_M \bar{d} + \begin{pmatrix} 0 \\ - \tilde{f}^{\ggb\gga}{}_\ggb\,v^m_\gga 
\end{pmatrix}, 
\end{align}
where we have used $a_\ggb{}^\gge\,a_\ggc{}^\ggf\,f_{\gge\ggf}{}^\gga = f_{\ggb\ggc}{}^\gge\,a_\gge{}^\gga$ and $\partial_m a_\gga{}^\ggb = a_\gga{}^\ggc\,f_{\ggc\ggd}{}^\ggb\,\ell_m^\ggd$\,. 

As we discuss below, for the covariance of the equation of motion under the PL $T$-plurality, $\cF_A$ needs to transform covariantly. 
However, even in the particular case $\bar{d}=0$\,, for example, we find that $\cF_A$ does not transform covariantly. 
Indeed, we have $\cF_A=0$ in a duality frame where $\tilde{f}^{\gga\ggb}{}_\gga=0$\,, while $\cF_A$ appears in a frame where $\tilde{f}^{\gga\ggb}{}_\gga\neq 0$\,. 
Therefore, in order to transform $\cF_A$ covariantly, we eliminate the non-covariant term by adding a vector field $\bX_M$ as
\begin{align}
 \partial_M d \ \to \ \partial_M d + \bX_M\,,\qquad 
 (\bX_M) \equiv \begin{pmatrix} 0 \\ I^m \end{pmatrix},\qquad 
 I^m = \frac{1}{2}\,\tilde{f}^{\ggb\gga}{}_\ggb\,v_\gga^m \,,
\label{eq:mDFT-shift-PL}
\end{align}
which was suggested in \cite{1810.11446}. 
This shift is a bit artificial, but without this procedure we need to abandon all Manin triples with non-unimodular dual algebra. 
In fact, this shift is precisely the modification of DFT equations of motion \eqref{eq:mDFT-shift} that reproduces the GSE after removing the dual-coordinate dependence. 
After this prescription, we obtain the simple flux
\begin{align}
 \cF_A = 2\,\cD_A \bar{d} \,. 
\end{align}
In fact, as we see later, $\cF_A = 2\,\cD_A \bar{d}$ are covariantly transformed under the PL $T$-plurality $\cF'_A = C_A{}^B\,\cF_B$\,,\footnote{This is non-trivial, because in general the derivative $\cD_A$ does not transform covariantly, $\cD'_A \neq C_A{}^B\,\cD_B$\,, which can be checked by performing the coordinate transformation $x'^M=x'^M(x)$ through \eqref{eq:Double-coordinate-change}. Therefore, at the present time, the covariance of $\cF_A$ needs to be checked on a case-by-case basis. Of course, when $\bar{d}$ is constant, the covariance is manifest because $\cF_A=0$ and $\cF'_A=0$\,.} and the prescription \eqref{eq:mDFT-shift-PL} works well in our examples. 

Now, let us discuss the covariance of the equations of motion. 
Since the derivative $\cD_A$ generally does not transform covariantly, we assume that $\cF_A = 2\,\cD_A \bar{d}$ is constant. 
Since $\cF_{ABC}$ is also constant in PL $T$-dualizable backgrounds, the DFT equations of motion become simple algebraic equations, \eqref{eq:flux-dilaton-eom} and \eqref{eq:flux-vielbein-eom}. 

Under the PL $T$-plurality $T'_A = C_A{}^B\,T_B$\,, the generalized fluxes are mapped as
\begin{align}
 \cF'_{ABC} = C_A{}^D\,C_B{}^E\,C_C{}^F\,\cF_{DEF} \,,\qquad
 \cF'_A = C_A{}^B\,\cF_B \,,
\end{align}
by introducing $\bX_M$ when the dual algebra is non-unimodular. 
According to \eqref{eq:plural}, the untwisted metric $\hat{\cH}_{AB}$ is also related covariantly,
\begin{align}
 \hat{\cH}'_{AB} = (C\,\hat{\cH}\,C^{\rmT})_{AB} \,.
\end{align}
Then, we find that the equations of motion in the original and the dual background are covariantly related by the $\OO(n,n)$ transformation $C$\,. 
Thus, as long as the original configuration is a DFT solution, the dual background also satisfies the DFT equations of motion. 

We note that this $\OO(n,n)$ transformation is totally different from the transformation \eqref{eq:gauged-ODD}, which is just a redefinition of $U$, and the generalized metric $\cH_{MN}$ is invariant. 
On the other hand, in the case of PL $T$-plurality, $U(x)$ in the original model and $U'(x')$ in the dual model are defined on a different manifold and there is no clear connection between $U(x)$ and $U'(x')$\,. 
Only the constant fluxes made out of $U(x)$ and $U'(x')$ are related by a constant $\OO(n,n)$ transformation, and this non-trivial relation connects the two equations of motion in a covariant manner. 

Before moving on to the R--R sector, we make a brief comment on the vector field $I^m$\,. 
In order to reproduce the (generalized) supergravity from (modified) DFT, we need to choose the standard section $\tilde{\partial}^m =0$\,. 
Therefore, when $\bar{d}$ has a dual-coordinate dependence, we need to make an additional field redefinition. 
Supposing that $\bar{d}$ only has a linear dual-coordinate dependence $\bar{d}=\bar{d}_0(x^m) + d^m\,\tilde{x}_m$\,, we make the field redefinition
\begin{align}
 \bar{d}\ \to\ \bar{d}' =\bar{d}_0(x^m) \,,\qquad 
 I^m \ \to \ I'^m = \frac{1}{2}\,\tilde{f}^{\ggb\gga}{}_\ggb\,v_\gga^m + d^m \,. 
\label{eq:I-modified}
\end{align}
Then, the dual-coordinate dependence disappears from the background. 
Note that this is different from the shift \eqref{eq:mDFT-shift-PL} and is just a field redefinition. 
In the following, when we display a (generalized) supergravity solution we always make this redefinition. 

Let us also make a brief comment on the Killing vector $I^m$\,. 
In the case of NATD, the Killing vector $I^m$ is given by \eqref{eq:I-in-NATD}, but \eqref{eq:I-in-NATD} is apparently different from the formula \eqref{eq:mDFT-shift-PL} by the factor $2$\,. 
Here, we will roughly sketch how to resolve the discrepancy by using the redefinition \eqref{eq:I-modified}. 
In the case of NATD, $\partial_m \abs{\det(v^m_\gga)}=0$ and $\partial_m \Phi=0$ are usually satisfied in the original background (under the gauge fixing $x^m=c^m$). 
Then, we have
\begin{align}
\begin{split}
 \partial_m \bar{d} &= \partial_m d = -\frac{1}{2}\,\partial_m \ln \sqrt{\abs{g}} = -\frac{1}{2}\,\partial_m \ln \abs{\det(r_m^\gga)}
\\
 &= \frac{1}{2}\,\partial_m \ln \abs{\det a} = \frac{1}{2}\, f_{\ggb\gga}{}^\ggb \,\ell^\gga_m \,. 
\end{split}
\end{align}
Namely, $\bar{d}$ has a linear coordinate dependence along the $v_\gga^m$ direction,
\begin{align}
 v_\gga^m\,\partial_m \bar{d} = \frac{1}{2}\, f_{\ggb\gga}{}^\ggb \,. 
\end{align}
After performing NATD, this gives a dual-coordinate dependence of $\bar{d}$ in the dual theory,
\begin{align}
 \bar{d} = \frac{1}{2}\, \tilde{f}^{\ggb\gga}{}_\ggb\,\tilde{x}_\gga \,,
\end{align}
where the dual structure constants $\tilde{f}^{\gga\ggb}{}_\ggc$ correspond to $f_{\gga\ggb}{}^\ggc$ in the original frame. 
Then, the modified $I^m$ in \eqref{eq:I-modified} recovers the formula \eqref{eq:I-in-NATD},
\begin{align}
 I^m = \frac{1}{2}\,\tilde{f}^{\ggb\gga}{}_\ggb\,v_\gga^m + d^m = \tilde{f}^{\ggb\gga}{}_\ggb\,\delta_{\gga}^m \,, 
\end{align}
where we have used $v_\gga^m=\delta_\gga^m$ in the dual theory. 
In a general setup \eqref{eq:I-in-NATD} does not work correctly, and we use the results discussed in this section. 

\subsubsection{Duality rule for R--R fields}

Now, let us determine the duality rule for the R--R fields. 
We will first find the duality rule from a heuristic approach, and then clarify the result in terms of the gauged DFT. 

In the presence of the R--R fields, the equations of motion for $\cH_{MN}$ and $d$ are
\begin{align}
 \cS_{MN} = \cE_{MN} \,,\qquad \cS = 0 \,,
\end{align}
and since $\cS_{MN}$ is transformed covariantly under the PL $T$-duality, the energy--momentum tensor $\cE_{MN}$ should also transform covariantly,
\begin{align}
 \cE'_{MN} = (h\,\cE\,h^{\rmT})_{MN}\,. 
\end{align}
The energy--momentum tensor $\cE_{MN}$ is a bilinear form of the combination $\cF\equiv \Exp{d} F$ and it does not contain a derivative of $\cF$\,. 
Therefore, we can covariantly transform $\cE_{MN}$ simply by rotating the combination $\cF$ covariantly, and this gives the transformation rule for the R--R fields. 

Under a PL $T$-plurality, $\cH'_{MN} = (h\,\cH\,h^{\rmT})_{MN}$ with $h = U' \,C \, U^{-1}$\,, the $\OO(n,n)$-covariant transformation rule for a scalar density $\Exp{-2\,d}$ is
\begin{align}
 \Exp{-2\,d^{(h)}} = \frac{\abs{\det(e^m_\gga)}}{\abs{\det(e'^m_\gga)}} \Exp{-2\,d} \,. 
\label{eq:covariant-dilaton}
\end{align}
Indeed, the twist matrix has the form $U = R \, \bm{\Pi}$\,, and the scalar density is invariant under the $\beta$-transformation $\bm{\Pi}$ while it is multiplied by $\abs{\det(e^m_\gga)}^{-1}$ under the twist $R$\,. 
Moreover, the scalar density is invariant under the $\OO(n,n)$ transformation $C$ by the definition of $C_A{}^B$\,. 
Thus, $\Exp{-2\,d^{(h)}}$ in \eqref{eq:covariant-dilaton} is the covariantly transformed DFT dilaton. 

On the other hand, let us denote the covariantly transformed R--R polyform as $\cF^{(h)}$\,. 
By denoting the action of an $\OO(n,n)$ transformation $h$ on the polyform as $F\to \mathbb{S}_h\,F$\,,\footnote{An explicit form of the operation $\mathbb{S}_h$ is given in Appendix \ref{app:DFT}.} we have
\begin{align}
 F^{(h)} = \mathbb{S}'_U\,\mathbb{S}_C\,\mathbb{S}_U^{-1}\, F\,.
\end{align}
Then, the energy--momentum tensor made of the combination $\Exp{d^{(h)}} F^{(h)}$ is the expected $\cE'_{MN}$\,. 
However, importantly, the actual DFT dilaton is given by
\begin{align}
 \abs{\det(a'^{-1})}\abs{\det(e'^m_\gga)} \Exp{-2\,[d'-\bar{d}(x')]} = \abs{\det(a^{-1})}\abs{\det(e^m_\gga)} \Exp{-2\,[d-\bar{d}(x)]} \,,
\end{align}
and $\Exp{d'}$ is related to the covariant one $\Exp{d^{(h)}}$ as
\begin{align}
 \Exp{d'} = \frac{\sqrt{\abs{\det a}}\Exp{-\bar{d}(x)}}{\sqrt{\abs{\det a'}}\Exp{-\bar{d}(x')}}\,\Exp{d^{(h)}}\,. 
\end{align}
Therefore, if we identify the dual R--R polyform as
\begin{align}
 F' \equiv \frac{\sqrt{\abs{\det a'}}\Exp{-\bar{d}(x')}}{\sqrt{\abs{\det a}}\Exp{-\bar{d}(x)}} \, F^{(h)}\,,
\label{eq:R-R-T-plurality-rule}
\end{align}
the energy--momentum tensor made from $\Exp{d'} F' = \Exp{d^{(h)}} F^{(h)}$ is $\cE'_{MN}$\,. 
Namely, \eqref{eq:R-R-T-plurality-rule} is the rule for the R--R fields. 

Now, as $\cE_{MN}$ is transformed covariantly, it is already clear that the equations of motion for $\cH_{MN}$ and $d$ are satisfied in the dual background. 
However, the equation of motion for the R--R fields is still not clear. 
To clarify the covariance, let us rewrite \eqref{eq:R-R-T-plurality-rule} as
\begin{align}
 \hat{\cF}' = \mathbb{S}_C \, \hat{\cF}\,,
\label{eq:PL-RR-transformation}
\end{align}
where we have defined
\begin{align}
 \hat{\cF} \equiv \frac{\Exp{\bar{d}}}{\sqrt{\abs{\det a}}}\, \mathbb{S}_{U^{-1}}\, F
 = \frac{\Exp{d}}{\sqrt{\abs{\det (e_\gga^m)}}}\, \mathbb{S}_{U^{-1}}\, F \,. 
\end{align}
Then, we find that the $\hat{\cF}$ is precisely the R--R field strength appearing in the gauged DFT or the flux formulation of DFT [see Eq.~\eqref{eq:cF-hat-form}],
\begin{align}
 \ket{\cF} = \sum_p \frac{1}{p!}\, \hat{\cF}_{\gga_1\cdots\gga_p}\,\Gamma^{\gga_1\cdots\gga_p}\,\ket{0}\,. 
\end{align} 
Here, $\Gamma^{\gga_1\cdots\gga_p}\equiv \Gamma^{[\gga_1}\,\cdots \Gamma^{\gga_p]}$ and $(\Gamma^A)\equiv (\Gamma^{\gga},\,\Gamma_{\gga})$ satisfy the algebra,
\begin{align}
 \{\Gamma^A,\,\Gamma^B\} = \eta^{AB} \,,
\end{align}
and the so-called Clifford vacuum $\ket{0}$ is defined by $\Gamma_{\gga}\,\ket{0}=0$\,. 
By using a nilpotent operator
\begin{align}
 \sla{\nabla} = \sla{\partial} - \frac{1}{2}\,\Gamma^A\,\cF_A + \frac{1}{3!}\,\Gamma^{ABC}\,\cF_{ABC} \qquad (\sla{\partial}\equiv \Gamma^A\,\cD_A)\,,
\end{align}
the Bianchi identity can be expressed as (see Appendix \ref{app:DFT})
\begin{align}
 \sla{\nabla} \ket{\cF} = \Bigl(\sla{\partial} - \frac{1}{2}\,\Gamma^A\,\cF_A + \frac{1}{3!}\,\Gamma^{ABC}\,\cF_{ABC}\Bigr)\,\ket{\cF} = 0\,.
\end{align}
As is well known in the democratic formulation \cite{hep-th/9907132,hep-th/0103233}, the Bianchi identity is equivalent to the equations of motion when the self-duality relation $G_p = (-1)^{\frac{p(p-1)}{2}} * G_{10-p}$ is satisfied. 

Now, we require the dualizability condition for the R--R fields,
\begin{align}
 \sla{\partial}\ket{\cF} = 0\,,
\end{align}
which will be the same as the proposal of \cite{1810.11446}. 
Then, the Bianchi identity or the equation of motion for the R--R fields becomes an algebraic equation:
\begin{align}
 \Bigl(\frac{1}{3!}\,\Gamma^{ABC}\,\cF_{ABC} - \frac{1}{2}\,\Gamma^A\,\cF_A\Bigr)\,\ket{\cF} = 0\,.
\label{eq:Bianchi-PL}
\end{align}
Note that when the dual algebra is non-unimodular, $\cF_A$ should be modified as $\cF_A + 2\,U_A{}^M\,\bX_M$ as we explained in the discussion of the NS--NS fields. 
By denoting the spinor representative of the $\OO(n,n)$ transformation by $S_C$\,, the duality relation \eqref{eq:PL-RR-transformation} becomes simply
\begin{align}
 \ket{\cF'} = S_{C}\, \ket{\cF} \,. 
\end{align}
Then, the equation of motion \eqref{eq:Bianchi-PL} after the $\OO(n,n)$ PL $T$-plurality transformation is
\begin{align}
 \Bigl(\frac{1}{3!}\,\Gamma^{ABC}\,C_A{}^D\,C_B{}^E\,C_C{}^F\,\cF_{DEF} - \frac{1}{2}\, \Gamma^A\, C_A{}^B\,\cF_B\Bigr)\,S_C\,\ket{\hat{\cF}} = 0 \,,
\end{align}
but from the relations $S_C^{-1}\,\Gamma_A\,S_C = C_A{}^B \, \Gamma_B$ and $C_A{}^C\,C_B{}^D\,\eta_{CD}=\eta_{AB}$ this is equivalent to the equation of motion in the original background \eqref{eq:Bianchi-PL}. 
In this manner, the equation of motion for the R--R fields \eqref{eq:Bianchi-PL} is also covariantly transformed.

We call the object $\hat{\cF}$ the untwisted R--R fields, and once $\hat{\cF}'_{\gga_1\cdots\gga_p}$ in the dual background is determined from \eqref{eq:PL-RR-transformation}, we can construct the Page form in the dual background as
\begin{align}
 F'= \Exp{-\bar{d}(x')} \sqrt{\abs{\det a'}}\, \mathbb{S}_{U'}\, \hat{\cF}' =
 \Exp{-\bar{d}(x')} \sqrt{\abs{\det a'}}\, \Exp{-\bm{\Pi}'\vee} \biggl(\sum_p\frac{1}{p!}\, \hat{\cF}'_{\gga_1\cdots \gga_p}\, r'^{\gga_1}\wedge\cdots\wedge r'^{\gga_p}\biggr) \,,
\end{align}
where $\bm{\Pi}'\vee \equiv \frac{1}{2}\,\Pi'^{\gga\ggb}\,\iota_{e'_\gga}\,\iota_{e'_\gga}$\,. 

\subsubsection{Spectator fields}
\label{sec:spectator}

In the following, we consider more general cases where spectator fields are also included. 
Namely, we suppose that the original model takes the form
\begin{align}
 S = -\frac{1}{4\pi\alpha'}\int_\Sigma\rmd^2\sigma \sqrt{-\gamma}\,(\gamma^{ab}\EPSplus \varepsilon^{ab}) 
 \begin{pmatrix} \partial_a y^\mu & r_i^\gga\,\partial_a x^i \end{pmatrix}
 \begin{pmatrix}
  E_{\mu\nu} & E_{\mu\ggb} \\
  E_{\gga\nu} & E_{\gga\ggb} 
 \end{pmatrix} \begin{pmatrix} \partial_b y^\nu\\ r_j^\ggb\,\partial_b x^j \end{pmatrix} . 
\end{align}
Here, we denote the coordinates as $(x^m)=(y^\mu,\,x^i)$ $(i=1,\dotsc,n)$\,. 
By assuming that the background field $(E_{mn})=\bigl(\begin{smallmatrix} E_{\mu\nu} & E_{\mu\ggb} \\ E_{\gga\nu} & E_{\gga\ggb} \end{smallmatrix}\bigr)$ satisfies the condition
\begin{align}
 \Lie_{v_\gga}E_{mn}= - \tilde{f}^{\ggb\ggc}{}_\gga\,E_{mp}\,v_\ggb^p \, v_\ggc^q\,E_{qn} \,,
\end{align}
we can again determine $E_{mn}$ as \cite{hep-th/9502122,hep-th/9509095}
\begin{align}
\begin{alignedat}{2}
 E_{\mu\nu} &= \hat{E}_{\mu\nu} + \hat{E}_{\mu\ggc}\,\hat{E}^{\ggc\ggd}\,N_{\ggd\gge}\,\Pi^{\gge\ggf}\,\hat{E}_{\ggf\nu}\,,&\qquad 
 E_{\mu\ggb} &= \hat{E}_{\mu\ggc}\,\hat{E}^{\ggc\ggd}\,N_{\ggd\ggb} \,,
\\
 E_{\gga\nu} &= N_{\gga\ggd}\,\hat{E}^{\ggd\gge}\,\hat{E}_{\gge\nu}\,,&\qquad 
 E_{\gga\ggb}&= N_{\gga\ggb} \,,
\end{alignedat}
\label{eq:spectators-E}
\end{align}
where $(N_{\gga\ggb}) \equiv (\hat{E}^{\gga\ggb}-\Pi^{\gga\ggb})^{-1}$\,.
This reduces to \eqref{eq:original-E} when there is no spectator field. 
Now, an important difference is that $\hat{E}_{mn}$ is not necessarily constant, but can depend on the spectator fields $y^\mu$\,, $\hat{E}_{mn}=\hat{E}_{mn}(y)$\,. 
The dependence should be determined from the DFT equations of motion and is independent of the structure of the Drinfel'd double. 

In terms of the generalized metric $\cH_{MN}$\,, we can clearly see that the relation \eqref{eq:spectators-E} is a straightforward generalization of \eqref{eq:PL-LM},
\begin{align}
\begin{split}
 &\cH_{MN} = \bigl[U(x)\, \hat{\cH}(y)\, U^{\rmT}(x)\bigr]_{MN}\,, \qquad
 U(x) \equiv R\, \bm{\Pi}\,,
\\
 &(R_{M}{}^{B}) \equiv \begin{pmatrix}
 \delta_\mu^\beta& 0 & 0 & 0 \\
 0 & r_i^\ggb & 0 & 0 \\
 0 & 0 & \delta^\mu_\beta & 0 \\
 0 & 0 & 0 & e^i_\ggb
\end{pmatrix} , \qquad
 (\bm{\Pi}_{A}{}^{B}) \equiv \begin{pmatrix}
 \delta_\alpha^\beta& 0 & 0 & 0 \\
 0 & \delta_\gga^\ggb & 0 & 0 \\
 0 & 0 & \delta^\alpha_\beta & 0 \\
 0 & -\Pi^{\gga\ggb} & 0 & \delta^\gga_\ggb
 \end{pmatrix},
\end{split}
\end{align}
where $(x^M)=(y^\mu,\,x^i,\,\tilde{y}_\mu,\,\tilde{x}_i)$\,. 
The $T$-plurality transformation of \eqref{eq:plural} is also generalized, in a natural manner, as an $\OO(n,n)$ transformation,
\begin{align}
 \hat{\cH}'_{AB} = (C\,\hat{\cH}\,C^{\rmT})_{AB} \,,\qquad 
 (C_{A}{}^{B}) = \begin{pmatrix}
 \delta_\alpha^\beta& 0 & 0 & 0 \\
 0 & \bm{p}_\gga{}^\ggb & 0 & \bm{q}_{\gga\ggb} \\
 0 & 0 & \delta^\alpha_\beta & 0 \\
 0 & \bm{r}^{\gga\ggb} & 0 & \bm{s}^\gga{}_\ggb
 \end{pmatrix}. 
\label{eq:def-CAB}
\end{align}
The dilaton can also have an additional dependence on the spectators similar to \eqref{eq:SS-ansatz},
\begin{align}
 \Exp{-2\,d} = \Exp{-2\,\hat{d}(y)} \Exp{-2\,\sfd(x)}\,,\qquad
 \Exp{-2\,\sfd(x)} \equiv \Exp{-2\,\bar{d}(x)} \, \abs{\det(\ell^\gga_i)}\,. 
\end{align}
We also suppose that the untwisted R--R fields can depend on the spectator fields $\hat{\cF}=\hat{\cF}(y)$\,. 

Then, by defining the fluxes $\cF_{ABC}$ and $\cF_A$ from $U_M{}^A(x)$ and $\sfd(x)$\,, we again obtain
\begin{align}
\begin{split}
 &\cF_{\gga\ggb}{}^\ggc = f_{\gga\ggb}{}^\ggc\,,\qquad 
 \cF^{\gga\ggb}{}_{\ggc} = \tilde{f}^{\gga\ggb}{}_\ggc \,,\qquad 
 \cF_{\gga\ggb\ggc} = \cF^{\gga\ggb\ggc} = \cF_{\alpha BC} = \cF^\alpha{}_{BC} = 0 \,,
\\
 &(\cF_A) = (\cF_\alpha,\,\cF_\gga,\,\cF^\alpha,\,\cF^\gga) = (0,\,2\,\cD_\gga \sfd,\,0,\,2\,\cD^\gga \sfd) \,. 
\end{split}
\end{align}
Here, we again need to perform the shift $\partial^{M}d \to \partial^{M}d + \bX^{M}$ \eqref{eq:mDFT-shift-PL} when the dual algebra is non-unimodular. 

The requirement in \eqref{eq:external-condition} is automatically satisfied with our twist matrix, and by using \eqref{eq:generalized-Ricci-gauged} the dilaton equation of motion becomes
\begin{align}
\begin{split}
 &\hat{\cS} + \frac{1}{12}\,\cF_{ABC}\,\cF_{DEF} \,\bigl(3\,\hat{\cH}^{AD}\,\eta^{BE}\,\eta^{CF}- \hat{\cH}^{AD}\,\hat{\cH}^{BE}\,\hat{\cH}^{CF}\bigr) - \hat{\cH}^{AB}\, \cF_A\,\cF_B
\\
 &- \frac{1}{2}\,\cF^A{}_{BC}\,\hat{\cH}^{BD}\,\hat{\cH}^{CE}\,\cD_D\hat{\cH}_{AE} + 2\,\cF_A\,\cD_B\hat{\cH}^{AB} - 4\,\cF_A\,\hat{\cH}^{AB}\,\cD_B \hat{d} = 0 \,. 
\end{split}
\label{eq:dilaton-EOM-spectator}
\end{align}
By requiring that the untwisted fields $\{\hat{\cH}_{AB}(y),\,\hat{d}(y),\,\hat{\cF}(y)\}$ in the original and the dual background are covariantly related by the $\OO(n,n)$ transformation,
\begin{align}
 \hat{\cH}_{AB} = (C\,\hat{\cH}\,C^\rmT)_{AB} \,,\qquad \hat{d}' =\hat{d}\,,\qquad \hat{\cF}' = \mathbb{S}_C \, \hat{\cF}\,,
\end{align}
we can easily see that $\cD_C \hat{\cH}_{AB}=\partial_C \hat{\cH}_{AB}$ and $\cD_A \hat{d}=\partial_A \hat{d}$ are also transformed covariantly,
\begin{align}
\begin{split}
 &\cD'_C \hat{\cH}'_{AB}(y) = C_A{}^D\,C_B{}^E\,\partial_C \hat{\cH}_{DE}(y) = C_C{}^F\,C_A{}^D\,C_B{}^E\,\cD_F \hat{\cH}_{DE}(y)\,,
\\
 &\cD'_A \hat{d}(y) = \partial_A \hat{d}(y) = C_A{}^B\,\cD_B \hat{d}(y) \,,\qquad 
 \cD'_A \hat{\cF}(y) = \partial_A \hat{\cF}(y) = C_A{}^B\,\cD_B \hat{\cF}(y) \,. 
\end{split}
\end{align}
Then, the dilaton equation of motion \eqref{eq:dilaton-EOM-spectator} is satisfied in the dualized background if it is satisfied in the original background. 
As long as the untwisted R--R field satisfies ``the Bianchi identity'' $\sla{\partial}\ket{\cF} = 0$\,, which is equivalent to $\rmd \hat{\cF}(y)=0$\,, the equation of motion for the R--R fields is again a simple algebraic equation,
\begin{align}
 \Bigl(\frac{1}{3!}\,\Gamma^{ABC}\,\cF_{ABC} - \frac{1}{2}\,\Gamma^A\,\cF_A\Bigr)\,\ket{\cF(y)} = 0\,,
\end{align}
and its covariance is manifest. 
The covariance of the equations of motion for the generalized metric $\cS_{MN}=0$ can also be shown in a similar manner. 
Since the computation is a little complicated, the details are discussed in Appendix \ref{app:DFT}. 

\section{PL $T$-plurality for \texorpdfstring{$\AdS_5 \times \rmS^5$}{AdS\textfiveinferior{x}S\textfivesuperior}}
\label{sec:PL-AdS5}

In this section we show an example of the Poisson--Lie $T$-plurality. 
As already mentioned, the Lie algebra $\mathfrak{d}$ of the Drinfel'd doubles can be realized as a direct sum of two maximally isotropic algebras $\cg$ and $\tilde{\cg}$\,, and $(\mathfrak{d},\,\cg,\,\tilde{\cg})$ is called the Manin triple. 
Following \cite{math/0202210}, we denote the pair simply as $(\cg|\tilde{\cg})$\,. 
The classification of six-dimensional real Drinfel'd doubles was worked out in \cite{math/0202210}, where the following series of Manin triples corresponding to a single Drinfel'd double $\mathfrak{d}$ was found:
\begin{align}
\begin{split}
 &(\mathbf{5}|\mathbf{1}) \cong (\mathbf{6_0}|\mathbf{1}) \cong (\mathbf{5}|\mathbf{2.i}) \cong (\mathbf{6_0}|\mathbf{5.ii})
\\
 \cong{}&{}(\mathbf{1}|\mathbf{5}) \cong (\mathbf{1}|\mathbf{6_0}) \cong (\mathbf{2.i}|\mathbf{5}) \cong (\mathbf{5.ii}|\mathbf{6_0})\,.
\end{split}
\label{eq:chain}
\end{align}
Here, the characters in each slot denote the Bianchi type of the three-dimensional Lie algebra,
\begin{align}
\begin{split}
 \mathbf{1}:&\quad [X_1,\,X_2] = 0\,,\qquad 
 [X_2,\,X_3]=0\,,\qquad
 [X_3,\,X_1]= 0\,,
\\
 \mathbf{2.i}:&\quad [X_1,\,X_2] = 0\,,\qquad 
 [X_2,\,X_3]=X_1\,,\qquad
 [X_3,\,X_1]= 0\,,
\\
 \mathbf{5}:&\quad [X_1,\,X_2] = -X_2\,,\qquad 
 [X_2,\,X_3]=0\,,\qquad
 [X_3,\,X_1]= X_3\,,
\\
 \mathbf{5.ii}:&\quad [X_1,\,X_2] = -X_1 + X_2\,,\qquad 
 [X_2,\,X_3]=X_3\,,\qquad
 [X_3,\,X_1]= -X_3\,,
\\
 \mathbf{6_0}:&\quad [X_1,\,X_2] = 0\,,\qquad 
 [X_2,\,X_3]= X_1\,,\qquad
 [X_3,\,X_1]= -X_2\,. 
\end{split}
\end{align}

Using an $\OO(3,3)$ transformation $T'_A=C_A{}^B\,T_B$\,,\footnote{As pointed out in \cite{1811.12235}, the matrix $C$ which connects two Manin triples may not be unique, and a different choice of $C$ may give a different background. We will use the matrices $C$ that are given in \cite{math/0202210}.}\footnote{Originally, the indices $A,B$ in $T_A$ and $C_A{}^B$ run from $1$ to $2\,n$ ($n=3$ here), but we extend the matrix $C_A{}^B$ as in \eqref{eq:def-CAB}\,; $T_A$ should then be understood as $(T_A)=(T_\alpha, T_\gga, \tilde{T}^\alpha, \tilde{T}^\gga)=(0, T_\gga, 0, \tilde{T}^\gga)$.} the PL $T$-plurality for this chain of Manin triples was studied in \cite{hep-th/0205245}. 
However, in \cite{hep-th/0205245}, since the initial background is the flat space (or the Bianchi type V universe) the R--R fields were absent in any of the dual backgrounds. 
Moreover, there has been an issue in the treatment of the dual-coordinate dependence of the dilaton, known as the dilaton puzzle (see also \cite{hep-th/0403164,hep-th/0408126,hep-th/0601172} for detailed discussion of the issue). 
Accordingly, the only three backgrounds discussed in \cite{hep-th/0205245} were
\begin{align}
 (\mathbf{5}|\mathbf{1}) \cong (\mathbf{6_0}|\mathbf{1}) \cong (\mathbf{5}|\mathbf{2.i}) \,.
\end{align}

In this section we identify the $\AdS_5\times\rmS^5$ solution as a background with the $(\mathbf{5}|\mathbf{1})$ symmetry, and write down all of the eight backgrounds associated with the Manin triples given in \eqref{eq:chain}. 

For convenience, we summarize the procedure of the PL $T$-plurality. 
\begin{align*}
\xymatrix@C=18pt@R=1pt{
 \text{Untwisted fields} & &
\\
 \bigl\{\hat{\cH}_{AB}(y),\,\hat{d}(y),\,\hat{\cF}(y)\bigr\} \ar@/^1pc/[r] \ar@/^2pc/[rr] \ar@/^3pc/[rrr]^{\OO(n,n)\text{ transformation }C} \ar[dddddd]_{\text{twist}}^{T_A} & \bigl\{\hat{\cH}'_{AB}(y),\,\hat{d}'(y),\,\hat{\cF}'(y)\bigr\} \ar[dddddd]_{\text{twist}}^{T'_A=C'_A{}^B\,T_B} & \bigl\{\hat{\cH}''_{AB}(y),\,\hat{d}''(y),\,\hat{\cF}''(y)\bigr\} \ar[dddddd]_{\text{twist}}^{T''_A=C''_A{}^B\,T_B} & \cdots
\\
\\
\\
\\
\\
\\
 \bigl\{\cH_{MN},\,d,\,F\bigr\} & \bigl\{\cH'_{MN},\,d',\,F'\bigr\} & \bigl\{\cH''_{MN},\,d'',\,F''\bigr\} & \cdots
}
\end{align*}
We first prepare the untwisted fields $\{\hat{\cH}_{AB}(y),\,\hat{d}(y),\,\hat{\cF}(y)\}$ that satisfy
\begin{align}
 \cD_A \hat{\cH}_{BC}(y) = \partial_A \hat{\cH}_{BC}(y)\,,\qquad
 \cD_A \hat{d}(y) = \partial_A \hat{d}(y) \,,\qquad
 \cD_A \hat{\cF}(y) = \partial_A \hat{\cF}(y) \,.
\end{align}
They are independent of the structure of the Drinfel'd double and can be chosen freely. 
Under the $\OO(n,n)$ PL $T$-plurality, they are transformed covariantly,
\begin{align}
 \hat{\cH}_{AB}\to (C\,\hat{\cH}\,C^{\rmT})_{AB}\,,\qquad 
 \hat{d} \to \hat{d}\,,\qquad
 \hat{\cF} \to \mathbb{S}_C\,\hat{\cF}\,. 
\end{align}
By using the generators $T_A$ in each frame, we construct the twist matrix $U$ as
\begin{align}
\begin{split}
 &U(x) \equiv R\, \bm{\Pi}\,,\qquad 
 \bm{\Pi}\vee \equiv \frac{1}{2}\,\Pi^{\gga\ggb}\,\iota_{e_\gga}\,\iota_{e_\gga}\,,\qquad 
 \text{Ad}_{g^{-1}} = \begin{pmatrix} \delta_\gga^\ggc & 0 \\
  \Pi^{\gga\ggc} & \delta^{\gga}_\ggc \end{pmatrix} 
 \begin{pmatrix} a_\ggc{}^\ggb & 0 \\
  0 & (a^{-\rmT})^{\ggc}{}_\ggb 
\end{pmatrix} ,
\\
 &(R_{M}{}^{B}) \equiv \begin{pmatrix}
 \delta_\mu^\beta& 0 & 0 & 0 \\
 0 & r_i^\ggb & 0 & 0 \\
 0 & 0 & \delta^\mu_\beta & 0 \\
 0 & 0 & 0 & e^i_\ggb
\end{pmatrix} , \qquad
 (\bm{\Pi}_{A}{}^{B}) \equiv \begin{pmatrix}
 \delta_\alpha^\beta& 0 & 0 & 0 \\
 0 & \delta_\gga^\ggb & 0 & 0 \\
 0 & 0 & \delta^\alpha_\beta & 0 \\
 0 & -\Pi^{\gga\ggb} & 0 & \delta^\gga_\ggb
 \end{pmatrix}.
\end{split} 
\end{align}
Then, by twisting the untwisted fields, we construct the DFT fields as
\begin{align}
\begin{split}
 &\cH_{MN} = \bigl[U(x)\, \hat{\cH}(y)\, U^{\rmT}(x)\bigr]_{MN}\,, 
\\
 &\Exp{-2\,d} = \Exp{-2\,\hat{d}(y)} \Exp{-2\,\bar{d}(x)} \, \abs{\det(\ell^\gga_i)}\,,\qquad (\bX^M) = \begin{pmatrix} \frac{1}{2}\,\tilde{f}^{\ggb\gga}{}_\ggb\,v_\gga^i\,\delta_i^m \\ 0 \end{pmatrix},
\\
 &F = \Exp{-\bar{d}(x)} \sqrt{\abs{\det a}}\, \Exp{-\bm{\Pi}(x)\vee} \biggl[\sum_p\frac{1}{p!}\, \hat{\cF}_{\gga_1\cdots \gga_p}(y)\, r^{\gga_1}\wedge\cdots\wedge r^{\gga_p}\biggr] \,. 
\end{split}
\end{align}
The function $\bar{d}(x)$ is given in the initial configuration, and after the PL $T$-plurality it is rewritten in the new coordinates determined through $g(x^i)\,\tilde{g}(\tilde{x}_i)=l=g'(x'^i)\,\tilde{g}'(\tilde{x}'_i)$\,. 
When $\bar{d}(x)$ has a linear dual-coordinate dependence $d^i\,\tilde{x}_i$\,, we make a redefinition and absorb the dependence into the Killing vector, $I^i=\frac{1}{2}\,\tilde{f}^{\ggb\gga}{}_\ggb\,v_\gga^i+d^i$\,. 

\subsection{$(\mathbf{5}|\mathbf{1})$: \texorpdfstring{$\AdS_5 \times \rmS^5$}{AdS\textfiveinferior{x}S\textfivesuperior}}

We start with the $\AdS_5\times\rmS^5$ background (in a non-standard coordinate system):
\begin{align}
\begin{split}
 \rmd s^2 &= \frac{-\rmd t^2 + t^2\, \bigl[\rmd x_1^2 + \Exp{-2\,x^1} (\rmd x_2^2 + \rmd x_3^2)\bigr] +\rmd z^2}{z^2} + \rmd s^2_{\rmS^5} \,, 
\\
 G_5 &= \frac{-4\Exp{-2\,x^1}t^3\, \rmd t \wedge \rmd x^1 \wedge \rmd x^2 \wedge \rmd x^3\wedge \rmd z}{z^5} + 4\,\omega_5 \,,
\end{split}
\end{align}
where
\begin{align}
\begin{split}
 \rmd s^2_{\rmS^5} &\equiv \rmd r^2 +\sin^2r\,\rmd\xi^2 +\cos^2\xi\sin^2r\,\rmd\phi_1^2 + \sin^2r\sin^2\xi\,\rmd\phi_2^2 +\cos^2r\,\rmd\phi_3^2\,,
\\
 \omega_5 &\equiv \sin^3r \cos r\sin\xi\cos\xi\,\rmd r\wedge\rmd\xi\wedge\rmd\phi_1\wedge\rmd\phi_2\wedge\rmd\phi_3\,.
\end{split}
\end{align}
This background has Killing vectors
\begin{align}
 v_1 \equiv \partial_1 + x^2\,\partial_2 + x^3\,\partial_3 \,,\qquad
 v_2 \equiv \partial_2 \,,\qquad
 v_3 \equiv \partial_3\,,
\end{align}
satisfying the $(\mathbf{5}|\mathbf{1})$ algebra,
\begin{align}
 [v_\gga,\,v_\ggb] = f_{\gga\ggb}{}^\ggc\,v_\ggc\,,\qquad 
 f_{12}{}^2 = f_{13}{}^3 = -1\,,\qquad 
 \Lie_{v_\gga} E_{mn} = 0\,.
\end{align}

We can reconstruct this background by providing the parameterization
\begin{align}
 l= g\,\tilde{g} \,,\qquad
 g= \Exp{x^1\,T_1} \Exp{x^2\,T_2} \Exp{x^3\,T_3}\,,\qquad
 \tilde{g}= \Exp{\tilde{x}_1\,\tilde{T}^1} \Exp{\tilde{x}_2\,\tilde{T}^2} \Exp{\tilde{x}_3\,\tilde{T}^3} \,,
\end{align}
where $(T_A)=(T_\gga,\,\tilde{T}^\gga)$ are generators of the Manin triple $(\mathbf{5}|\mathbf{1})$\,. 
We obtain
\begin{align}
\begin{split}
 &\ell = \rmd x^1\,T_1 + (\rmd x^2 - x^2\,\rmd x^1)\,T_2 + (\rmd x^3 - x^3\,\rmd x^1)\,T_3 \,, 
\\
 &r = \rmd x^1\,T_1 + \Exp{-x^1} \bigl(\rmd x^2\,T_2 + \rmd x^3\,T_3\bigr) \,, 
\end{split}
\\
\begin{split}
 &a = {\footnotesize\begin{pmatrix}
 1 & -x^2 & -x^3 \\
 0 & \Exp{x^1} & 0 \\
 0 & 0 & \Exp{x^1} 
\end{pmatrix}} , \qquad
 \Pi^{\gga\ggb} = 0\,,
\end{split}
\end{align}
and they give the twist matrix $U_{M}{}^{A}$\,. 
We can easily determine the untwisted metric from the relation $\hat{\cH}_{MN}=(U^{-1}\,\cH\,U^{-\rmT})_{MN}$\,, and the result is
\begin{align}
 (\hat{E}_{mn}) = \diag\bigl(-\tfrac{1}{z^2},\,\tfrac{t^2}{z^2},\,\tfrac{t^2}{z^2},\,\tfrac{t^2}{z^2},\,\tfrac{1}{z^2},\,1,\,\sin^2 r,\, \sin^2 r\,\cos^2\xi,\, \sin^2 r\,\sin^2 \xi,\, \cos^2 r\bigr)\,,
\end{align}
in the coordinate system $(x^m)= (t, x^1, x^2, x^3, z, r, \xi, \phi_1, \phi_2, \phi_3)$\,. 
Since the dilaton is absent, $\Phi=0$\,, the DFT dilaton becomes
\begin{align}
 \Exp{-2\,d} = \sqrt{\abs{g}} = \frac{t^3 \Exp{-2\,x^1} \sin^3 r \cos r \sin\xi \cos \xi}{z^5} \,. 
\end{align}
We also have $\abs{\det(\ell^\gga_m)}=1$\,, and we can identify $\hat{d}(y)$ and $\bar{d}(x)$ as
\begin{align}
 \Exp{-2\,d} = \Exp{-2\,\hat{d}(y)}\Exp{-2\,\bar{d}(x)},\quad
 \Exp{-2\,\hat{d}(y)} \equiv \frac{t^3 \sin^3 r \cos r \sin\xi \cos \xi}{z^5} \,,\quad
 \Exp{-2\,\bar{d}(x)} \equiv \Exp{-2\,x^1} .
\end{align}
From this, we obtain the $(x^1, x^2, x^3, \tilde{x}_1, \tilde{x}_2, \tilde{x}_3)$-components of the single-index flux as
\begin{align}
 \cF_A = (2,0,0,0,0,0) \equiv \cF_A^{(\mathbf{5}|\mathbf{1})} \,.
\end{align}
In addition, from $\Exp{-\bar{d}(x)} \sqrt{\abs{\det a}}=1$\,, the untwisted R--R fields become
\begin{align}
 \hat{\cF} \equiv \sum_p\frac{1}{p!}\, \hat{\cF}_{\gga_1\cdots \gga_p}\,\rmd x^{\gga_1}\wedge\cdots\wedge \rmd x^{\gga_p}
 = -\frac{4\,t^3\, \rmd t \wedge \rmd x^1 \wedge \rmd x^2 \wedge \rmd x^3\wedge \rmd z}{z^5} + 4\,\omega_5 \,, 
\end{align}
which is a function of the spectator fields $(y^\mu)=(t, z, r, \xi, \phi_1, \phi_2, \phi_3)$\,, as expected. 

Note that if we choose the untwisted fields as
\begin{align}
 (\hat{E}_{mn}) = \diag\bigl(-1,\, t^2,\, t^2 ,\, t^2 ,\,1,\,1,\,1,\, 1,\, 1,\, 1\bigr)\,,\qquad 
 \Exp{-2\,\hat{d}} = t^3 \,,\qquad
 \hat{\cF} = 0\,,
\label{eq:NS-NS-untwisted}
\end{align}
the purely NS--NS solutions studied in \cite{hep-th/0205245} can be recovered. 

\subsection{$(\mathbf{1}|\mathbf{5})$: Type IIA GSE}

In order to consider the NATD background, we perform a redefinition of generators,
\begin{align}
 T'_A = C_A{}^B\,T_B^{(\mathbf{5}|\mathbf{1})}\,,\qquad 
 C = {\footnotesize\begin{pmatrix}
  0 & 0 & 0 & 1 & 0 & 0 \\
  0 & 0 & 0 & 0 & 1 & 0 \\
  0 & 0 & 0 & 0 & 0 & 1 \\
  1 & 0 & 0 & 0 & 0 & 0 \\
  0 & 1 & 0 & 0 & 0 & 0 \\
  0 & 0 & 1 & 0 & 0 & 0 
\end{pmatrix}}\,,
\end{align}
and give a parameterization,
\begin{align}
 l= g'\,\tilde{g}' \,,\qquad
 g'= \Exp{x'^1\,T'_1} \Exp{x'^2\,T'_2} \Exp{x'^3\,T'_3}\,,\qquad
 \tilde{g}'= \Exp{\tilde{x}'_1\,\tilde{T}'^1} \Exp{\tilde{x}'_2\,\tilde{T}'^2} \Exp{\tilde{x}'_3\,\tilde{T}'^3} \,. 
\end{align}
Then, from the identification with the original background,
\begin{align}
 g(x)\,\tilde{g}(\tilde{x}) = l= g'(x')\,\tilde{g}'(\tilde{x}') \,,
\end{align}
we find the following relation between the coordinates:
\begin{align}
\begin{split}
 x^1 &= \tilde{x}'_1\,,\quad 
 x^2 = \tilde{x}'_2 \,,\quad 
 x^3 = \tilde{x}'_3 \,,
\\
 \tilde{x}_1 &= x'^1 + x'^2\Exp{-\tilde{x}'_1} \tilde{x}'_2 + x'^3\Exp{-\tilde{x}'_1} \tilde{x}'_3\,,\quad 
 \tilde{x}_2 = \Exp{-\tilde{x}'_1} x'^2 \,,\quad
 \tilde{x}_3 = \Exp{-\tilde{x}'_1} x'^3 \,.
\end{split}
\end{align}
From this relation, we can identify $\bar{d}$ as
\begin{align}
 \Exp{-2\,\bar{d}} = \Exp{-2\,x^1} = \Exp{-2\,\tilde{x}'_1}. 
\end{align}
For notational simplicity, in the following we drop the prime. 

The untwisted fields in this frame become
\begin{align}
\begin{split}
 &(\hat{E}_{mn}) = \diag\bigl(-\tfrac{1}{z^2},\,\tfrac{z^2}{t^2},\,\tfrac{z^2}{t^2},\,\tfrac{z^2}{t^2},\,\tfrac{1}{z^2},\,1,\,\sin^2 r,\, \sin^2 r\,\cos^2\xi,\, \sin^2 r\,\sin^2 \xi,\, \cos^2 r\bigr)\,,
\\
 &\Exp{-2\,\hat{d}} = \frac{t^3 \sin^3 r \cos r \sin\xi \cos \xi}{z^5} \,, \qquad
 \hat{\cF} = -\frac{4\,t^3\, \rmd t \wedge \rmd z}{z^5} + 4\,\omega_5\wedge \rmd x^1 \wedge \rmd x^2 \wedge \rmd x^3 \,,
\end{split}
\end{align}
and we twist them by using the quantities
\begin{align}
\begin{split}
 &\ell = \rmd x^1\,T_1 + \rmd x^2\,T_2 + \rmd x^3\,T_3 \,, \quad
 r = \rmd x^1\,T_1 + \rmd x^2\, T_2 + \rmd x^3\,T_3 \,,
\\
 &v_1 = \partial_1 \,,\qquad 
 v_2 = \partial_2 \,,\qquad 
 v_3 = \partial_3 \,,
\end{split}
\\
\begin{split}
 &a = {\footnotesize\begin{pmatrix}
 1 & 0 & 0 \\
 0 & 1 & 0 \\
 0 & 0 & 1 
\end{pmatrix}} ,\qquad 
 \Pi^{\gga\ggb} = - \tilde{f}^{\gga\ggb}{}_\ggc\,x^\ggc\,. 
\end{split}
\end{align}
The resulting metric and the $B$-field are
\begin{align}
 &\rmd s^2 = \frac{-\rmd t^2 +\rmd z^2}{z^2} 
 + \frac{z^2\,\bigl[t^4\,(\rmd x_1^2 + \rmd x_2^2 + \rmd x_3^2) + z^4\,(x^3\,\rmd x^2 - x^2\,\rmd x^3)^2\bigr]}{t^2\,\bigl[t^4+ (x_2^2+x_3^2)\,z^4\bigr]} + \rmd s^2_{\rmS^5}\,,
\nn\\
 &B_2 = \frac{z^4\,\rmd x^1\wedge \bigl(x^2\,\rmd x^2+x^3\,\rmd x^3\bigr)}{t^4+ (x_2^2+x_3^2)\,z^4}\,. 
\end{align}
Since the dual algebra $\mathbf{5}$ is non-unimodular, we need to introduce the Killing vector
\begin{align}
 I = \frac{1}{2}\,\tilde{f}^{\ggb\gga}{}_\ggb\,v_\gga^i\,\partial_i = \partial_1\,. 
\end{align}
We can check that the flux $\cF_A$ is transformed covariantly from the original one, $\cF_A^{(\mathbf{5}|\mathbf{1})}$,
\begin{align}
 \cF_A = (0,0,0,2,0,0) = C_A{}^B\,\cF_B^{(\mathbf{5}|\mathbf{1})} \,,
\end{align}
which ensures that the equations of motion are transformed covariantly. 
In order to make the background a solution of GSE, we make the redefinition \eqref{eq:I-modified}, which gives
\begin{align}
 \bar{d} = 0\,,\qquad 
 I = \Bigl(\frac{1}{2}\,\tilde{f}^{\ggb\gga}{}_\ggb\,v_\gga^i + \tilde{\partial}^i\bar{d}\Bigr)\,\partial_i = 2\,\partial_1 \,. 
\end{align}
After this redefinition, the dual geometry becomes
\begin{align}
 &\rmd s^2 = \frac{-\rmd t^2 +\rmd z^2}{z^2} 
 + \frac{z^2\,\bigl[t^4\,(\rmd x_1^2 + \rmd x_2^2 + \rmd x_3^2) + z^4\,(x^3\,\rmd x^2 - x^2\,\rmd x^3)^2\bigr]}{t^2\,\bigl[t^4+ (x_2^2+x_3^2)\,z^4\bigr]} + \rmd s^2_{\rmS^5}\,,
\nn\\
 &\Exp{-2\,\Phi} = \frac{t^2\, \bigl[t^4+ (x_2^2+x_3^2)\,z^4\bigr]}{z^6}\,,\qquad 
 B_2 = \frac{z^4\,\rmd x^1\wedge \bigl(x^2\,\rmd x^2+x^3\,\rmd x^3\bigr)}{t^4+ (x_2^2+x_3^2)\,z^4}\,,
\\
 &G_2 = -\frac{4\,t^3\,\rmd t \wedge \rmd z}{z^5}\,,\qquad
 G_4 = -\frac{4\,t^3\,\rmd t \wedge \rmd x^1 \wedge \bigl(x^2\,\rmd x^2 + x^3\, \rmd x^3\bigr) \wedge \rmd z}{\bigl[t^4+ (x_2^2+x_3^2)\,z^4\bigr]\,z}\,,\qquad
 I = 2\, \partial_1\,,
\nn
\end{align}
which is a solution of type IIA GSE. 
We can explicitly check that this background has the $(\mathbf{1}|\mathbf{5})$ symmetry,
\begin{align}
 [v_\gga,\,v_\ggb] = f_{\gga\ggb}{}^\ggc\,v_\ggc =0\,,\qquad 
 \Lie_{v_\gga} E^{mn} = \tilde{f}^{\ggb\ggc}{}_\gga\,v_{\ggb}^m \,v_{\ggc}^n \,.
\end{align}

A formal $T$-duality along the $x^1$-direction gives a simple solution of type IIB supergravity
\begin{align}
\begin{split}
 &\rmd s^2 = \frac{-\rmd t^2 +\rmd z^2 + t^2\,\rmd x_1^2}{z^2} 
 + \frac{z^2\,\bigl[\bigl(\rmd x^2 - x^2\,\rmd x^1)^2 + \bigl(\rmd x^3 - x^3\,\rmd x^1)^2\bigr]}{t^2} + \rmd s^2_{\rmS^5}\,,
\\
 &\Phi = \ln \Bigl(\frac{z^2}{t^2}\Bigr) + 2\,x^1 \,,\qquad
 G_3 = \frac{4\,t^3 \Exp{-2\,x^1} \rmd t\wedge \rmd x^1 \wedge \rmd z}{z^5}\,.
\end{split}
\end{align}

\subsection{\texorpdfstring{$(\mathbf{6_0}|\mathbf{1})$}{(6\textzeroinferior|1)}: Type IIA SUGRA}

We next perform the following redefinition of the original $(\mathbf{5}|\mathbf{1})$ generators:
\begin{align}
 T'_A = C_A{}^B\,T_B^{(\mathbf{5}|\mathbf{1})}\,,\qquad 
 C = {\footnotesize\begin{pmatrix}
  0 & 0 & -\frac{1}{2} & 0 & 1 & 0 \\
  0 & 0 & \frac{1}{2} & 0 & 1 & 0 \\
  -1 & 0 & 0 & 0 & 0 & 0 \\
  0 & \frac{1}{2} & 0 & 0 & 0 & -1 \\
  0 & \frac{1}{2} & 0 & 0 & 0 & 1 \\
  0 & 0 & 0 & -1 & 0 & 0 
\end{pmatrix}}\,. 
\end{align}
This time, we provide the parameterization
\begin{align}
 l= g'\,\tilde{g}' \,,\qquad
 g' = \Exp{-x'^3\,T'_3}\Exp{x'^2\,T'_2}\Exp{x'^1\,T'_1}\,,\qquad 
 \tilde{g}' = \Exp{\tilde{x}'_1\,\tilde{T}'^1}\Exp{\tilde{x}'_2\,\tilde{T}'^2}\Exp{-\tilde{x}'_3\,\tilde{T}'^3}\,, 
\end{align}
and the coordinates are related to the original ones as
\begin{align}
\begin{split}
 x^1 &= x'^3\,,\quad 
 x^2 = \frac{\tilde{x}'_1 + \tilde{x}'_2}{2}\,,\quad 
 x^3 = \frac{x'^2 - x'^1}{2} \,,
\\
 \tilde{x}_1 &= \tilde{x}'_3+\frac{(x'^1 + x'^2)\,(\tilde{x}'_1 + \tilde{x}'_2)}{2}\,,\quad 
 \tilde{x}_2 = x'^1 + x'^2 \,,\quad
 \tilde{x}_3 = \tilde{x}'_2-\tilde{x}'_1 \,.
\end{split}
\end{align}
Then, in this frame, $\bar{d}$ becomes
\begin{align}
 \Exp{-2\,\bar{d}} = \Exp{-2\,x^1} = \Exp{-2\,x'^3}. 
\end{align}
Again we remove the prime, and then the $(t,x^1,x^2,x^3,z)$-part of the untwisted metric becomes
\begin{align}
 (\hat{E}_{mn})
 = \begin{pmatrix}
 -\frac{1}{z^2} & 0 & 0 & 0 & 0 \\
 0 & \frac{t^2}{4 z^2}+\frac{z^2}{t^2} & \frac{z^2}{t^2}-\frac{t^2}{4 z^2} & 0 & 0 \\
 0 & \frac{z^2}{t^2}-\frac{t^2}{4 z^2} & \frac{t^2}{4 z^2}+\frac{z^2}{t^2} & 0 & 0 \\
 0 & 0 & 0 & \frac{t^2}{z^2} & 0 \\
 0 & 0 & 0 & 0 & \frac{1}{z^2}
\end{pmatrix} .
\end{align}
In order to obtain the untwisted R--R fields, it may be useful to decompose the matrix $C$ into products of $\GL(D)$ transformation, $B$-transformation, $T$-dualities, and $\beta$-transformation. 
In this case, for example, we can use the decomposition
\begin{align}
 C = {\footnotesize\begin{pmatrix}
 0 & 1 & -\frac{1}{2} & 0 & 0 & 0 \\
 0 & 1 & \frac{1}{2} & 0 & 0 & 0 \\
 -1 & 0 & 0 & 0 & 0 & 0 \\
 0 & 0 & 0 & 0 & \frac{1}{2} & -1 \\
 0 & 0 & 0 & 0 & \frac{1}{2} & 1 \\
 0 & 0 & 0 & -1 & 0 & 0 
\end{pmatrix} \begin{pmatrix}
 1 & 0 & 0 & 0 & 0 & 0 \\
 0 & 0 & 0 & 0 & 1 & 0 \\
 0 & 0 & 1 & 0 & 0 & 0 \\
 0 & 0 & 0 & 1 & 0 & 0 \\
 0 & 1 & 0 & 0 & 0 & 0 \\
 0 & 0 & 0 & 0 & 0 & 1 
\end{pmatrix}}.
\end{align}
Then, the $T$-duality along the $x^2$-direction and the $\GL(3)$ transformation give
\begin{align}
 \hat{\cF} &= \frac{2\,t^3\,\rmd t \wedge (\rmd x^1 - \rmd x^2)\wedge \rmd x^3\wedge \rmd z}{z^5} - 4\, (\rmd x^1 + \rmd x^2)\wedge \omega_5 \,. 
\end{align}
In order to obtain the twist matrix, we compute
\begin{align}
\begin{split}
 &\ell =(\rmd x^1+x^2\,\rmd x^3)\,T_1 + (\rmd x^2+x^2\,\rmd x^3)\,T_2 - \rmd x^3\,T_3 \,,
\\
 &r = (\cosh x^3\,\rmd x^1+\sinh x^3\,\rmd x^2)\,T_1 + (\sinh x^3\,\rmd x^1+\cosh x^3\,\rmd x^2)\,T_2 - \rmd x^3\,T_3 \,,
\\
 &v_1 = \partial_1 \,,\qquad
 v_2 = \partial_2\,,\qquad
 v_3 = x^2\,\partial_1 + x^1\,\partial_2 - \partial_3 \,,
\end{split}
\\
\begin{split}
 &a = {\footnotesize\begin{pmatrix}
 \cosh x^3 & -\sinh x^3 & 0 \\
 -\sinh x^3 & \cosh x^3 & 0 \\
 -x^2 & -x^1 & 1 
\end{pmatrix}} , \qquad \Pi^{\gga\ggb} = 0\,. 
\end{split}
\end{align}
Again, the flux $\cF_A$ is transformed covariantly,
\begin{align}
 (\cF_A)=(0,\,0,\,-2,\,0,\,0,\,0) = C_A{}^B\,\cF_B^{(\mathbf{5}|\mathbf{1})} \,.
\end{align}
The background fields are determined as
\begin{align}
\begin{split}
 &\rmd s^2 = \frac{-\rmd t^2 + t^2\,\rmd x_3^2 + \rmd z^2}{z^2} 
 + \frac{\Exp{-2\,x^3}t^2\,(\rmd x^1 -\rmd x^2)^2}{4\,z^2}+\frac{\Exp{2\,x^3} z^2\,(\rmd x^1+\rmd x^2)^2}{t^2} + \rmd s^2_{\rmS^5}\,,
\\
 &\Exp{-2\,\Phi}= \frac{\Exp{-2\,x^3} t^2}{z^2}\,, \qquad 
 G_4 = \frac{2 \Exp{-2\,x^3} t^3\, \bigl( \rmd x^1 - \rmd x^2 \bigr)\wedge \rmd t\wedge \rmd x^3 \wedge \rmd z}{z^5} \,,
\end{split}
\end{align}
and this is a solution of type IIA supergravity. 

\subsection{\texorpdfstring{$(\mathbf{1}|\mathbf{6_0})$}{(1|6\textzeroinferior)}: Type IIB GSE}

The NATD of the $(\mathbf{6_0}|\mathbf{1})$ background, namely $(\mathbf{1}|\mathbf{6_0})$ can be realized by
\begin{align}
 T'_A = C_A{}^B\,T_B^{(\mathbf{5}|\mathbf{1})}\,,\qquad 
 C = {\footnotesize\begin{pmatrix}
 0 & \frac{1}{2} & 0 & 0 & 0 & -1 \\
 0 & \frac{1}{2} & 0 & 0 & 0 & 1 \\
 0 & 0 & 0 & -1 & 0 & 0 \\
 0 & 0 & -\frac{1}{2} & 0 & 1 & 0 \\
 0 & 0 & \frac{1}{2} & 0 & 1 & 0 \\
 -1 & 0 & 0 & 0 & 0 & 0 \\
\end{pmatrix}}.
\end{align}
We give the parameterization
\begin{align}
 l= g'\,\tilde{g}' \,,\qquad
 g' = \Exp{x'^1\,T'_1}\Exp{x'^2\,T'_2}\Exp{-x'^3\,T'_3}\,,\qquad 
 \tilde{g}' = \Exp{-\tilde{x}'_3\,\tilde{T}'^3}\Exp{\tilde{x}'_2\,\tilde{T}'^2}\Exp{\tilde{x}'_1\,\tilde{T}'^1}\,. 
\end{align}
In order to determine $\bar{d}$\,, it is enough to identify the coordinate $x^1$\,, and we find
\begin{align}
 \Exp{-2\,\bar{d}} = \Exp{-2\,x^1} = \Exp{-2\,\tilde{x}'_3}. 
\end{align}
Note that the appearance of the dual-coordinate dependence was discussed in \cite{hep-th/0205245}, but at that time DFT had not been developed and the interpretation was not clear.

We can construct the twist matrix $U$ from
\begin{align}
\begin{split}
 &\ell = \rmd x^1 \,T_1 + \rmd x^2\,T_2 - \rmd x^3 \,T_3 \,,\qquad
 r = \rmd x^1 \,T_1 + \rmd x^2\,T_2 - \rmd x^3 \,T_3 \,,
\\
 &v_1 = \partial_1 \,,\qquad
 v_2 = \partial_2\,,\qquad
 v_3 = - \partial_3 \,.
\end{split}
\\
\begin{split}
 &a = {\footnotesize\begin{pmatrix}
 1 & 0 & 0 \\
 0 & 1 & 0 \\
 0 & 0 & 1 
\end{pmatrix}} \,, \qquad 
 (\Pi^{\gga\ggb}) = {\footnotesize\begin{pmatrix}
 0 & 0 & -x^2 \\
 0 & 0 & -x^1 \\
 x^2 & x^1 & 0 
\end{pmatrix}} ,
\end{split}
\end{align}
and the flux $\cF_A$ becomes
\begin{align}
 (\cF_A)=(0,\,0,\,0,\,0,\,0,\,-2) = C_A{}^B\,\,\cF_B^{(\mathbf{5}|\mathbf{1})}\,.
\end{align}
Thus, the DFT equations of motion are covariantly transformed. 

Although the dual algebra is unimodular, in order to absorb the dual coordinate dependence in $\bar{d}$\,, we make a field redefinition \eqref{eq:I-modified}, and obtain
\begin{align}
 \Exp{-2\,\bar{d}} = 1\,,\qquad I = \partial_3\,. 
\end{align}
After the redefinition we obtain a solution of type IIB GSE,
\begin{align}
\begin{split}
 &\rmd s^2 = \frac{-\rmd t^2 + \rmd z^2}{z^2} + \rmd s^2_{\rmS^5}
\\
 &\qquad +\frac{t^6\,(\rmd x^1+\rmd x^2)^2+4\,t^2\,z^4\,\bigl[(\rmd x^1-\rmd x^2)^2+ (x^1 \, \rmd x^1 - x^2 \, \rmd x^2)^2+\rmd x_3^2\bigr]}{t^4\,z^2\,\bigl[(x^1+x^2)^2+4\bigr] +4\,z^6\,(x^1-x^2)^2}\,,
\\
 &B_2 = \frac{t^4\, (x^1+x^2) \,(\rmd x^1+\rmd x^2) - 4\,z^4\,(x^1-x^2)\,(\rmd x^1-\rmd x^2)}{t^4\, \bigl[(x^1+x^2)^2+4\bigr] +4\,z^4\,(x^1-x^2)^2} \wedge \rmd x^3 \,,\qquad
 I = \partial_3\,,
\\
 &\Exp{-2\,\Phi} = \frac{t^4\,\bigl[(x^1+x^2)^2+4\bigr] + 4\, z^4\,(x^1-x^2)^2}{4\,z^4}\,, \qquad
 G_3 = \frac{2\,t^3\,(\rmd x^1 +\rmd x^2)\wedge \rmd t \wedge \rmd z}{z^5}\,,
\\
 &G_5 =2\,(x^1-x^2)\,\biggl[\frac{8\,t^3\,z^{-1}\,\rmd t \wedge \rmd x^1\wedge \rmd x^2\wedge \rmd x^3\wedge \rmd z}{t^4\, \bigl[(x^1+x^2)^2+4\bigr] +4\,z^4\,(x^1-x^2)^2} - 2\,\omega_{\rmS^5}\biggr] \,.
\end{split}
\label{eq:RR-61}
\end{align}

It is important to note that the duality $(\mathbf{6_0}|\mathbf{1})\to (\mathbf{1}|\mathbf{6_0})$ is a NATD for traceless structure constants. 
In the literature, it has been discussed that if the structure constants are traceless, the NATD background satisfies the supergravity equations of motion, but here we obtained a solution of GSE. 
The consistency is to be clarified in a future study. 
Of course, the existence of the R--R fields is not important here. 
As already mentioned, we can obtain a purely NS--NS solution, by starting with the untwisted fields \eqref{eq:NS-NS-untwisted}. 
The $(\mathbf{6_0}|\mathbf{1})$ background is
\begin{align}
\begin{split}
 &\rmd s^2 = -\rmd t^2 + t^2\,\rmd x_3^2 
 + \frac{1}{4}\Exp{-2\,x^3}t^2\,(\rmd x^1 -\rmd x^2)^2+ \Exp{2\,x^3}t^{-2}\, (\rmd x^1+\rmd x^2)^2 + \rmd s^2_{\TT^6}\,,
\\
 &\Exp{-2\,\Phi}= \Exp{-2\,x^3} t^2 \,,
\end{split}
\end{align}
while its NATD, namely the $(\mathbf{1}|\mathbf{6_0})$ background, is a GSE solution,
\begin{align}
\begin{split}
 &\rmd s^2 = -\rmd t^2 + \rmd s^2_{\TT^6}
\\
 &\qquad +\frac{t^6\,(\rmd x^1+\rmd x^2)^2+4\,t^2\, \bigl[(\rmd x^1-\rmd x^2)^2+ (x^1 \, \rmd x^1 - x^2 \, \rmd x^2)^2+\rmd x_3^2\bigr]}{t^4\, \bigl[(x^1+x^2)^2+4\bigr] +4\, (x^1-x^2)^2}\,,
\\
 &B_2 = \frac{t^4\, (x^1+x^2) \,(\rmd x^1+\rmd x^2) - 4\, (x^1-x^2)\,(\rmd x^1-\rmd x^2)}{t^4\, \bigl[(x^1+x^2)^2+4\bigr] +4\, (x^1-x^2)^2} \wedge \rmd x^3 \,,\qquad
 I = \partial_3\,,
\\
 &\Exp{-2\,\Phi} = \frac{t^4\,\bigl[(x^1+x^2)^2+4\bigr] + 4\, (x^1-x^2)^2}{4} \,.
\end{split}
\end{align}
It will be interesting to study string theory on these backgrounds in detail. 

We also note that in the $(\mathbf{1}|\mathbf{6_0})$ background \eqref{eq:RR-61}, if we perform a formal $T$-duality along the $x^3$-direction we obtain a solution of type IIA supergravity,
\begin{align}
\begin{split}
 &\rmd s^2 = \frac{-\rmd t^2 + \rmd z^2}{z^2}
 +\frac{z^2\,[\rmd x^1 -\rmd x^2-(x^1-x^2)\,\rmd x^3]^2}{t^2}
\\
 &\qquad\ + \frac{t^2\,[(\rmd x^1 +\rmd x^2)^2 + 2\,(x^1+x^2)\,(\rmd x^1 +\rmd x^2)\,\rmd x^3 + [(x^1+x^2)^2+4]\,\rmd x_3^2]}{4\,z^2} + \rmd s^2_{\rmS^5}\,,
\\
 &\Exp{-2\,\Phi} = \frac{\Exp{-2\,x^3} t^2}{z^2}\,,\qquad
 G_4 = - \frac{2\Exp{-x^3} t^3 \,(\rmd x^1 + \rmd x^2)\wedge \rmd t \wedge \rmd x^3 \wedge \rmd z}{z^5}\,.
\end{split}
\end{align}

\subsection{$(\mathbf{5}|\mathbf{2.i})$: Type IIB SUGRA}

In order to obtain the Manin triple $(\mathbf{5}|\mathbf{2.i})$, we perform the redefinition
\begin{align}
 T'_A = C_A{}^B\,T_B^{(\mathbf{5}|\mathbf{1})}\,,\qquad 
 C = {\footnotesize\begin{pmatrix}
 -1 & 0 & 0 & 0 & 0 & 0 \\
 0 & 0 & 0 & 0 & 1 & 0 \\
 0 & 0 & 0 & 0 & 0 & 1 \\
 0 & 0 & 0 & -1 & 0 & 0 \\
 0 & 1 & 0 & 0 & 0 & -\frac{1}{2} \\
 0 & 0 & 1 & 0 & \frac{1}{2} & 0 
\end{pmatrix}}\,.
\end{align}
Again we consider the parameterization
\begin{align}
 l= g'\,\tilde{g}' \,,\qquad
 g' = \Exp{x'^1\,T'_1}\Exp{x'^2\, T'_2} \Exp{-x'^3\,T'_3}\,,\qquad
 \tilde{g}' = \Exp{\tilde{x}'_1\,\tilde{T}'^1} \Exp{\tilde{x}'_2\,\tilde{T}'^2} \Exp{-\tilde{x}'_3\,\tilde{T}'^3}\,,
\end{align}
and from the coordinate transformation, we obtain
\begin{align}
 \Exp{-2\,\bar{d}} = \Exp{-2\,x^1} = \Exp{2\,x'^1}. 
\end{align}
The necessary quantities are obtained as
\begin{align}
\begin{split}
 &\ell = \rmd x^1\,T_1 + (\rmd x^2 -x^2\,\rmd x^1)\,T_2 - (\rmd x^3-x^3\,\rmd x^1)\,T_3 \,,
\\
 &r = \rmd x^1\,T_1 + \Exp{-x^1} \bigl(\rmd x^2\,T_2 - \rmd x^3\,T_3\bigr) \,,
\\
 &v_1 = \partial_1+ x^2\,\partial_2 + x^3\,\partial_3 \,,\qquad 
 v_2 = \partial_2 \,,\qquad 
 v_3 = - \partial_3 \,,
\end{split}
\\
\begin{split}
 &a = {\footnotesize\begin{pmatrix}
 1 & -x^2 & x^3 \\
 0 & \Exp{x^1} & 0 \\
 0 & 0 & \Exp{x^1} \end{pmatrix}} , \qquad
 (\Pi^{\gga\ggb}) = {\footnotesize\begin{pmatrix}
 0 & 0 & 0 \\
 0 & 0 & -\Exp{-x^1}\sinh x^1 \\
 0 & \Exp{-x^1}\sinh x^1 & 0 
\end{pmatrix}} \,,
\end{split}
\end{align}
and again the flux $\cF_A$ is covariantly transformed,
\begin{align}
 (\cF_A)=(-2,\,0,\,0,\,0,\,0,\,0) = C_A{}^B\,\,\cF_B^{(\mathbf{5}|\mathbf{1})}\,.
\end{align}

A straightforward computation gives
\begin{align}
\begin{split}
 \rmd s^2 &= \frac{-\rmd t^2 + t^2\,\rmd x_1^2 + \rmd z^2}{z^2} + \frac{4 \Exp{2\,x^1} t^2\, z^2\,(\rmd x_2^2+\rmd x_3^2)}{4\Exp{4\,x^2} t^4 +z^4} + \rmd s^2_{\rmS^5}\,,
\\
 B_2&= - \frac{2\,z^4\,\rmd x^2\wedge\rmd x^3}{4\Exp{4\,x^1}t^4 + z^4} \,, \qquad
 \Exp{-2\,\Phi} = \frac{4\Exp{4\,x^1}t^4+ z^4}{4\,z^4}\,, 
\\
 G_3 &=-\Exp{2\,x^1} \frac{4\,t^3\,\rmd t \wedge \rmd x^1 \wedge \rmd z}{z^5}\,,
\\
 G_5 &= - \frac{8\Exp{2\,x^1}t^3}{z}\,\frac{\rmd t\wedge \rmd x^1 \wedge \rmd x^2\wedge \rmd x^3\wedge \rmd z}{4\Exp{4\,x^1} t^4 + z^4} + 2\,\omega_{\rmS^5} \,,
\end{split}
\end{align}
and this is a solution of type IIB supergravity.

\subsection{$(\mathbf{2.i}|\mathbf{5})$: Type IIA SUGRA}

We next consider the transformation
\begin{align}
 T'_A = C_A{}^B\,T_B^{(\mathbf{5}|\mathbf{1})}\,,\qquad 
 C = {\footnotesize\begin{pmatrix}
 0 & 0 & 0 & -1 & 0 & 0 \\
 0 & 1 & 0 & 0 & 0 & -\frac{1}{2} \\
 0 & 0 & 1 & 0 & \frac{1}{2} & 0 \\
 -1 & 0 & 0 & 0 & 0 & 0 \\
 0 & 0 & 0 & 0 & 1 & 0 \\
 0 & 0 & 0 & 0 & 0 & 1 
\end{pmatrix}},
\end{align}
and provide the parameterization
\begin{align}
 l= g'\,\tilde{g}' \,,\qquad
 g' = \Exp{x'^1\,T'_1}\Exp{x'^2\, T'_2} \Exp{-x'^3\,T'_3}\,,\qquad
 \tilde{g}' = \Exp{\tilde{x}'_1\,\tilde{T}'^1} \Exp{\tilde{x}'_2\,\tilde{T}'^2} \Exp{-\tilde{x}'_3\,\tilde{T}'^3}\,. 
\end{align}
The coordinate transformation gives
\begin{align}
 \Exp{-2\,\bar{d}} = \Exp{-2\,x^1} = \Exp{2\,\tilde{x}'_1}. 
\end{align}

Again, we compute
\begin{align}
\begin{split}
 &\ell = (\rmd x^1 - x^3\,\rmd x^2\bigr)\,T_1 + \rmd x^2\,T_2 - \rmd x^3\,T_3 \,,
\\
 &r = (\rmd x^1 - x^2\,\rmd x^3\bigr)\,T_1 + \rmd x^2\,T_2 - \rmd x^3\,T_3 \,,
\\
 &v_1 = \partial_1 \,,\qquad
 v_2 = x^3\, \partial_1 + \partial_2 \,,\qquad
 v_3 = - \partial_3\,,
\end{split}
\\
\begin{split}
 &a = {\footnotesize\begin{pmatrix}
 1 & 0 & 0 \\
 - x^3 & 1 & 0 \\
 - x^2 & 0 & 1
\end{pmatrix}} , \qquad
 (\Pi^{\gga\ggb}) = {\footnotesize\begin{pmatrix}
 0 & x^2 & -x^3 \\
 -x^2 & 0 & 0 \\
 x^3 & 0 & 0 
\end{pmatrix}} \,,
\end{split}
\end{align}
and we can check that the flux is covariantly transformed,
\begin{align}
 (\cF_A)=(0,\,0,\,0,\,-2,\,0,\,0) = C_A{}^B\,\,\cF_B^{(\mathbf{5}|\mathbf{1})}\,.
\end{align}

Since the dual algebra $\mathbf{5}$ is non-unimodular, we have
\begin{align}
 I = \frac{1}{2}\,\tilde{f}^{\ggb\gga}{}_\ggb\,v_\gga^m\,\partial_m = \partial_1 \,. 
\end{align}
We thus expect that this background is a solution of the GSE. 
However, according to the field redefinition \eqref{eq:I-modified}, we obtain
\begin{align}
 \Exp{-2\,\bar{d}} = 1\,,\qquad 
 I = \partial_1 - \partial_1 = 0 \,.
\end{align}
As the result, we obtain a solution of the conventional type IIA supergravity,
\begin{align}
\begin{split}
 \rmd s^2 &= \frac{-\rmd t^2 + \rmd z^2}{z^2} 
+ \frac{z^2\,\bigl[4\, \rmd x^1\,(\rmd x^1 - x^3\,\rmd x^2 - x^2\,\rmd x^3) + ( x^3 \,\rmd x^2 + x^2 \,\rmd x^3)^2\bigr]}{4\,t^2\,(1+ x_2^2+ x_3^2)}
\\
 &\quad + \frac{t^2\,\bigl[\rmd x_2^2 + \rmd x_3^2 + ( x^3 \,\rmd x^2 - x^2 \,\rmd x^3)^2\bigr]}{z^2\,(1+ x_2^2+ x_3^2)}
 + \rmd s^2_{\rmS^5}\,,
\\
 B_2&= \frac{\rmd x^1\wedge ( x^2\,\rmd x^2+ x^3\,\rmd x^3)}{1+ x_2^2+ x_3^2}
 + \frac{(1+2\, x_2^2)\,\rmd x^2\wedge\rmd x^3}{2\,(1+ x_2^2+ x_3^2)}\,,
\\
 \Exp{-2\,\Phi} &= \frac{t^2\,(1+ x_2^2+ x_3^2)}{z^2}\,,\qquad 
 G_4 = -\frac{4\,t^3\,\rmd t \wedge \rmd x^2 \wedge \rmd x^3 \wedge \rmd z}{z^5}\,.
\end{split}
\end{align}
Namely, even if the dual algebra is non-unimodular, the background can satisfy the usual supergravity equations of motion. 
This is a remarkable example of such unusual cases. 

\subsection{\texorpdfstring{$(\mathbf{5.ii}|\mathbf{6_0})$}{(5.ii|6\textzeroinferior)}: Type IIB GSE}

We next consider
\begin{align}
 T'_A = C_A{}^B\,T_B^{(\mathbf{5}|\mathbf{1})}\,,\qquad 
 C = {\footnotesize\begin{pmatrix}
 1 & 0 & 0 & 0 & -1 & 0 \\
 1 & 0 & 0 & 0 & 0 & 0 \\
 0 & 0 & 0 & 0 & 0 & 1 \\
 0 & -1 & 0 & 0 & 0 & \frac{1}{2} \\
 0 & 1 & 0 & 1 & 0 & \frac{1}{2} \\
 -1 & 0 & 1 & 0 & \frac{1}{2} & 0 \\
\end{pmatrix}},
\end{align}
and give the parameterization,
\begin{align}
 l= g'\,\tilde{g}' \,,\qquad
 g'= \Exp{x'^1\,T'_1}\Exp{(x'^2-x'^1)\,T'_2}\Exp{x'^3\,T'_3}\,,\qquad
 \tilde{g}'= \Exp{\tilde{x}'_3\,\tilde{T}'^3}\Exp{\tilde{x}'_2 \,\tilde{T}'^2}\Exp{(\tilde{x}'_1+\tilde{x}'_2)\,\tilde{T}'^1}\,.
\end{align}
We then obtain $\bar{d}$ as
\begin{align}
 \Exp{-2\,\bar{d}} = \Exp{-2\,x^1} = \Exp{-2\,(x'^2-\tilde{x}'_3)}. 
\end{align}
From a straightforward computation,
\begin{align}
\begin{split}
 &\ell = \Exp{x^1-x^2}\,\rmd x^1\,T_1+(\rmd x^2-\Exp{x^1-x^2}\rmd x^1)\,T_2 +(\rmd x^3+x^3\,\rmd x^2)\,T_3 \,,
\\
 &r = \bigl[\Exp{x^1}\,\rmd x^1 +(1-\Exp{x^1})\,\rmd x^2\bigr]\,T_1+\Exp{x^1}(\rmd x^2-\rmd x^1)\,T_2 +\Exp{x^2} \rmd x^3\,T_3 \,,
\\
 &v_1 = \Exp{x^2-x^1} \partial_1 + \partial_2 - x^3\, \partial_3\,,\qquad
 v_2 = \partial_2 - x^3\, \partial_3\,,\qquad
 v_3 = \partial_3\,,
\end{split}
\\
\begin{split}
 &a ={\footnotesize\begin{pmatrix}
 \Exp{x^1-x^2} & 1-\Exp{x^1-x^2} & x^3 \\
 \Exp{-x^2} (\Exp{x^1}-1) & 1+\Exp{-x^2} (1-\Exp{x^1}) & x^3 \\
 0 & 0 & \Exp{-x^2} \end{pmatrix}} ,
\\
 &(\Pi^{\gga\ggb}) = {\footnotesize\begin{pmatrix}
 0 & 0 & \frac{1-\Exp{x^2} (2-2 \Exp{x^1}+\Exp{x^2})}{2} \\
 0 & 0 & \frac{1+\Exp{2 x^2}-2 \Exp{x^1+x^2}}{2} \\
 -\frac{1-\Exp{x^2} (2-2 \Exp{x^1}+\Exp{x^2})}{2} & -\frac{1+\Exp{2 x^2}-2 \Exp{x^1+x^2}}{2} & 0 
\end{pmatrix}}\,, 
\end{split}
\end{align}
we obtain the twist matrix $U$, and the flux is covariantly transformed
\begin{align}
 (\cF_A)=(2,\,2,\,0,\,0,\,0,\,-2) = C_A{}^B\,\,\cF_B^{(\mathbf{5}|\mathbf{1})}\,.
\end{align}

Since the dual algebra is unimodular, originally we have $I^m=0$\,. 
However, due to the dual-coordinate dependence of $\bar{d}$\,, we make the field redefinition \eqref{eq:I-modified} and obtain
\begin{align}
 \Exp{-2\,\bar{d}} = \Exp{-2\, x^2},\qquad 
 I = - \partial_3 \,.
\end{align}
After the redefinition we obtain a solution of type IIB GSE,
\begin{align}
\begin{split}
 \rmd s^2 &= \frac{-\rmd t^2 + \rmd z^2}{z^2} + \rmd s^2_{\rmS^5}
\\
 &+\frac{t^2\,\bigl\{ 4 \Exp{2\,x^2} z^4\,(\Exp{2\,x^1} \rmd x_1^2 +\rmd x_3^2) + [4\,(t^4+z^4) + \Exp{4\,x^2} z^4]\,\rmd x_2^2\bigr\}}{\Delta^2} 
\\
 &+\frac{4\Exp{x^1} t^2\,z^4\,[\Exp{x^1} (\rmd x^1 -\rmd x^2)^2- \Exp{3\,x^2} \rmd x^1\,\rmd x^2+2\,(\rmd x^1 -\rmd x^2)\,\rmd x^2]}{\Delta^2} \,,
\\
 B_2&= -\frac{2 \Exp{2\,x^2} z^2\,\bigl\{2\,t^4\,\rmd x^2 -z^4\,(2 \Exp{x^1}-\Exp{x^2}-2)\,\bigl[\Exp{x^1} \rmd x^1 - (\Exp{x^1}-1)\,\rmd x^2\bigr]\bigr\}\wedge \rmd x^3}{\Delta^2} \,, 
\\
 \Exp{-2\,\Phi} &= \frac{\Exp{-4\,x^2}\Delta^2}{4\,z^4}\,,\qquad
 I= - \partial_3\,, \qquad
 G_3 = \frac{4\Exp{-2\,x^2} t^3\,\rmd t \wedge \rmd x^2 \wedge \rmd z}{z^5}\,, 
\\
 G_5 &= (2\Exp{x^1}-\Exp{x^2}-2)\,\Bigl[\frac{8\,t^3 \Exp{x^1} z\,\rmd t \wedge \rmd x_1 \wedge \rmd x^2 \wedge \rmd x^3 \wedge \rmd z}{\Delta^2} - 2 \Exp{-x^2} \omega_{\rmS^5}\Bigr] \,,
\end{split}
\end{align}
which is defined on the region
\begin{align}
 \Delta^2 &\equiv 4\,t^4\,(\Exp{2\,x^2}+1)\,z^2 + \Exp{2\,x^2} z^6\,(2-2\Exp{x^1}+\Exp{x^2})^2 \geq 0 \,.
\end{align}

A formal $T$-duality along the $x^3$-direction gives a solution of type IIA supergravity,
\begin{align}
\begin{split}
 &\rmd s^2 = \frac{-\rmd t^2 + \rmd z^2}{z^2} 
 +\frac{(t^4+z^4)\,(\rmd x^2 -\rmd x^3)^2 + z^4\Exp{2\,x^1} (\rmd x^1 -\rmd x^2 +\rmd x^3)^2}{t^2\,z^2}
\\
 &\qquad +z^2 \Exp{x^1} \frac{2\, (\rmd x^2 -\rmd x^3)\,(\rmd x^1 -\rmd x^2 +\rmd x^3)- \Exp{x^2} (\rmd x^1 -\rmd x^2 +\rmd x^3)\,\rmd x^3}{t^2}
\\
 &\qquad + z^2 \Exp{x^2} \frac{(\rmd x^3 - \rmd x^2)\,\rmd x^3}{t^2} + \frac{\Exp{-2\,x^2} (4\,t^4+\Exp{4\,x^2} z^4)\,\rmd x_3^2}{4\,t^2\,z^2} + \rmd s^2_{\rmS^5}\,,
\nn\\
 &\Exp{-2\,\Phi} =\frac{t^2 \Exp{2\,(x^3 - x^2)}}{z^2}\,,\qquad 
 G_4 = -\frac{4\Exp{x^3 -2\,x^2} t^3\,\rmd t \wedge \rmd x^2 \wedge \rmd x^3 \wedge \rmd z}{z^5}\,.
\end{split}
\end{align}

\subsection{\texorpdfstring{$(\mathbf{6_0}|\mathbf{5.ii})$}{(6\textzeroinferior|5.ii)}: Type IIA SUGRA}

Finally, we consider the redefinition
\begin{align}
 T'_A = C_A{}^B\,T_B^{(\mathbf{5}|\mathbf{1})}\,,\qquad 
 C = {\footnotesize\begin{pmatrix}
 0 & -1 & 0 & 0 & 0 & \frac{1}{2} \\
 0 & 1 & 0 & 1 & 0 & \frac{1}{2} \\
 -1 & 0 & 1 & 0 & \frac{1}{2} & 0 \\
 1 & 0 & 0 & 0 & -1 & 0 \\
 1 & 0 & 0 & 0 & 0 & 0 \\
 0 & 0 & 0 & 0 & 0 & 1 \\
\end{pmatrix}}\,.
\end{align}
This time, we consider the parameterization\footnote{We note that, in general, the parameterization should be carefully chosen such that the resulting twist matrix $U$ does not break the section condition.}
\begin{align}
 l= g'\,\tilde{g}' \,,\qquad
 g'= \Exp{x'^3\,T'_3}\Exp{x'^2 \,T'_2}\Exp{(x'^1+x'^2)\,T'_1}\,,\qquad
 \tilde{g}'= \Exp{\tilde{x}'_1\,\tilde{T}'^1}\Exp{(\tilde{x}'_2-\tilde{x}'_1)\,\tilde{T}'^2}\Exp{\tilde{x}'_3\,\tilde{T}'^3}\,,
\end{align}
which leads to
\begin{align}
 \Exp{-2\,\bar{d}} = \Exp{-2\,x^1} = \Exp{-2\,(\tilde{x}'_2-x'^3)}. 
\end{align}
By using
\begin{align}
\begin{split}
 &\ell = (\rmd x^1+\rmd x^2-x^2\,\rmd x^3)\,T_1+\bigl[\rmd x^2-(x^1+x^2)\,\rmd x^3\bigr]\,T_2 + \rmd x^3 \,T_3 \,,
\\
 &r = (\cosh x^3 \,\rmd x^1 + \Exp{-x^3} \rmd x^2)\,T_1+(-\sinh x^3 \,\rmd x^1 + \Exp{-x^3} \rmd x^2)\,T_2 + \rmd x^3\,T_3 \,,
\\
 &v_1 = \partial_1 \,,\qquad
 v_2 = \partial_2 - \partial_1 \,,\qquad
 v_3 = \partial_3 - x^1\,\partial_1 + (x^1+x^2)\, \partial_2 \,,
\end{split}
\\
\begin{split}
 &a ={\footnotesize\begin{pmatrix}
 \cosh x^3 & \sinh x^3 & 0 \\
 \sinh x^3 & \cosh x^3 & 0 \\
 -x^2 & -x^1-x^2 & 1
\end{pmatrix}} ,
\qquad
 (\Pi^{\gga\ggb}) = {\footnotesize\begin{pmatrix}
 0 & x^1 & \Exp{-x^3}-1 \\
 -x^1 & 0 & \Exp{-x^3}-1 \\
 1 -\Exp{-x^3} & 1 -\Exp{-x^3} & 0 
\end{pmatrix}}\,,
\end{split}
\end{align}
we can check the covariance of the flux,
\begin{align}
 (\cF_A)=(0,\,0,\,-2,\,2,\,2,\,0) = C_A{}^B\,\,\cF_B^{(\mathbf{5}|\mathbf{1})}\,.
\end{align}

Since the dual algebra $\mathbf{5.ii}$ is non-unimodular, the Killing vector becomes
\begin{align}
 I = \frac{1}{2}\, \tilde{f}^{\ggb\gga}{}_\ggb\,\tilde{v}_\gga = - (v_1+v_2) = - \partial_2 \qquad
 \bigl(\tilde{f}^{\ggb 1}{}_\ggb = -2 \,,\quad 
 \tilde{f}^{\ggb 2}{}_\ggb = -2\bigr) \,,
\end{align}
but by absorbing the dual-coordinate dependence of $\bar{d}$\,, we obtain
\begin{align}
 \Exp{-2\,\bar{d}} = \Exp{2\, x^3},\qquad 
 I^m = 0\,. 
\end{align}
Then, after the redefinition, we obtain a solution of type IIA supergravity,
\begin{align}
\begin{split}
 \rmd s^2 &= \frac{-\rmd t^2 + \rmd z^2}{z^2} 
 + t^2 \,\frac{\Exp{4\,x^3} \rmd x_1^2 - 2 \Exp{3\,x^3} (\rmd x^1 + x^1\,\rmd x^3)\,\rmd x^1}{z^2\,[2 -2\Exp{x^3} + (x_1^2+1) \Exp{2\,x^3}]}
\\
 &\quad +z^2\,\frac{\bigl[(1- \Exp{x^3})\,\rmd x^1 +2\,\rmd x^2 - \Exp{x^3}x^1\,\rmd x^3\bigr]^2}{4\,t^2\,[2 -2\Exp{x^3} + (x_1^2+1) \Exp{2\,x^3}]}\,,
\\
 &\quad +\frac{\Exp{2\,x^3} t^2\, \bigl[2\,\rmd x_1^2 +4\,x^1\,\rmd x^1\,\rmd x^3 + (2\,x_1^2+1)\,\rmd x_3^2 \bigr]}{z^2\,[2 -2\Exp{x^3} + (1+x_1^2) \Exp{2\,x^3}]} + \rmd s^2_{\rmS^5}\,,
\\
 B_2 &=\frac{\Exp{2\,x^3} x^1\,\rmd x^1 \wedge \rmd x^2 + [1+\Exp{2\,x^3} (\sinh x^3 -\tfrac{1}{2})]\,\rmd x^1 \wedge \rmd x^3 +(2-\Exp{x^3})\,\rmd x^2 \wedge \rmd x^3}{2 -2\Exp{x^3} + (x_1^2+1) \Exp{2\,x^3}}\,,
\\
 \Exp{-2\,\Phi}&=\frac{t^2\,\bigl[2-2\Exp{x^3} + \Exp{2\,x^3}(x_1^2+1)\bigr]}{z^2}\,,
\qquad
 G_4 = -\frac{4\, t^3 \Exp{2\,x^3} \rmd t \wedge \rmd x^1 \wedge \rmd x^3 \wedge \rmd z}{z^5}\,. 
\end{split}
\end{align}

\section{Conclusion and outlook}
\label{sec:conclusion}

\subsection*{Summary of results}

We have discussed two approaches to the non-Abelian $T$-duality. 
One is the traditional NATD, obtained by integrating out the gauge fields associated with non-Abelian isometries, and the other is the PL $T$-duality/plurality, which is based on the Drinfel'd double. 

In NATD, a closed-form expression for the duality rules including the R--R fields was explicitly written down only for a certain isometry group, $\SU(2)$\,, but we proposed a general formula by assuming that the isometry group freely acts on the target space. 
The duality rules, under the setup \eqref{eq:setup1}, are summarized in \eqref{eq:NATD-summary1} and \eqref{eq:NATD-summary2}. 
In order to check the formula, we studied many examples, particularly the NATD for non-unimodular isometry groups. 

For the PL $T$-duality, the treatments of the R--R fields have been discussed in recent papers \cite{1707.08624,1810.07763,1810.11446}, but concrete examples have not been well studied. 
We first considered the case without spectator fields, and translated the known transformation rules for $\{g_{mn},\,B_{mn},\,\Phi\}$ into the rules for the generalized metric $\cH_{MN}$ and the DFT dilaton $d$\,. 
Then, using a result of the gauged DFT, we showed that the equations of motion are transformed covariantly under the PL $T$-plurality (by introducing a Killing vector $I^m$ appropriately). 
We also introduced the R--R fields, and determined their transformation rule under the $\OO(n,n)$ PL $T$-plurality transformation such that the equations of motion are covariantly transformed. 
We further considered the case with spectator fields and, requiring some dualizability conditions, we showed that the DFT equations of motion are indeed satisfied even in the presence of spectators. 
Finally, we studied a concrete example of the PL $T$-plurality. 
Starting with the $\AdS_5\times \rmS^5$ solution, we obtained the following family of solutions. 
\begin{align*}
\xymatrix@C=18pt@R=1pt{
 \AdS_5\times\rmS^5 & \text{type IIA SUGRA} & \text{type IIB SUGRA} & \text{type IIB \textcolor{red}{GSE}}
\\
 \bigl(\bar{d}=x^1,\ I=0\bigr) & \bigl(\bar{d}=x^3,\ I=0\bigr) & \bigl(\bar{d}=-x^1,\ I=0\bigr) & \bigl(\bar{d}=x^2-\textcolor{red}{\tilde{x}_3},\ I=0\bigr)
\\
 (\mathbf{5}|\mathbf{1}) \ar@{<->}[ddddd]^{\text{NATD}} \ar@{-}[r] & (\mathbf{6_0}|\mathbf{1}) \ar@{<->}[ddddd]^{\text{NATD}} \ar@{-}[r] & (\mathbf{5}|\mathbf{2.i}) \ar@{<->}[ddddd]^{\text{PL $T$-dual}} \ar@{-}[r] & (\mathbf{5.ii}|\mathbf{6_0}) \ar@{<->}[ddddd]^{\text{PL $T$-dual}}
\\
\\
\\
\\
\\
 (\mathbf{1}|\mathbf{5}) \ar@{-}[r] & (\mathbf{1}|\mathbf{6_0}) \ar@{-}[r] & (\mathbf{2.i}|\mathbf{5}) \ar@{-}[r] & (\mathbf{6_0}|\mathbf{5.ii}) 
\\
 \text{type IIA \textcolor{red}{GSE}} & \text{type IIB \textcolor{red}{GSE}} & \text{type IIA SUGRA} & \text{type IIA SUGRA}
\\
 \bigl(\bar{d}=\textcolor{red}{\tilde{x}_1},\ I=\textcolor{red}{\partial_1}\bigr) & \bigl(\bar{d}=\textcolor{red}{\tilde{x}_3},\ I=0\bigr) & \bigl(\bar{d}=\textcolor{blue}{-\tilde{x}_1},\ I=\textcolor{blue}{\partial_1}\bigr) & \bigl(\bar{d}=\textcolor{blue}{\tilde{x}_2}-x^3,\ I=\textcolor{blue}{-\partial_2}\bigr)
}
\end{align*}
Three of these are solutions of GSE. 
There are two origins of GSE: one is the Killing vector $I^i=\frac{1}{2}\,\tilde{f}^{\ggb\gga}{}_\ggb\,v_\gga^i$ that appears when the dual algebra is non-unimodular, and the other is the dual-coordinate dependence in $\bar{d}$\,. 
In the examples $(\mathbf{2.i}|\mathbf{5})$ and $(\mathbf{6_0}|\mathbf{5.ii})$, the two contributions are canceled with each other, and they are solutions of the usual supergravity even though their dual algebras are non-unimodular. 
In the literature, when $\bar{d}$ has a dual-coordinate dependence, since its interpretation is not clear in string theory or supergravity, such Manin triple was ignored. 
However, in DFT, we can treat the dual coordinates in the same ways as the physical coordinates, and we can lift the restriction. 
In this way, the PL $T$-plurality is a solution-generating technique of the DFT, rather than the usual supergravity. 

\subsection*{Discussion and outlook}

As we discussed, if we consider a supergravity solution that contains a four-dimensional Minkowski spacetime, $\rmd s^2 = f^2(y)\,\eta_{\mu\nu}\,\rmd x^\mu\,\rmd x^\nu + \cdots$\,, we can choose the coordinates such that the $(\mathbf{5}|\mathbf{1})$ symmetry is manifest. 
Then, as long as the $B$-field is isometric along the three Killing vectors, we will obtain a family of eight solutions similar to the case of $\AdS_5\times\rmS^5$\,. 
Moreover, low-dimensional Drinfel'd doubles have already been classified in \cite{hep-th/0110139,math/0202209,math/0202210}, and a useful list is given in section 3 of \cite{math/0202210}. 
If we have a DFT solution with an isometry algebra $\cg$\,, we may find a series of Manin triples,
\begin{align}
 (\cg|\mathbf{1}) \cong (\cg'|\cg'') \cong \cdots \,, 
\end{align}
and obtain a chain of DFT solutions. 
We may also start from a background with a $(\cg|\tilde{\cg})$ symmetry. 
For example, as discussed in \cite{hep-th/0210095}, the Yang--Baxter deformed backgrounds are also PL $T$-dualizable. 
Indeed, a Yang--Baxter deformed background has the form
\begin{align}
 E^{mn}=\tilde{g}^{mn} -\beta^{mn}\,,\qquad \beta^{mn} \equiv 2\,\eta\,r^{\gga\ggb}\,v_\gga^m\,v_\ggb^n\qquad (r^{\gga\ggb}=-r^{\ggb\gga})\,,
\end{align}
where $\eta$ and $r^{\gga\ggb}$ are constant, and $\Lie_{v_\gga}\tilde{g}_{mn}=0$ and $[v_\gga,\,v_\ggb]=f_{\gga\ggb}{}^\ggc\,v_\ggc$ are satisfied. 
Then, we can show that $\Lie_{v_\gga}E^{mn} = \tilde{f}^{\ggb\ggc}{}_\gga\,v_\ggb^m \, v_\ggc^n$ with $\tilde{f}^{\ggb\ggc}{}_\gga = 2\,\eta\,\bigl(r^{\ggb\ggd}\,f_{\ggd\gga}{}^\ggc - r^{\ggc\ggd}\,f_{\ggd\gga}{}^\ggb\bigr)$\,, and this is a dualizable background with the $(\cg|\tilde{\cg})$ symmetry. 
Then, by finding a group element $g(x)$\,, which realizes the set of Killing vectors $v_\gga^m$ as the left-invariant vector fields and $\beta^{mn}$ as $\beta^{mn} =e^m_\gga\,e^n_\ggb\,\Pi^{\gga\ggb}$ [\,i.e.~$(a^\rmT\,b)^{\gga\ggb}=2\,\eta\,r^{\gga\ggb}$\,], we can perform the PL $T$-plurality transformations of the Yang--Baxter-deformed background. 
In this way, from a given solution, we can find new solutions one after another, and the PL $T$-plurality is a useful solution-generating technique. 

In the traditional approach to NATD, we introduced the generalized Killing vector $(V_\gga^M)=(v_\gga^m,\,\tilde{v}_{\gga m})$\,. 
When the dual components $\tilde{v}_{\gga m}$ are present, we cannot regard the NATD as a particular case of the PL $T$-plurality. 
Also when the generalized Killing vectors depend on the spectator fields $y^\mu$\,, we cannot realize them as the left-invariant vector fields. 
In this sense, the traditional NATD is not completely contained in the PL $T$-plurality discussed here. 
It is interesting to study whether it is possible to generalize the PL $T$-plurality such that the traditional NATD can be realized as a particular case. 
In the realm of NATD that is going beyond the PL $T$-plurality, it is not ensured that the dual background is a solution of DFT. 
By the definition of the NATD, the duality rules for the metric and $B$-field should not be modified, but the transformation rule for the dilaton and $I^m$ may be modified from \eqref{eq:NATD-summary1}. 
It will be an important task to determine the general rule for the dilaton and $I^m$ that is consistent with the DFT equations of motion. 
Once the modification of the rule for the dilaton is determined, the modification of the rule for the R--R fields (by an overall factor) can also be determined, and then we can check the equation of motion for the R--R fields. 

In the two approaches studied in this paper, we have assumed that the isometry group acts on the target space freely, or without isotropy. 
If the assumption is not satisfied, we cannot take a gauge $x^i=c^i$ and we need to consider a more non-trivial gauge fixing. 
Treatments in such cases are discussed, for example, in \cite{hep-th/9210021,hep-th/9709071,1104.5196,1301.6755} for the NATD, and in \cite{hep-th/9602162,hep-th/9904188,1105.0162} for the PL $T$-duality. 
It is an interesting future direction to check whether the DFT equations of motion are covariantly rotated even in such general cases. 

In the study of the PL $T$-plurality, we have checked the covariance of the flux $\cF_A=2\,\cD_A\bar{d}$ on a case-by-case basis. 
The covariance is highly non-trivial but it was indeed transformed covariantly in all of the examples, and we suspect that there is some mechanism to be clarified. 
To show the covariance of $\cF_A$\,, clear understanding of the (finite) coordinate transformation $(x^i,\,\tilde{x}_i) \to (x'^i,\,\tilde{x}'_i)$ on the Drinfel'd double will be indispensable. 
The 2$D$ diffeomorphism in DFT$_{\text{WZW}}$ \cite{1410.6374,1502.02428,1509.04176} may be useful for this purpose. 

\subsection*{Toward non-Abelian $U$-duality}

Another important future direction is an investigation of the non-Abelian $U$-duality. 
As an attempt toward this, let us first consider an extension of the traditional NATD. 
As a natural extension of \eqref{eq:setup2}, let us consider the following setup \cite{Hull:1990ms}:
\begin{align}
 \Lie_{v_\gga} g_{ij}=0\,,\quad \iota_{v_\gga} F_4 + \rmd \hat{v}^{(2)}_\gga = 0 \,,\quad 
 \Lie_{v_\gga} v_\ggb = f_{\gga\ggb}{}^\ggc\,v_\ggc\,,\quad 
 \Lie_{v_\gga} \hat{v}^{(2)}_{\ggb} = f_{\gga\ggb}{}^\ggc\,\hat{v}^{(2)}_{\ggc}\,,
\label{eq:M2-setup}
\end{align}
where $F_4 \equiv \rmd C_3$ is the 4-form field strength in the eleven-dimensional supergravity. 
The 2-form $\hat{v}^{(2)}_{\gga}$ is the generalization of the 1-form $\hat{v}_{\gga m}$ appearing in \eqref{eq:setup2}, and the first two relations in \eqref{eq:M2-setup} are understood as a kind of generalized Killing equations. 
The remaining two equations are generalizations of the C-brackets between the generalized Killing vectors. 

We define $\hat{v}_{\gga\ggb}^{(1)} \equiv \iota_{v_\ggb} \hat{v}_\gga^{(2)}$\,, and assume the following relation for simplicity:
\begin{align}
 \hat{v}_{(\gga\ggb)}^{(1)} = 0 \,,\qquad
 \iota_{v_\gga}\hat{v}_{[\ggb\ggc]}^{(1)} = \iota_{v_{[\gga}}\hat{v}_{\ggb\ggc]}^{(1)} \,.
\label{eq:M2-condition}
\end{align}
We also assume the existence of the 1-forms $\ell^\gga \equiv \ell^\gga_i\,\rmd x^i$ that are dual to $v_\gga$ ($\iota_{v_\gga}\ell^\ggb=\delta_\gga^\ggb$), and then we find that the action\footnote{In the string action \eqref{eq:gauged-NATD}, by adding a total-derivative term the Lagrangian multiplier was introduced with derivative $\rmd \tilde{x}_\gga$ (see \cite{hep-th/9308154} for the Abelian case), but here we only discuss the classical equations of motion without investigating such a total-derivative term.}
\begin{align}
\begin{split}
 S &= \int_\Sigma \Bigl[\,\frac{1}{2}\,\bigl(g_{ij}\,Dx^i\wedge * Dx^j + *1\bigr) + C_3 + 2\, y_{\gga\ggb}\,F^\gga\wedge \bigl(\ell^\ggb - A^\ggb\bigr)\,\Bigr]
\\
 &\quad + \int_\Sigma\Bigl[-\wA^{\gga}\wedge \hat{v}^{(2)}_{\gga} + \frac{1}{2}\,\wA^{\gga}\wedge\wA^{\ggb}\wedge \hat{v}_{\gga\ggb}^{(1)}-\frac{1}{3!}\,\wA^{\gga}\wedge\wA^{\ggb}\wedge\wA^{\ggc}\,\iota_{v_\gga}\hat{v}_{\ggb\ggc}^{(1)} \Bigr] \,,
\end{split}
\label{eq:M2-action}
\end{align}
is invariant under
\begin{align}
\begin{split}
 \delta_\epsilon x^i(\sigma) &= \epsilon^\gga(\sigma)\,v_\gga^i(x)\,,\qquad 
 \delta_\epsilon \wA^\gga(\sigma) = \rmd \epsilon^\gga(\sigma) + f_{\ggb\ggc}{}^\gga\,\wA^\ggb(\sigma)\,\epsilon^\ggc(\sigma) \,, 
\\
 \delta_\epsilon y_{\gga\ggb} &= \epsilon^\ggc\,\bigl(f_{\ggc\gga}{}^\ggd\,y_{\ggd\ggb} + f_{\ggc\ggb}{}^\ggd\,y_{\gga\ggd}\bigr)\,. 
\end{split}
\end{align}
Here, by following the approach of \cite{Duff:1990hn} (see also \cite{1008.1763}), we have introduced antisymmetric Lagrange multipliers $y_{\gga\ggb}=-y_{\ggb\gga}$ that will ensure $F^\gga=0$\,. 

In the Abelian limit we can realize $v_\gga^i=\delta_\gga^i$ and $\ell^\gga=\delta_i^\gga\,\rmd x^i$\,, and then we can always choose a gauge $x^i=0$\,. 
By further assuming $\hat{v}_{\gga}^{(2)} = - \iota_{v_{\gga}} C_3$\,, the action reduces to
\begin{align}
 S = \int_\Sigma \Bigl[ \frac{1}{2}\,\bigl(g_{ij}\,\wA^i\wedge * \wA^j + *1\bigr) 
 + \frac{1}{3!}\,C_{\gga\ggb\ggc}\, \wA^{\gga}\wedge\wA^{\ggb}\wedge\wA^{\ggc} 
 + \rmd y_{\gga\ggb}\wedge\wA^\gga\wedge \wA^\ggb \Bigr]\,.
\end{align}
This is precisely the action discussed in \cite{Duff:1990hn,1008.1763} and \eqref{eq:M2-action} can be regarded as a natural extension. 
However, unlike the case of the string action, it is not clear how to eliminate the gauge fields $\wA^\gga$\,, and at the present time, we do not know how to obtain the dual action. 

A more promising approach may be the following one based on a generalization of DFT. 
The $U$-dual version of DFT is known as the exceptional field theory (EFT) \cite{hep-th/0104081,hep-th/0307098,0902.1509,1008.1763,1111.0459,1208.5884,1206.7045,1308.1673} and it is actively been studied. 
In DFT, the generalized coordinates are $(x^M)=(x^m,\,\tilde{x}_m)$ and the dual coordinates $\tilde{x}_m$ are associated with the string winding number. 
On the other hand, in EFT we introduce the dual coordinates for all of the wrapped branes that are connected by $U$-duality transformations. 
For example, in M-theory on a $n$-torus we have the M2-brane, the M5-brane, and the Kaluza--Klein monopole, and more exotic branes in general. 
Correspondingly, we introduce the generalized coordinates as
\begin{align}
 (x^I)= (x^i,\,y_{i_1i_2},\,y_{i_1\cdots i_5},\,y_{i_1\cdots i_7,\,i},\dotsc) \qquad (i=1,\dotsc,n)\,.
\label{eq:M-coordinates}
\end{align}
By understanding that the multiple indices separated by commas are totally antisymmetrized, we can easily see that the number of dimensions of the extended space $x^I$ is the same as the dimension $D$ of a fundamental representation of the $E_{n(n)}$ $U$-duality group:
\begin{align}
 \begin{array}{|c|c|c|c|c|c|} \hline
 n & 4 & 5 & 6 & 7 & 8 \\\hline
 \text{$U$-duality group $E_{n(n)}$} & \text{SL}(5) & \SO(5,5) & E_{6(6)} & E_{7(7)} & E_{8(8)} \\\hline
 \text{dimension $D$} & 10 & 16 & 27 & 56 & 248 \\\hline
\end{array}
\end{align}
In such extended space, the generalized metric $\cM_{IJ}$ has been constructed in \cite{1008.1763,1111.0459}, and it contains the bosonic fields, such as the metric $g_{ij}$, and the 3-form and 6-form potentials, $C_{i_1i_2i_3}$ and $C_{i_1\cdots i_6}$\,. 
It is a natural generalization of the generalized metric $\cH_{MN}$ in DFT. 

In DFT, the section condition $\eta^{IJ}\,\partial_I\,\partial_J=0$ reduces the doubled space to the physical subspace. 
The section condition in EFT (for $n\leq 6$) also has a similar form $\eta^{IJ;\,\hat{K}}\,\partial_I\,\partial_J = 0$\,, where $\eta^{IJ;\,\hat{K}}$ is known as the $\eta$-symbol and it has an additional index $\hat{K}$ transforming in another representation (see \cite{1708.06342} for the explicit form of the $\eta$-symbol). 
When all of the fields depend only on the coordinates $x^i$ of \eqref{eq:M-coordinates}, we find
\begin{align}
 \eta^{IJ;\,\hat{K}}\,\partial_I\,\partial_J = \eta^{ij;\,\hat{K}}\,\partial_i\,\partial_j = 0\qquad \bigl(\because\ \, \eta^{ij;\,\hat{K}}=0\bigr)\,,
\end{align}
and the section condition is satisfied. 
This $n$-dimensional solution is called the M-theory section. 
Another solution, called the type IIB section, was found in \cite{1311.5109}, and in order to discuss the type IIB section, it is convenient to reparameterize the coordinates as\footnote{The explicit relation between $x^I$ and $x^M$ was determined in \cite{1701.07819}.}
\begin{align}
 (x^M) = (x^m,\,y_m^\alpha,\,y_{m_1m_2m_3},\,y^\alpha_{m_1\cdots m_5},\,y_{m_1\cdots m_6,m},\dotsc)\quad (m=1,\dotsc,n-1\,,\ \alpha =1,2)\,,
\end{align}
where the dual coordinates are associated with the type IIB branes. 
If the fields depend only on the $x^m$\,, the section condition is again satisfied because $\eta^{mn;\,\hat{P}}=0$\,.
Since we cannot introduce any more coordinate dependence, the subspace spanned by $x^m$ is also a maximally isotropic subspace, although it is $(n-1)$ dimensional unlike the M-theory section. 
In this way, a single EFT can be understood from two viewpoints: M-theory and type IIB theory. 

One of the key relations in the PL $T$-duality is the self-duality relation,
\begin{align}
 \eta_{AB}\,\hat{\cP}^B = \EPSneg \hat{\cH}_{AB} \,* \hat{\cP}^B \,,\qquad 
 \hat{\cP}(\sigma)= \rmd l\,l^{-1} \,. 
\end{align}
This is a covariant rewriting of the string equations of motion, but a similar equation for the M2- or M5-brane theory has been discussed in \cite{1208.1232,1305.2258} for the $\text{SL}(5)$ and $\SO(5,5)$ case, and in \cite{1712.10316} for higher exceptional groups. 
For the M$p$-brane ($p=2,5$), it has a similar form
\begin{align}
 \bm{\eta}_{IJ} \wedge\cP^J = \EPSneg \cM_{IJ}\,* \cP^J \,,
\end{align}
where $\bm{\eta}_{IJ}$ is some $(p-1)$-form that contains $\rmd x^i$ and the field strengths of the worldvolume gauge fields. 
In the case of the flat torus, the equations of motion give $\rmd\cP^I=0$ and we find the on-shell expression $\cP^I=\rmd x^I$\,. 
On the other hand, by requiring a certain ``dualizability condition'' on $\cM_{IJ}$ appropriately, the equations of motion may lead to $\cP=\rmd l\,l^{-1}$\,, where $l$ is an element of a certain large group $\mathfrak{E}$ with dimension $D$\,. 
The corresponding algebra $\mathfrak{e}$ will be endowed with a bilinear form, corresponding to the $\eta$-symbol. 
Then, the $U$-dual version of the PL $T$-plurality may be the equivalence between sigma models with $n$- or $(n-1)$-dimensional target spaces that have an isometry algebra $[T_\gga,\,T_\ggb]=f_{\gga\ggb}{}^\ggc\,T_\ggc$ satisfying $\eta^{\gga\ggb;\,\hat{A}}=0$\,. 
The identification of the detailed structure of the group $\cE$ and the systematic construction of the twist matrix $U$, whose flux gives the structure constant of $\mathfrak{e}$, are interesting future directions. 

\medskip

\subsection*{Note added}

After this paper appeared on arXiv, an interesting paper \cite{1904.00362} appeared, which also discussing NATD from the perspective of the gauged DFT. 

\medskip

\subsection*{Acknowledgments}

We would like to thank Falk Hassler for elucidating the approach of \cite{1707.08624,1810.11446}. 
We also would like to thank Yolanda Lozano, Jeong-Hyuck Park, Jun-ichi Sakamoto, and Shozo Uehara for helpful discussions and comments. 
We are also grateful to the organizers and the participants of the workshop ``String: T-duality, Integrability and Geometry'' held at Tohoku University. 
This work is supported by JSPS Grant-in-Aids for Scientific Research (C) 18K13540 and (B) 18H01214. 

\appendix

\section{Conventions}
\label{app:conventions}

The symmetrization and antisymmetrization are normalized as
\begin{align}
 A_{(m_1\cdots m_n)} \equiv \frac{1}{n!}\,\bigl(A_{m_1\cdots m_n} + \cdots \bigr) \,,\qquad
 A_{[m_1\cdots m_n]} \equiv \frac{1}{n!}\,\bigl(A_{m_1\cdots m_n} \pm \cdots \bigr) \,.
\end{align}
Our conventions for differential forms are as follows, both for the spacetime and the worldsheet:
\begin{align}
\begin{split}
 &(* \alpha_q)_{m_1\cdots m_{D-q}} =\frac{1}{q!}\,\varepsilon^{n_1\cdots n_q}{}_{m_1\cdots m_{D-q}}\,\alpha_{n_1\cdots n_q} \,,\qquad 
 \rmd^{D}x = \rmd x^1\wedge\cdots\wedge\rmd x^D \,,
\\
 &* (\rmd x^{m_1}\wedge \cdots \wedge \rmd x^{m_q}) = \frac{1}{(D-q)!}\,\varepsilon^{m_1\cdots m_q}{}_{n_1\cdots n_{D-q}}\,\rmd x^{n_1}\wedge \cdots \wedge \rmd x^{n_{D-q}} \,,
\\
 &(\iota_v \alpha_n) = \frac{1}{(n-1)!}\,v^n\,\alpha_{nm_1\cdots m_{n-1}}\,\rmd x^{m_1}\wedge\cdots\wedge \rmd x^{m_{n-1}}\,. 
\end{split}
\end{align}
The epsilon tensors on the spacetime and the worldsheet are defined as follows:
\begin{align}
 \varepsilon^{01}= \EPSneg \frac{1}{\sqrt{\abs{\gamma}}}\,,\qquad 
 \varepsilon_{01}= \EPSpos \sqrt{\abs{\gamma}} \,, \qquad 
 \varepsilon^{1\cdots D}=-\frac{1}{\sqrt{\abs{g}}}\,,\qquad 
 \varepsilon_{1\cdots D}= \sqrt{\abs{g}} \,.
\end{align}

For the R--R fields, we have the R--R potential in the A-basis $A_{m_1\cdots m_p}$ and the C-basis $C_{m_1\cdots m_p}$ \cite{hep-th/0103233}. 
In terms of the polyform,
\begin{align}
 A \equiv \sum_p \frac{1}{p!}\,A_{m_1\cdots m_p}\,\rmd x^{m_1}\wedge\cdots\wedge\rmd x^{m_p}\,,\qquad
 C \equiv \sum_p \frac{1}{p!}\,C_{m_1\cdots m_p}\,\rmd x^{m_1}\wedge\cdots\wedge\rmd x^{m_p}\,, 
\end{align}
they are related as
\begin{align}
 A = \Exp{B_2\wedge} C \,, \qquad C = \Exp{-B_2\wedge} A\,.
\end{align}
Their field strengths are defined as
\begin{align}
 F = \rmd A \,,\qquad G = \rmd C + H_3\wedge C \,,
\label{eq:RR-field-strength}
\end{align}
and they are also related as
\begin{align}
 F = \Exp{B_2\wedge} G \,, \qquad G = \Exp{-B_2\wedge} F\,.
\end{align}
For simplicity, in this paper we call the field strength $F$ the Page form. 
In our convention, the $G$ satisfies the self-duality relation
\begin{align}
 * G_p = (-1)^{\frac{p(p+1)}{2}+1} G_{10-p} \,,\qquad 
 G_p = (-1)^{\frac{p(p-1)}{2}} * G_{10-p} \,. 
\label{eq:RR-self}
\end{align}
In the presence of the Killing vector $I^m$ in the GSE, which satisfies
\begin{align}
 \Lie_I g_{mn} = \Lie_I B_2 = \Lie_I \Phi = \Lie_I F = \Lie_I G = 0\,,
\end{align}
the relations \eqref{eq:RR-field-strength} are modified as
\begin{align}
 F = \rmd A - \iota_I A \,,\qquad 
 G = \rmd C + H_3 \wedge C - \iota_I B_2 \wedge C - \iota_I C \,,
\label{eq:GSE-F=dA}
\end{align}
and the Bianchi identities, which are equivalent to the equations of motion under \eqref{eq:RR-self}, become
\begin{align}
 \rmd F - \iota_I F = 0 \,,\qquad 
 \rmd G + H_3 \wedge G - \iota_I B_2 \wedge G - \iota_I G = 0 \,.
\end{align}
The GSE for the fields in the NS--NS sector can be summarized as
\begin{align}
 &R + 4\,D^m \partial_m \Phi - 4\,\abs{\partial \Phi}^2 - \frac{1}{2}\,\abs{H_3}^2 - 4\,\bigl(I^m I_m+U^m U_m + 2\,U^m\,\partial_m \Phi - D_m U^m\bigr) =0 \,,
\nn\\
 &R_{mn}-\frac{1}{4}\,H_{mpq}\,H_n{}^{pq} + 2 D_m \partial_n \Phi + D_m U_n +D_n U_m = T_{mn} \,,
\\
 &-\frac{1}{2}\,D^k H_{kmn} + \partial_k\Phi\,H^k{}_{mn} + U^k\,H_{kmn} + D_m I_n - D_n I_m = K_{mn} \,,
\nn
\end{align}
where $U_1\equiv U_m\,\rmd x^m$ is defined as $U_1\equiv \iota_I B_2$\,, and $T_{mn}$ and $K_{mn}$ are
\begin{align}
\begin{split}
 T_{mn}&\equiv \frac{\Exp{2\,\Phi}}{4} \sum_p \Bigl[ \frac{1}{(p-1)!}\, G_{(m}{}^{q_1\cdots q_{p-1}} G_{n) q_1\cdots q_{p-1}} - \frac{1}{2}\,g_{mn}\,\abs{G_p}^2 \Bigr] \,,
\\
 K_{mn}&\equiv \frac{\Exp{2\,\Phi}}{4} \sum_p \frac{1}{(p-2)!}\, G_{q_1\cdots q_{p-2}}\, G_{mn}{}^{q_1\cdots q_{p-2}} \,. 
\end{split}
\end{align}

In the presence of the Killing vector $(I^m)=(I^i,\,I^z)$\,, if we perform a formal $T$-duality along the $x^z$-direction, the supergravity fields are transformed as follows \cite{1703.09213}:
\begin{align}
\begin{split}
 &g'_{ij} = g_{ij} - \frac{g_{iz}\, g_{jz}- B_{iz}\, B_{jz}}{g_{zz}}\,,\qquad 
 g'_{iz} = \frac{B_{iz}}{g_{zz}}\,,\qquad 
 g'_{zz}=\frac{1}{g_{zz}}\,,
\\
 &B'_{ij} = B_{ij} - \frac{B_{iz}\, g_{jz}- g_{iz}\, B_{jz}}{g_{zz}}\,,\qquad 
 B'_{iz} = \frac{g_{iz}}{g_{zz}} \,, 
\\
 &\Phi' =\Phi + \frac{1}{4}\,\ln\Bigl\lvert \frac{\det (g'_{mn})}{\det (g_{mn})} \Bigr\rvert 
+ I^z z \,,\qquad 
 I'^i = I^i\,,\qquad I'^z = 0 \,,
\\
 &A'_{i_1\cdots i_{p-1}z} = \Exp{-I^z z} A_{i_1\cdots i_{p-1}} \,,
\qquad
 A'_{i_1\cdots i_p} = \Exp{-I^z z} A_{i_1\cdots i_pz} \,,
\\
 &C'_{i_1\cdots i_{p-1}z} = \Exp{-I^z z} \Bigl[C_{i_1\cdots i_{p-1}} - (p-1)\,\frac{C_{[i_1\cdots i_{p-2}|z|}\, g_{i_{p-1}]z}}{g_{zz}} \Bigr] \,,
\\
 &C'_{i_1\cdots i_p} = \Exp{-I^z z} \Bigl[ C_{i_1\cdots i_pz} + p\, C_{[i_1\cdots i_{p-1}}\, B_{i_p]z} + p\,(p-1)\,\frac{C_{[i_1\cdots i_{p-2}|z|}\, B_{i_{p-1}|z|}\, g_{i_p]z}}{g_{zz}} \Bigr] \,. 
\end{split}
\label{eq:T-duality-rule}
\end{align}

\section{Technical details of DFT}
\label{app:DFT}

In this appendix, we explain the technical details of (gauged) DFT and show the covariance of the DFT equations of motion under the PL $T$-plurality with spectator fields. 

\subsection*{NS--NS sector}

For convenience, let us introduce the double vielbein $(V_{\hat{A}}{}^M)\equiv (V_a{}^M,\,V_{\bar{a}}{}^M)$ as
\begin{align}
 \cH_{MN} = V_M{}^{\hat{A}}\,V_N{}^{\hat{B}}\,\cH_{\hat{A}\hat{B}} \,,\qquad 
 \eta_{MN} = V_M{}^{\hat{A}}\,V_N{}^{\hat{B}}\,\eta_{\hat{A}\hat{B}} \,,\qquad 
 V_M{}^{\hat{A}} \,V_{\hat{A}}{}^M = \delta_M^N\,,
\end{align}
where $V_M{}^{\hat{A}}$ is an $\OO(D,D)$ matrix and we have defined
\begin{align}
 (\cH_{\hat{A}\hat{B}}) = \begin{pmatrix} \eta_{ab} & 0 \\ 0 & \eta_{\bar{a}\bar{b}} \end{pmatrix},\qquad 
 (\eta_{\hat{A}\hat{B}}) = \begin{pmatrix} \eta_{ab} & 0 \\ 0 & -\eta_{\bar{a}\bar{b}} \end{pmatrix},
\end{align}
and $(\eta_{ab})\equiv(\eta_{\bar{a}\bar{b}})\equiv\diag(-1,1,\dotsc,1)$\,. 
They can be parameterized as
\begin{align}
 (V_a{}^M) = \frac{1}{\sqrt{2}}\begin{pmatrix} e_a^m \\ (g+B)_{mn}\,e_a^n \end{pmatrix} \,,\qquad 
 (V_{\bar{a}}{}^M) = \frac{1}{\sqrt{2}}\begin{pmatrix} e_{\bar{a}}^m \\ (-g+B)_{mn}\,e_{\bar{a}}^n \end{pmatrix} \,, 
\end{align}
where $e_a^m=e_{\bar{a}}^m$ is the vielbein satisfying $g_{mn}=e_m^a\,e_n^b\,\eta_{ab}=e_m^{\bar{a}}\,e_n^{\bar{b}}\,\eta_{\bar{a}\bar{b}}$\,. 

The equations of motion for the DFT dilaton and the generalized metric are
\begin{align}
\begin{split}
 \cR &\equiv -2\,\Pm^{\hat{A}\hat{B}}\, \bigl(2\,\bm{\cD}_{\hat{A}} \bm{\cF}_{\hat{B}} - \bm{\cF}_{\hat{A}}\,\bm{\cF}_{\hat{B}}\bigr) 
 - \frac{1}{3}\,\Pm^{\hat{A}\hat{B}\hat{C}\hat{D}\hat{E}\hat{F}}\,\bm{\cF}_{\hat{A}\hat{B}\hat{C}}\,\bm{\cF}_{\hat{D}\hat{E}\hat{F}} = 0\,,
\\
 \cG^{\hat{A}\hat{B}} &\equiv -4\,\Pm^{\hat{D}[\hat{A}}\, \bm{\cD}^{\hat{B}]} \bm{\cF}_{\hat{D}} + 2\,\bigl(\bm{\cF}_{\hat{D}}-\bm{\cD}_{\hat{D}} \bigr)\,\check{\bm{\cF}}^{\hat{D}[\hat{A}\hat{B}]} - 2\,\check{\bm{\cF}}^{\hat{C}\hat{D}[\hat{A}}\,\bm{\cF}_{\hat{C}\hat{D}}{}^{\hat{B}]} = 0\,, 
\end{split}
\end{align}
where $\bm{\cD}_{\hat{A}}\equiv V_{\hat{A}}{}^M\,\partial_M$ and $\bm{\cF}_{\hat{A}}$ and $\bm{\cF}_{\hat{A}\hat{B}\hat{C}}$ are defined by
\begin{align}
 \bm{\cF}_{\hat{A}\hat{B}\hat{C}} \equiv 3\,\bm{\Omega}_{[\hat{A}\hat{B}\hat{C}]}\,, \qquad 
 \bm{\cF}_{\hat{A}} \equiv \bm{\Omega}^{\hat{B}}{}_{\hat{A}\hat{B}} + 2\, \bm{\cD}_{\hat{A}} d \,, \qquad
 \bm{\Omega}_{\hat{A}\hat{B}\hat{C}} \equiv -\bm{\cD}_{\hat{A}} V_{\hat{B}}{}^M\, V_{M\hat{C}} \,,
\end{align}
and $\check{\bm{\cF}}^{\hat{A}\hat{B}\hat{C}}$ is defined similarly to \eqref{eq:P-check}. 
We can show that $\cR=\cS$ under the section condition, but the equivalence of $\cG^{\hat{A}\hat{B}}=0$ and $\cS_{MN}=0$ is non-trivial. 
To see the equivalence, we show
\begin{align}
 V^{\bar{a} M}\,V^{bN}\,\cS_{MN} = e^{\bar{a}m}\,e^{bm}\, s_{mn} \,,\qquad 
 V^{a M}\,V^{bN}\,\cS_{MN} = 0 = V^{\bar{a} M}\,V^{\bar{b}N}\,\cS_{MN} \,.
\end{align}
Under the section condition, we can also find
\begin{align}
 \cG^{\bar{a}b} = V^{\bar{a} M}\,V^{bN}\,\cS_{MN} \,,\qquad 
 \cG^{ab}=0 = \cG^{\bar{a}\bar{b}} \,,
\end{align}
and they clearly show the equivalence of $\cG^{AB}=0$ and $\cS_{MN}=0$\,. 

By using the identities \cite{1304.1472}
\begin{align}
\begin{split}
 \cZ &\equiv \bm{\cD}^{\hat{A}} \bm{\cF}_{\hat{A}} - \frac{1}{2}\,\bm{\cF}^{\hat{A}}\,\bm{\cF}_{\hat{A}} + \frac{1}{12}\, \bm{\cF}_{\hat{A}\hat{B}\hat{C}}\,\bm{\cF}^{\hat{A}\hat{B}\hat{C}} = 0\,,
\\
 \cZ_{\hat{A}\hat{B}} &\equiv 2\,\bm{\cD}_{[\hat{A}} \bm{\cF}_{\hat{B}]} + \bm{\cF}^{\hat{C}}\,\bm{\cF}_{\hat{C}\hat{A}\hat{B}} - \bm{\cD}^{\hat{C}} \bm{\cF}_{\hat{C}\hat{A}\hat{B}} = 0\,,
\end{split}
\end{align}
which hold under the section condition, we can simplify the expressions for $\cR$ and $\cG^{\hat{A}\hat{B}}$ as
\begin{align}
 \cR &= \cH^{\hat{A}\hat{B}}\, \bigl(2\,\bm{\cD}_{\hat{A}} \bm{\cF}_{\hat{B}} - \bm{\cF}_{\hat{A}}\,\bm{\cF}_{\hat{B}}\bigr) 
\nn\\
 &\quad + \frac{1}{12}\, \cH^{\hat{A}\hat{D}}\,\bigl(3\,\eta^{\hat{B}\hat{E}}\,\eta^{\hat{C}\hat{F}} - \cH^{\hat{B}\hat{E}}\, \cH^{\hat{C}\hat{F}}\bigr) \,\bm{\cF}_{\hat{A}\hat{B}\hat{C}}\,\bm{\cF}_{\hat{D}\hat{E}\hat{F}} \,,
\label{eq:cR-total}
\\
 \cG^{\hat{A}\hat{B}} &= 2\,\cH^{\hat{D}[\hat{A}}\, \bm{\cD}^{\hat{B}]} \bm{\cF}_{\hat{D}} 
 -\frac{1}{2}\,\cH^{\hat{D}\hat{E}}\,(\eta^{\hat{A}\hat{F}}\,\eta^{\hat{B}\hat{G}}-\cH^{\hat{A}\hat{F}}\,\cH^{\hat{B}\hat{G}})\,\bigl(\bm{\cF}_{\hat{D}}-\bm{\cD}_{\hat{D}}\bigr)\, \bm{\cF}_{\hat{E}\hat{F}\hat{G}} 
\nn\\
 &\quad -\cH_{\hat{E}}{}^{[\hat{A}}\,\bigl(\bm{\cF}_{\hat{D}}-\bm{\cD}_{\hat{D}}\bigr)\, \bm{\cF}^{\hat{B]}\hat{D}\hat{E}} 
  +\frac{1}{2}\,\bigl(\eta^{\hat{C}\hat{E}}\,\eta^{\hat{D}\hat{F}} - \cH^{\hat{C}\hat{E}}\, \cH^{\hat{D}\hat{F}}\bigr)\, \cH^{\hat{G}[\hat{A}}\,\bm{\cF}_{\hat{C}\hat{D}}{}^{\hat{B}]}\,\bm{\cF}_{\hat{E}\hat{F}\hat{G}} \,. 
\label{eq:cG-total}
\end{align}
As a side remark, we note that the equations of motion $\cG^{\hat{A}\hat{B}}=0$ can also be expressed as
\begin{align}
 \widetilde{\cG}^{\hat{A}\hat{B}} &\equiv \cH^{\hat{A}}{}_{\hat{C}}\,\cG^{\hat{C}\hat{B}}
\nn\\
 &= \bar{P}^{(\hat{A}\hat{B})\hat{C}\hat{D}}\,\bigl(\bar{P}^{\hat{E}\hat{F}\hat{G}\hat{H}}\,\bm{\cF}_{\hat{C}\hat{E}\hat{F}}\,\bm{\cF}_{\hat{D}\hat{G}\hat{H}} + 2\, \bm{\cD}_{(\hat{C}}\,\bm{\cF}_{\hat{D})}\bigr)
 +2\,\bar{P}^{\hat{C}\hat{D}\hat{E}(\hat{A}}\,\bigl(\bm{\cF}_{\hat{E}}-\bm{\cD}_{\hat{E}}\bigr)\,\bm{\cF}^{\hat{B})}{}_{\hat{C}\hat{D}} 
\nn\\
 &\quad - P^{[\hat{A}\hat{B}]\hat{C}\hat{D}}\, \bigl[\bigl(\bm{\cF}^{\hat{E}}-\bm{\cD}^{\hat{E}}\bigr)\,\bm{\cF}_{\hat{E}\hat{C}\hat{D}}
 +2\, \bm{\cD}_{[\hat{C}}\,\bm{\cF}_{\hat{D}]} \bigr] = 0\,,
\end{align}
where we have defined the projectors
\begin{align}
 P^{\hat{A}\hat{B}\hat{C}\hat{D}} \equiv \frac{1}{2}\,\bigl(\eta^{\hat{A}\hat{C}}\,\eta^{\hat{B}\hat{D}} + \cH^{\hat{A}\hat{C}}\,\cH^{\hat{B}\hat{D}}\bigr)\,,\qquad
 \bar{P}^{\hat{A}\hat{B}\hat{C}\hat{D}} \equiv \frac{1}{2}\,\bigl(\eta^{\hat{A}\hat{C}}\,\eta^{\hat{B}\hat{D}} - \cH^{\hat{A}\hat{C}}\,\cH^{\hat{B}\hat{D}}\bigr)\,.
\end{align}

Now, let us decompose the double vielbein and the DFT dilaton as
\begin{align}
 V_M{}^{\hat{A}} = U_M{}^B(x^I)\,\hat{V}_B{}^{\hat{A}}(y^\mu)\,,\qquad 
 d=\hat{d}(y^\mu) + \sfd(x^I)\,,
\end{align}
where the twist matrix $U_M{}^A$ is an $\OO(D,D)$ matrix and the untwisted metric is defined by
\begin{align}
 \hat{\cH}_{AB}(y) \equiv \hat{V}_A{}^{\hat{C}}(y)\,\hat{V}_B{}^{\hat{D}}(y)\,\cH_{\hat{C}\hat{D}}\,. 
\label{eq:untwisted-vielbein}
\end{align}
Then, by requiring
\begin{align}
 \cD_A \hat{V}_B{}^{\hat{C}} = \partial_A \hat{V}_B{}^{\hat{C}}\,,\qquad 
 \cD_A \hat{d} = \partial_A \hat{d} \qquad \bigl(\cD_A\equiv U_A{}^M\,\partial_M\bigr)\,, 
\label{eq:untwisted-condition}
\end{align}
the generalized fluxes can be decomposed as
\begin{align}
\begin{split}
 \bm{\cF}_{\hat{A}} &= \hat{\cF}_{\hat{A}}(y) + \hat{V}_{\hat{A}}{}^{B}(y) \,\cF_{B}\,,
\\
 \bm{\cF}_{\hat{A}\hat{B}\hat{C}} &= \hat{\cF}_{\hat{A}\hat{B}\hat{C}}(y) + \hat{V}_{\hat{A}}{}^{D}(y)\,\hat{V}_{\hat{B}}{}^{E}(y)\,\hat{V}_{\hat{C}}{}^{F}(y)\,\cF_{DEF}\,,
\end{split}
\label{eq:fluxes-decomp}
\end{align}
where $\hat{\cF}_{\hat{A}}(y)$ and $\hat{\cF}_{\hat{A}\hat{B}\hat{C}}(y)$ are the generalized fluxes associated with $\{\hat{V}_A{}^{\hat{B}},\,\hat{d}\}$\,,
\begin{align}
 \hat{\cF}_{\hat{A}\hat{B}\hat{C}} \equiv 3\,\hat{\Omega}_{[\hat{A}\hat{B}\hat{C}]}\,, \qquad 
 \hat{\cF}_{\hat{A}} \equiv \hat{\Omega}^{\hat{B}}{}_{\hat{A}\hat{B}} + 2\, \hat{\cD}_{\hat{A}} \hat{d} \,, \qquad
 \hat{\Omega}_{\hat{A}\hat{B}\hat{C}} \equiv -\hat{\cD}_{\hat{A}} \hat{V}_{\hat{B}}{}^D\, \hat{V}_{D\hat{C}} \,,
\end{align}
and $\hat{\cD}_A \equiv \hat{V}_{\hat{A}}{}^{B}\,\partial_B$\,. 
Then, the generalized Ricci scalar can be decomposed as
\begin{align}
 \cR &= \hat{\cR} + \frac{1}{12}\,\hat{\cH}^{AD}\,\bigl(3\,\eta^{BE}\,\eta^{CF}-\hat{\cH}^{BE}\,\hat{\cH}^{CF} \bigr) \,\cF_{ABC}\,\cF_{DEF} 
 - \hat{\cH}^{AB}\, \cF_A\,\cF_B 
\nn\\
 &\quad - \frac{1}{2}\, \cF^A{}_{BC}\,\hat{\cH}^{BD}\,\hat{\cH}^{CE} \, \cD_B \hat{\cH}_{AE} 
 + 2\,\cF_A\,\cD_B \hat{\cH}^{AB} 
 - 4\,\hat{\cH}^{AB}\,\cF_A\, \cD_B\hat{d} \,.
\label{eq:cR-decomp}
\end{align}
Here, we have assumed that $\cF_A$ and $\cF_{ABC}$ are constant and have used $\cF^{A}{}_{DE}\,\partial_A \hat{E}_{\hat{B}}{}^C(y)=0$\,, which is satisfied under our setup $\cF^{\alpha}{}_{BC}=0$\,. 
In addition, $\hat{\cR}$ is the generalized Ricci scalar associated with the untwisted fields $\{\hat{\cH}_{AB},\,\hat{d}\}$\,,
\begin{align}
 \hat{\cR} &\equiv \cH^{\hat{A}\hat{B}}\, \bigl(2\,\hat{\cD}_{\hat{A}} \hat{\cF}_{\hat{B}} - \hat{\cF}_{\hat{A}}\,\hat{\cF}_{\hat{B}}\bigr) 
 - \frac{1}{12}\, \cH^{\hat{A}\hat{D}}\,\bigl(\cH^{\hat{B}\hat{E}}\, \cH^{\hat{C}\hat{F}}
 - 3\,\eta^{\hat{B}\hat{E}}\,\eta^{\hat{C}\hat{F}}\bigr) \,\hat{\cF}_{\hat{A}\hat{B}\hat{C}}\,\hat{\cF}_{\hat{D}\hat{E}\hat{F}} \,.
\end{align}

Now, let us show the covariance of the equations of motion under the $\OO(n,n)$ PL $T$-plurality transformation,
\begin{align}
 \hat{\cH}_{AB}\to (C\,\hat{\cH}\,C^{\rmT})_{AB}\,,\quad 
 \hat{d} \to \hat{d}\,,\quad
 \cF_A \to (C\,\cF)_A\,,\quad 
 \cF_{ABC} \to C_A{}^D\,C_B{}^E\,C_C{}^F\,\cF_{DEF}\,. 
\end{align}
From the relation \eqref{eq:untwisted-vielbein}, the first rule implies the following rule for the untwisted vielbein:
\begin{align}
 \hat{V}_A{}^{\hat{B}}\to C_A{}^C\,\hat{V}_C{}^{\hat{B}}\,,\qquad 
 \hat{V}_{\hat{A}}{}^B \to \hat{V}_{\hat{A}}{}^C\,(C^{-1})_C{}^B\,. 
\end{align}
Since the untwisted fields satisfy \eqref{eq:untwisted-condition}, we can show for an arbitrary untwisted field $g(y)$
\begin{align}
 \hat{\cD}'_{\hat{A}} g'(y) = \hat{V}_{\hat{A}}{}^B\,(C^{-1})_B{}^C\,\partial_C g'(y)
 = \hat{V}_{\hat{A}}{}^B\,\partial_B g'(y) = \hat{\cD}_{\hat{A}} g'(y)\,,
\label{eq:cD-invariant}
\end{align}
and $\hat{\cF}_{\hat{A}}(y)$ and $\hat{\cF}_{\hat{A}\hat{B}\hat{C}}(y)$ are invariant under the PL $T$-plurality transformation. 
Then, from \eqref{eq:fluxes-decomp}, the fluxes $\bm{\cF}_{\hat{A}}$ and $\bm{\cF}_{\hat{A}\hat{B}\hat{C}}$ are also invariant,
\begin{align}
 \bm{\cF}'_{\hat{A}} = \bm{\cF}_{\hat{A}}\,,\qquad \bm{\cF}'_{\hat{A}\hat{B}\hat{C}}=\bm{\cF}_{\hat{A}\hat{B}\hat{C}}\,.
\label{eq:bmF-invariant}
\end{align}
Moreover, according to the constancy of $\cF_A$ and $\cF_{ABC}$\,, $\bm{\cF}_{\hat{A}}$ and $\bm{\cF}_{\hat{A}\hat{B}\hat{C}}$ depends only on the spectator fields, and from \eqref{eq:untwisted-condition}, \eqref{eq:cD-invariant}, and \eqref{eq:bmF-invariant} we have
\begin{align}
\begin{split}
 \bm{\cD}'_{\hat{A}} \bm{\cF}'_{\hat{B}} &= \hat{\cD}'_{\hat{A}} \bm{\cF}'_{\hat{B}} = \hat{\cD}_{\hat{A}} \bm{\cF}'_{\hat{B}} = \bm{\cD}_{\hat{A}} \bm{\cF}_{\hat{B}} \,,
\\
 \bm{\cD}'_{\hat{A}} \bm{\cF}'_{\hat{B}\hat{C}\hat{D}} &= \hat{\cD}'_{\hat{A}} \bm{\cF}'_{\hat{B}\hat{C}\hat{D}} = \hat{\cD}_{\hat{A}} \bm{\cF}'_{\hat{B}\hat{C}\hat{D}} = \bm{\cD}_{\hat{A}} \bm{\cF}_{\hat{B}\hat{C}\hat{D}} \,. 
\end{split}
\end{align}
Then, as is clear from \eqref{eq:cR-total} and \eqref{eq:cG-total}, $\cR$ and $\cG^{\hat{A}\hat{B}}$ are also invariant,
\begin{align}
 \cR' =\cR \,,\qquad \cG'^{\hat{A}\hat{B}}=\cG^{\hat{A}\hat{B}}\,.
\end{align}

Note that if we define the quantity
\begin{align}
 \cG^{AB} \equiv \hat{V}_{\hat{C}}{}^A\,\hat{V}_{\hat{D}}{}^B\,\cG^{\hat{C}\hat{D}}\,,
\end{align}
we can clearly see that it transforms as $\cG^{AB} \to (C^{-\rmT}\,\cG\,C^{-1})^{AB}=C^A{}_C\,\cG^{CD}\,C_D{}^B$\,. 
This is precisely the $\cG^{AB}$ discussed in \eqref{eq:flux-vielbein-eom} when the untwisted fields are constant. 

In order to show the covariance of $S_{MN}$\,, it is convenient to use the relation
\begin{align}
 \hat{V}^{\bar{a} C}\,V^{b D}\,U_C{}^M\,U_D{}^N\,\cS_{MN} 
 = \cG^{\bar{a}b}
 = \cG'^{\bar{a}b}
 = \hat{V}'^{\bar{a} C}\,V'^{b D}\,U'_C{}^M\,U'_D{}^N\,\cS'_{MN} \,.
\end{align}
From $\hat{V}'^{\bar{a} C}=\hat{V}^{\bar{a} D}\,(C^{-1})_D{}^C$\,, we find that
\begin{align}
 U_A{}^M\,U_B{}^N\,\cS_{MN} = (C^{-1})_A{}^C\,(C^{-1})_B{}^D\,U'_C{}^M\,U'_D{}^N\,\cS'_{MN}\,. 
\end{align}
Namely, we obtain
\begin{align}
 \cS'_{MN} = (h\,\cS\,h^\rmT)_{MN}\,,\qquad 
 h_M{}^N \equiv U'_M{}^A\,C_A{}^B\,U_B{}^N \,. 
\end{align}
Therefore, the generalized Ricci tensor transforms covariantly in the same manner as $\cH_{MN}$\,. 

\subsection*{R--R sector}

The R--R fields in the approach of \cite{1107.0008} are defined as
\begin{align}
 \ket{F} = \sum_p \frac{1}{p!}\, F_{m_1\cdots m_p}\,\Gamma^{m_1\cdots m_p}\,\ket{0}\,,\qquad 
 \Gamma^{m_1\cdots m_p}\equiv \Gamma^{[m_1}\,\cdots \Gamma^{m_p]}\,.
\end{align} 
Here, the gamma matrix $(\Gamma^M)\equiv (\Gamma^m,\,\Gamma_m)$ is real and satisfies $(\Gamma^M)^\rmT = \Gamma_M$ and
\begin{align}
 \{\Gamma^M,\,\Gamma^N\} = \eta^{MN} \qquad \bigl(\ \Leftrightarrow \{\Gamma^m,\,\Gamma_n\} = \delta^m_n\,,\quad \{\Gamma^m,\,\Gamma^n\} = 0 = \{\Gamma_m,\,\Gamma_n\}\ \bigr) \,. 
\end{align}
By considering $\Gamma^m$ and $\Gamma_n$ as the creation and annihilation operator, we define the Clifford vacuum $\ket{0}$ as $\Gamma_m\,\ket{0}=0$ that is normalized as $\langle 0\vert 0\rangle = 1$ where $\bra{0}\equiv \ket{0}^\rmT$. 
We define the charge conjugation matrix as
\begin{align}
\begin{split}
 &\cC \equiv (\Gamma^0\pm \Gamma_0) \cdots (\Gamma^{D-1}\pm \Gamma_{D-1})\qquad (D:\text{even/odd})\,, 
\\
 &\cC\,\Gamma^A\,\cC^{-1} = -(\Gamma^A)^\rmT \,,\qquad \cC^{-1}= (-1)^{\frac{D(D+1)}{2}}\cC = \cC^\rmT \,,
\end{split}
\end{align}
and introduce the notations
\begin{align}
 \bra{F} \equiv (\ket{F})^\rmT = \sum_p \frac{1}{p!}\, F_{m_1\cdots m_p}\,\bra{0}\,(\Gamma^{m_p})^\rmT \cdots (\Gamma^{m_1})^\rmT \,,\qquad
 \overline{\bra{F}} \equiv \bra{F}\,\cC^\rmT\,.
\end{align}

In type IIA/IIB theory, the R--R field strength satisfies
\begin{align}
 \Gamma^{11}\,\ket{F} = \pm \ket{F} \qquad (\text{type IIA/IIB})\,,
\end{align}
where the chirality operator is defined by
\begin{align}
 \Gamma^{11} \equiv (-1)^{N_F}\,,\qquad N_F\equiv \Gamma^m\,\Gamma_m \,. 
\end{align}
The Bianchi identity is given by
\begin{align}
 \sla{\partial} \ket{F} = 0\,,\qquad \sla{\partial} \equiv \Gamma^M\,\partial_M\,,
\end{align}
where the nilpotency $\sla{\partial}^2 =0$ is ensured by the section condition.
The R--R potential (in the A-basis) is defined through
\begin{align}
 \ket{F} = \sla{\partial} \ket{A} \,,\qquad
 \ket{A} = \sum_p \frac{1}{p!}\, A_{m_1\cdots m_p}\,\Gamma^{m_1\cdots m_p}\,\ket{0}\,,
\end{align}
and in terms of differential form we have\footnote{In GSE, the R--R fields have the dual-coordinate dependence as $A=\Exp{-I^m\,\tilde{x}_m}\bar{A}(x^m)$ and $F=\Exp{-I^m\,\tilde{x}_m}\bar{F}(x^m)$\,, and the relation $F=\bm{d}A$ reproduces $\bar{F}= \Exp{I^m\,\tilde{x}_m}\bm{d}A = \rmd \bar{A}-\iota_I\bar{A}$\,. By considering $\{\bar{A},\,\bar{F}\}$ as the dynamical fields, we obtain the relation \eqref{eq:GSE-F=dA}. See \cite{1703.09213} for more detail.}
\begin{align}
 F = \bm{d} A \,,\qquad \bm{d} \equiv \rmd x^m\wedge \partial_m + \iota_m\,\tilde{\partial}^m\,.
\label{eq:bmd-def}
\end{align}

Under an $\OO(D,D)$ transformation,
\begin{align}
 \cH_{MN} \to \cH'_{MN} = (h\,\cH\,h^\rmT)_{MN} \,,
\end{align}
the $\OO(D,D)$ spinors, $\ket{F}$ and $\ket{A}$\,, transform as
\begin{align}
 \ket{F} \to \ket{F'} = S_h\,\ket{F}\,,\qquad 
 \ket{A} \to \ket{A'} = S_h\,\ket{A}\,,
\end{align}
where $S_h$ is defined through
\begin{align}
 S_h\,\Gamma_M\,S_h^{-1} = (h^{-1})_M{}^N \, \Gamma_N \,. 
\end{align}
We also define the corresponding operation $\mathbb{S}_h$ acting on the polyform $F$ as
\begin{align}
 S_h\,\ket{F} = \ket{\mathbb{S}_h\,F}\,. 
\end{align}
The concrete expressions of $S_h$ and $\mathbb{S}_h$ for the $\GL(D)$-, $B$-, and $\beta$-transformation are as follows:
\begin{alignat}{2}
 &S_{h_M} = \Exp{\frac{1}{2}\,\rho_m{}^n\,[\Gamma^m,\,\Gamma_n]} = \frac{1}{\sqrt{\abs{\det M}}}\,\Exp{\rho_m{}^n\,\Gamma^m\,\Gamma_n} \qquad \bigl(\rho \equiv \ln M \bigr)
\\
 &\qquad\! \leftrightarrow\ 
 h_M = \begin{pmatrix}
 M &0 \\
 0& M^{-\rmT} 
\end{pmatrix} \ \leftrightarrow\ \mathbb{S}_{h_M} \, F \equiv F^{(M)} \quad \bigl(F^{(M)}_{m_1\cdots m_p} \equiv M_{m_1}{}^{n_1}\cdots M_{m_p}{}^{n_p}\,F_{n_1\cdots n_p}\bigr) \,,
\nn\\
 &S_{h_\omega} = \Exp{\frac{1}{2}\,\omega_{mn}\,\Gamma^{mn}} \ \leftrightarrow\ 
 h_\omega = \begin{pmatrix}
 \bm{1}_d & \omega \\
 0& \bm{1}_d 
\end{pmatrix} \ \leftrightarrow\ \mathbb{S}_{h_\omega} = \Exp{(\frac{1}{2}\,\omega_{mn}\,\rmd x^m\wedge\rmd x^n) \wedge}\,,
\\
 &S_{h_\chi} = \Exp{\frac{1}{2}\,\chi^{mn}\,\Gamma_{mn}} \ \leftrightarrow\ 
 h_\chi = \begin{pmatrix}
 \bm{1}_d & 0 \\
 \chi & \bm{1}_d 
\end{pmatrix} \ \leftrightarrow\ \mathbb{S}_{h_\chi} = \Exp{\frac{1}{2}\,\chi^{mn}\,\iota_m\iota_n}\,. 
\end{alignat}
The factorized $T$-duality along the $x^z$-direction is generated by
\begin{align}
 S_{h_z} = \bigl(\Gamma^z - \Gamma_z\bigr)\,\Gamma^{11} \ &\leftrightarrow \ 
 h_z = \begin{pmatrix}
 \bm{1}_d-e_z & e_z \\
 e_z & \bm{1}_d-e_z 
\end{pmatrix}
\nn\\
 &\leftrightarrow \ \mathbb{S}_{h_z}\, F = F\, \wedge \rmd x^z + F \vee \rmd x^z \,. 
\end{align}
In fact, the R--R field $\ket{F}$ is as an $\OO(D,D)$ spinor density with weight $1/2$\,.\footnote{This can also be observed from the definition of the generalized Lie derivative
\begin{align*}
 \gLie_V \ket{F} = \bigl(V^M\,\partial_M + \partial_M V_N \,\Gamma^{MN}\bigr)\,\ket{F} + \frac{1}{2}\,\partial_M V^M \,\ket{F}\,.
\end{align*}}
Correspondingly, under the $\GL(D)$ transformation, the above $S_{h_M}$ needs to be corrected as
\begin{align}
 \widetilde{S}_{h_M}\, \ket{F} \equiv \sqrt{\abs{\det M}}\, S_{h_M}\, \ket{F} = \ket{\mathbb{S}_{h_M}F} 
\end{align}
when acting on the $\OO(D,D)$ spinor density. 
We can absorb the extra factor $\sqrt{\abs{\det M}}$ into the DFT dilaton by considering a weightless $\OO(D,D)$ spinor, $\ket{\cF} \equiv \Exp{d}\, \ket{F}$\,. 

For later convenience, we define $S_{\bm{g}}$ and $\cK$ as
\begin{align}
\begin{split}
 &S_{\bm{g}}\,\Gamma_M\, S_{\bm{g}}^{-1} = -\bm{g}_M{}^N \, \Gamma_N \,,\qquad 
 (\bm{g}_{MN})\equiv \begin{pmatrix} g_{mn} & 0 \\ 0 & g^{mn} \end{pmatrix},
\\
 &\cK\,\Gamma_M\, \cK^{-1} = - \cH_M{}^N\, \Gamma_N \,,\qquad 
 \cK = \Exp{\bm{B}} S_{\bm{g}} \Exp{-\bm{B}}\,, \qquad \bm{B} \equiv \frac{1}{2}\,B_{mn}\,\Gamma^{mn}\,. 
\end{split}
\end{align}
By using the property $S_{\bm{g}}\,\ket{0}=\sqrt{\abs{g}}\,\ket{0}$\,, we can show that
\begin{align}
 \overline{\bra{\alpha}}\,S_{\bm{g}}\,\ket{\beta}
 = -\sqrt{\abs{g}}\,\sum_p \frac{1}{p!}\,g^{m_1n_1}\cdots g^{m_pn_p}\,\alpha_{m_1\cdots m_p}\,\beta_{n_1\cdots n_p} = \overline{\bra{\beta}}\,S_{\bm{g}}\,\ket{\alpha}\,,
\end{align}
for $\OO(D,D)$ spinors $\ket{\alpha}$ and $\ket{\beta}$\,. 
Moreover, the self-duality relation \eqref{eq:RR-self} can be expressed as
\begin{align}
 \ket{F} = \cK \, \ket{F} \,. 
\label{eq:self-duality}
\end{align}
We also define the correspondent of $\cK$ for the untwisted metric as
\begin{align}
 \cK = S_U\,\hat{\cK}\, S_U^{-1} \,, \qquad 
 \hat{\cK}\,\Gamma_A\, \hat{\cK}^{-1} = -\hat{\cH}_A{}^B \, \Gamma_B \,,
\end{align}
where $\Gamma_A\equiv U_A{}^M\, S_{U}^{-1}\,\Gamma_M\,S_U=\delta_A^M\,\Gamma_M$\,. 

The bosonic part of the Lagrangian in type II DFT is
\begin{align}
 \cL = \Exp{-2\,d} \cS + \frac{1}{4}\,\overline{\bra{F}}\, \cK\, \ket{F} \,,
\end{align}
and the equations of motion for $\{\cH_{MN},\,d,\,\ket{A}\}$ are summarized as
\begin{align}
\begin{split}
 &\cS_{MN} = \cE_{MN}\,,\qquad \cS=0\,,\qquad \sla{\partial}\,\cK\,\ket{F} = 0\,,
\\
 &\cE_{MN} \equiv -\frac{1}{4} \Exp{2\,d} \Bigl[\overline{\bra{F}} \,\Gamma_{(M}\,\cK\,\Gamma_{N)}\, \ket{F} + \frac{1}{2}\,\cH_{MN}\,\overline{\bra{F}}\,\cK\,\ket{F}\Bigr]\,.
\end{split}
\end{align}
Under the self-duality relation \eqref{eq:self-duality}, the equation of motion for the R--R field $\sla{\partial}\,\cK\,\ket{F} = 0$ is precisely the Bianchi identity $\sla{\partial}\,\ket{F} = 0$\,. 

In the gauged DFT, we consider the reduction ansatz (see \cite{1109.4280,1304.1472,1705.08181,1706.08883}),
\begin{align}
 \ket{F} = \Exp{-\sfd(x^I)}S_{U(x^I)} \ket{\hat{\cF}(y)}\,,
\end{align}
and assume that $\ket{\hat{\cF}(y)}$ satisfies the condition $\cD_A \ket{\hat{\cF}(y)} = \partial_A \ket{\hat{\cF}(y)}$\,, similar to \eqref{eq:untwisted-condition}. 
In the case of the twist matrix $U = R \, \bm{\Pi}$\,, $\ket{\hat{\cF}}$ is explicitly given by
\begin{align}
 \ket{\hat{\cF}} &= \Exp{\sfd} S_{\bm{\Pi}^{-1}} S_{R^{-1}} \ket{F} 
\nn\\
 &= \frac{\Exp{\sfd}}{\sqrt{\abs{\det(e_\gga^m)}}}\, \Exp{\frac{1}{2}\,\Pi^{\gga\ggb}\,\Gamma_{\gga\ggb}} \sum_p \frac{1}{p!}\, e_{\gga_1}^{m_1}\cdots e_{\gga_p}^{m_p}\,F_{m_1\cdots m_p}\,\Gamma^{\gga_1\cdots \gga_p}\,\ket{0} \,.
\end{align}
In terms of the differential form, this reads as
\begin{align}
 \hat{\cF} = \frac{\Exp{\sfd}}{\sqrt{\abs{\det(e_\gga^m)}}}\,\mathbb{S}_{U^{-1}}\,F\,.
\label{eq:cF-hat-form}
\end{align}

In terms of the untwisted field, the self-duality relation can be expressed as
\begin{align}
 \ket{\hat{\cF}} = \hat{\cK}\, \ket{\hat{\cF}} \,.
\end{align}
We can clearly see that this relation is preserved under the PL $T$-plurality transformation
\begin{align}
 \hat{\cK} \to S_{C} \,\hat{\cK}\,S_{C^{-1}}\,,\qquad 
 \ket{\hat{\cF}} \to S_C\,\ket{\hat{\cF}}\,. 
\label{eq:T-plurality-RR}
\end{align}
Now, let us show the covariance of the Bianchi identity. 
From the reduction ansatz, we have
\begin{align}
 &0= \sla{\partial} \ket{F} = \Exp{-\sfd} S_{U} \, \bigl(\sla{\partial} - \Gamma^A\, \cD_A \sfd + S_{U}^{-1} \, \sla{\partial} S_{U} \bigr)\,\ket{\hat{\cF}}
\nn\\
 &\Leftrightarrow\ \Bigl(\sla{\partial} - \frac{1}{2}\, \Gamma^A\, \cF_A + \frac{1}{3!}\,\Gamma^{ABC}\,\cF_{ABC} \Bigr)\,\ket{\hat{\cF}} = 0 \,,
\label{eq:BI}
\end{align}
where we have used the identity
\begin{align}
 S_{U}^{-1} \, \sla{\partial} S_{U} = \frac{1}{3!}\,\cF_{ABC}\,\Gamma^{ABC} - \frac{1}{2}\,\Omega^B{}_{AB}\,\Gamma^A\,.
\end{align}
We require ``the Bianchi identity'' for the untwisted field $\sla{\partial} \ket{\hat{\cF}} = 0$\,, and then the Bianchi identity \eqref{eq:BI} becomes an algebraic relation
\begin{align}
 \Bigl(\frac{1}{3!}\,\Gamma^{ABC}\,\cF_{ABC} - \frac{1}{2}\, \Gamma^A\, \cF_A\Bigr)\,\ket{\hat{\cF}} = 0 \,.
\end{align}
This is manifestly covariant under the PL $T$-plurality transformation. 
Since the Bianchi identity and the self-duality relation are covariantly transformed, the equation of motion for the R--R field is also satisfied in the dualized background. 

Finally, we show the covariance of the energy--momentum tensor. 
To this end, we define
\begin{align}
 \hat{\cE}_{AB} \equiv U_A{}^M \,\,U_B{}^N\, \cE_{MN} 
 = -\frac{1}{4}\Exp{2\,\hat{d}}\,\Bigl[\overline{\bra{\hat{\cF}}} \,\Gamma_{(A} \,\hat{\cK}\,\Gamma_{B)} \, \ket{\hat{\cF}} + \frac{1}{2}\,\hat{\cH}_{AB}\,\overline{\bra{\hat{\cF}}}\,\hat{\cK}\,\ket{\hat{\cF}}\Bigr]\,.
\end{align}
Under the PL $T$-plurality \eqref{eq:T-plurality-RR}, we can easily show that
\begin{align}
 U'_A{}^M\,U'_B{}^N\,\cE'_{MN} = \hat{\cE}'_{AB} = C_A{}^C\,C_B{}^D \,\hat{\cE}_{CD} = C_A{}^C\,C_B{}^D \,U_C{}^M\,U_D{}^N\, \cE_{MN}\,,
\end{align}
and, similar to the generalized Ricci tensor, we obtain
\begin{align}
 \cE'_{MN} = (h\,\cE\,h^\rmT)_{MN}\,,\qquad 
 h_M{}^N \equiv U'_M{}^A\,C_A{}^B\,U_B{}^N \,. 
\end{align}
This complete the proof of the covariance of the equations of motion.

\end{document}